\documentclass[sigplan,screen]{acmart}
\usepackage{subcaption}
\acmSubmissionID{27}
\usepackage{todonotes}
\usepackage{url}
\usepackage{enumitem}
\usepackage{graphicx}
\usepackage{caption}
\usepackage{amsmath}
\usepackage{tabu}
\usepackage{multirow}
\usepackage{hyperref}
\usepackage{adjustbox}
\usepackage{mathptmx}
\usepackage{amsfonts}
\usepackage{algpseudocode}
\usepackage{pgfplots}
\pgfplotsset{compat=1.12}
\usepackage{soul}
\usepackage{footnote}
\usepackage{lipsum}
\usepackage[ruled,vlined, linesnumbered]{algorithm2e}
\usepackage{subcaption}
\usepackage{listings}
\colorlet{punct}{red!60!black}
\definecolor{background}{HTML}{FFFFFF}
\definecolor{delim}{RGB}{20,105,176}
\colorlet{numb}{magenta!60!black}

\lstdefinelanguage{json}{
    basicstyle=\footnotesize\ttfamily,
    numbers=left,
    numberstyle=\scriptsize,
    xleftmargin=2.3em,
    xrightmargin=0.5em,
    framexleftmargin=1.9em,
    stepnumber=1,
    numbersep=8pt,
    showstringspaces=false,
    breaklines=true,
    frame=single,
    backgroundcolor=\color{background},
    literate=
     *{0}{{{\color{numb}0}}}{1}
      {1}{{{\color{numb}1}}}{1}
      {2}{{{\color{numb}2}}}{1}
      {3}{{{\color{numb}3}}}{1}
      {4}{{{\color{numb}4}}}{1}
      {5}{{{\color{numb}5}}}{1}
      {6}{{{\color{numb}6}}}{1}
      {7}{{{\color{numb}7}}}{1}
      {8}{{{\color{numb}8}}}{1}
      {9}{{{\color{numb}9}}}{1}
      {:}{{{\color{punct}{:}}}}{1}
      {,}{{{\color{punct}{,}}}}{1}
      {\{}{{{\color{delim}{\{}}}}{1}
      {\}}{{{\color{delim}{\}}}}}{1}
      {[}{{{\color{delim}{[}}}}{1}
      {]}{{{\color{delim}{]}}}}{1},
}

\DeclareRobustCommand*{\IEEEauthorrefmark}[1]{\raisebox{0pt}[0pt][0pt]{\textsuperscript{\footnotesize\ensuremath{\ifcase#1\or *\or \dagger\or \ddagger\or%
    \mathsection\or \mathparagraph\or \|\or **\or \dagger\dagger%
    \or \ddagger\ddagger \else\textsuperscript{\expandafter\romannumeral#1}\fi}}}}

\AtBeginDocument{%
  \providecommand\BibTeX{{%
    \normalfont B\kern-0.5em{\scshape i\kern-0.25em b}\kern-0.8em\TeX}}}

\copyrightyear{2023} 
\acmYear{2023} 
\setcopyright{acmcopyright}\acmConference[PPoPP '23]{The 28th ACM SIGPLAN Annual Symposium on Principles and Practice of Parallel Programming}{February 25-March 1, 2023}{Montreal, QC, Canada}
\acmBooktitle{The 28th ACM SIGPLAN Annual Symposium on Principles and Practice of Parallel Programming (PPoPP '23), February 25-March 1, 2023, Montreal, QC, Canada}
\acmPrice{15.00}
\acmISBN{979-8-4007-0015-6/23/02}

\begin{document}

\title{Exploring the Use of WebAssembly in HPC}

\author{Mohak Chadha, Nils Krueger, Jophin John, Anshul Jindal, Michael Gerndt}
\email{  {firstname.lastname}@tum.de}
\affiliation{%
 \institution{Chair of Computer Architecture and Parallel Systems,  Technische Universit{\"a}t M{\"u}nchen, Germany}
}

\author{Shajulin Benedict}
\email{shajulin@iiitkottayam.ac.in}
\affiliation{%
 \institution{Department of Computer Science and Engg., Indian Institute of Information Technology Kottayam, Kerala}
}

\renewcommand{\shortauthors}{Mohak Chadha et al.}

\begin{abstract}

Containerization approaches based on \emph{namespaces} offered by the Linux kernel have seen an increasing popularity in the HPC community both as a means to isolate applications and as a format to package and distribute them. However, their adoption and usage in HPC systems faces several challenges. These include difficulties in unprivileged running and building of scientific application container images directly on HPC resources, increasing heterogeneity of HPC architectures, and access to specialized networking libraries available only on HPC systems. These challenges of container-based HPC application development closely align with the several advantages that a new universal intermediate binary format called \emph{WebAssembly} (Wasm) has to offer. These include a lightweight userspace isolation mechanism and portability across operating systems and processor architectures. In this paper, we explore the usage of Wasm as a distribution format for MPI-based HPC applications. To this end, we present \emph{MPIWasm}, a novel Wasm embedder for MPI-based HPC applications that enables high-performance execution of Wasm code, has low-overhead for MPI calls, and supports high-performance networking interconnects present on HPC systems. We evaluate the performance and overhead of \emph{MPIWasm}  on a production HPC system and AWS Graviton2 nodes using standardized HPC benchmarks. Results from our experiments demonstrate that \emph{MPIWasm} delivers competitive native application performance across all scenarios. Moreover, we observe that Wasm binaries are $139.5$x smaller on average as compared to the statically-linked binaries for the different standardized benchmarks.





\end{abstract}

\ccsdesc[500]{Software and its engineering~Process management}

\keywords{WebAssembly, Wasmer, Wasm, MPI, HPC}

\maketitle
\vspace{-1em}

\section{Introduction}
\label{sec:intro}
Linux containers, due to their portability and high availability, have become the de-facto standard for developing, testing, and deploying a wide range of applications from enterprise to web services in cloud environments~\cite{cloud_container}. This is because containers enable users to package their application along with its custom software dependencies as a single unit into easy-to-deploy images. Motivated by their popularity in the cloud, containers have also seen a growing interest in the HPC community~\cite{xavier2013performance, abraham2020use, rudyy2019containers}. For HPC systems, containers provide flexibility to users and allow them to define custom software stacks, i.e., \emph{user-defined software stack} (UDSS) for their large-scale scientific applications. Moreover, they enable easy, reliable, and verifiable environments that can be reproduced in the future. To this end, several HPC-focused containerization solutions, such as Charliecloud~\cite{priedhorsky2017charliecloud}, 
Shifter~\cite{gerhardt2017shifter}, Singularity~\cite{kurtzer2017singularity}, Podman~\cite{gantikow2020rootless}, and Sarus~\cite{benedicic2019sarus} 
have been introduced. In contrast to previous approaches, this paper investigates using a new novel technology called \emph{WebAssembly} (Wasm)~\cite{webassembly}, \textit{dubbed} as an alternative to Linux containers~\cite{alttocontainers}, for packaging and distributing HPC applications.

Despite their increasing popularity, the adoption and usage of containers in HPC systems is still significantly limited~\cite{nersc}. This can be attributed to the several challenges commonly faced by users in running and building container images for their applications on HPC systems. For executing containers, most containerization solutions require \texttt{root} privileges which are not possible for normal HPC users due to shared filesystems and their UNIX permissions in HPC. While HPC-focused containerization solutions such as Singularity~\cite{kurtzer2017singularity} and Podman~\cite{gantikow2020rootless} support \textit{rootless-containers} through \texttt{fakeroot}~\cite{fakeroot}, their current implementations do not support distributed filesystems such as GPFS commonly found on HPC systems~\cite{podmanlimitations, singularitylimitatsion}. Moreover, as argued by~\cite{minimizingpriv}, building Open Container Initiative (OCI)~\cite{oci} compliant  container images on HPC resources by unprivileged (normal) users where the applications will eventually run is significantly hard and requires support from the supercomputing center. This is because most container building solutions such as Docker~\cite{docker} also require root privileges. As a result, most users use their own local systems for building/developing their application container images and then transfer the built image to a login/front-end node of an HPC system for execution. However, this scenario leads to several problems in container-based HPC application development. First, HPC nodes are becoming more heterogeneous~\cite{top500_june} with different processor architectures such as \texttt{x86\_64} or \texttt{aarch64} and have specialized accelerators such as GPUs.  As application performance is critical in HPC, compiling an application using the specific microarchitectural features of a  particular processor is significantly important.  While building container images for multiple platforms either by cross-compiling HPC applications or by emulation with QEMU is possible with plugins such as \texttt{build-x}~\cite{dockerbuildx}, it is not widely supported by HPC application build procedures and requires the presence of specific Linux kernel features (\texttt{binfmt\_misc}~\cite{binft_misc}). Moreover, testing and developing HPC applications offers insights only on the target system. In addition, most container images can range from several MiBs to several GiBs. As a result, frequent network transfers from the local to the HPC system can be cumbersome. Second, building HPC applications requires access to specialized networking libraries and licenses to compilers that are not available on the local user systems. Finally, while  different containerization solutions have almost no impact on the performance of the containerized application~\cite{priedhorsky2017charliecloud, containerhpcnoperf, ruhela2021characterizing}, building high-performant HPC application container images is non-trivial, involves a steep 
learning curve, and requires knowledge about specific MPI library versions (e.g., OpenMPI~\cite{openmpi} 4.0) and high performance network interconnect hardware (e.g., Intel OmniPath~\cite{omnipath}) and libraries (e.g., Intel Performance Scaled Messaging~\cite{psm2}) present on the target system. These challenges of container-based HPC application development closely align with the several advantages and core problems that Wasm~\cite{webassembly} aims to solve.  

Wasm is a low-level, statically typed universal binary instruction format for memory-safe, sandboxed execution in a virtual machine. It offers portability across modern processor architectures and operating systems, fast execution, and a low-level memory model~\cite{webassembly}. Although originally meant for execution in Web browsers, due to its simplicity and generality, Wasm has seen widespread adoption and usage in non-Web domains such as serverless computing~\cite{faasm}, edge computing~\cite{serverlessedge, sledge}, and Internet of Things~\cite{warduino}.  It does not require garbage collection and is designed to be a universal compilation target with mature support for programming languages with an LLVM~\cite{llvm} front-end such as C, C++, C\#, and Rust~\cite{Blazor, rusttowasm, emscripten, gcc_wasm, clang_wasm}.

\begin{figure}[t]
\centering
\includegraphics[width=0.70\columnwidth]{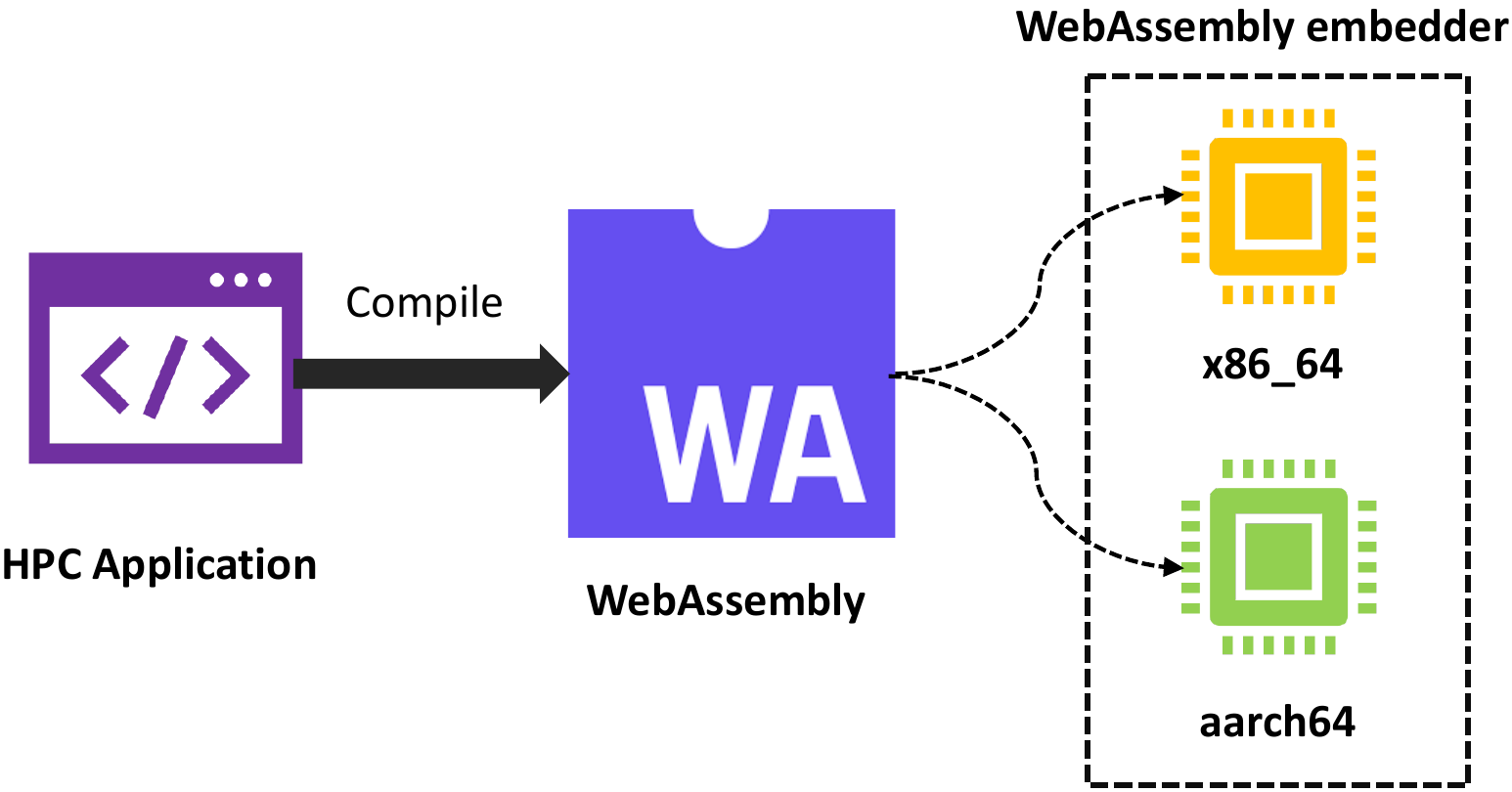}
\caption{An HPC application can be compiled to \emph{WebAssembly} and distributed to multiple platforms where it can be executed efficiently by a supporting \emph{WebAssembly embedder}.}
\label{fig:mainidea}
\end{figure}

Figure~\ref{fig:mainidea} demonstrates a general workflow for using Wasm in HPC. Developers can compile their HPC applications to Wasm once on their local systems ahead-of-time (AoT) and distribute it across multiple platforms instead of distributing source code or building application containers. Typically, Wasm binaries have a smaller size as compared to native \texttt{x86\_64} binaries~\cite{webassembly, notsofast, understandingperf}. Following this, the resulting binary can be executed on any platform using a standalone \emph{Wasm embedder}~\cite{webassembly}. The \emph{Wasm embedder} serves two major purposes. First, it provides an isolated execution environment for running a Wasm binary on a platform. In contrast to container-based approaches that utilize different Linux \emph{namespaces}~\cite{minimizingpriv} for isolation and security, Wasm provides lightweight isolation at the application level based on software fault isolation (SFI)~\cite{sfi} and control flow integrity (\S\ref{sec:sec_model}). Second, it is responsible for compiling Wasm binaries to native machine code, either by using Just-in-Time (JIT) engines at the time of execution, or AoT by using the same JIT engines or AoT compilers. Note that, Wasm binaries can be executed by normal users and are completely unprivileged.  Several open-source standalone embedders such as \texttt{Wasmer}~\cite{wasmer}, \texttt{Wasmtime}~\cite{wasmtime}, and \texttt{Wasm3}~\cite{wasm3} are currently available. However, none of them support the execution of HPC applications.


As the first step towards bringing Wasm to the HPC ecosystem, we only focus on MPI-based~\cite{mpi_standard} HPC applications in this paper. We chose MPI due to its understanding and influence in the HPC community~\cite{bernholdt2017survey}. Towards this, our key contributions are:
\begin{itemize}
    \item We implement and present \emph{MPIWasm}, a novel \emph{Wasm embedder} for MPI-based HPC applications based on \texttt{Wasmer}~\cite{wasmer}. \emph{MPIWasm} enables high performance execution of Wasm code, has low-overhead for MPI calls through zero-copy memory operations, and supports high-performance networking interconnects such as Intel OmniPath~\cite{omnipath}.
    \item We demonstrate with extensive experiments the low-overhead and performance of \emph{MPIWasm} using standardized HPC benchmarks on a production HPC system and AWS Graviton2~\cite{graviton} nodes based on the \texttt{x86\_64} and the \texttt{aarch64} architectures respectively.
    \item We elaborate on the different possible future directions for using Wasm in the HPC ecosystem.
\end{itemize}

The rest of this paper is structured as follows. \S\ref{sec:background} provides a detailed overview on Wasm. In \S\ref{sec:mpiwasm}, we describe our embedder \emph{MPIWasm} in detail. Our experimental results are presented in \S\ref{sec:results}. In \S\ref{sec:futurehpcwasm}, we describe the different possible directions for using Wasm in the HPC ecosystem. \S\ref{sec:related_work} describes some previous approaches related to our work. In \S\ref{sec:conclusion}, we conclude the paper and present an outlook.


\section{A primer on WebAssembly}
\label{sec:background}

\lstset{ %
    basicstyle=\ttfamily\footnotesize,
    commentstyle=\color{blue}\ttfamily,
    frame=single,
    keywordstyle=\color{red}\ttfamily,
    language=Bash,
    showstringspaces=false,
    numbers=left,
    xleftmargin=2.6em,
    framexleftmargin=2.7em,
    morekeywords={blue},
    morestring=[s][\color{Gray}]{<}{>},
    morestring=[s][\color{OrangeRed}]{\ -}{\ },
    morestring=[s][\color{OrangeRed}]{*}{\ },
    morestring=[s][\color{OrangeRed}]{|}{\ },
    morestring=[s][\color{OrangeRed}]{\&}{\ },
}

\definecolor{bluekeywords}{rgb}{0.13,0.13,1}
\definecolor{greencomments}{rgb}{0,0.5,0}
\definecolor{turqusnumbers}{rgb}{0.17,0.57,0.69}
\definecolor{redstrings}{rgb}{0.5,0,0}

\lstdefinelanguage{WebAssembly}{
  sensitive=true,
  otherkeywords={},
  morekeywords=[1]{i32,f32,i64,f64},
  keywordstyle={[1]\color{violet}},
  morekeywords=[2]{0},
  keywordstyle={[2]\color{violet}},
  morekeywords=[3]{add,const}
  keywordstyle={[3]\color{bluemunsell}},
  morekeywords=[4]{},
  keywordstyle={[4]\color{candypink}},
  morekeywords=[5]{module, func, param, result, global, get_global, mut, set_global, export, import, memory, data, get_local, set_local, elem, table, call,call_indirect, type},
  keywordstyle={[5]\color{bluekeywords}},
  morekeywords=[6]{=,;},
  keywordstyle={[6]\color{britishracinggreen}},
  morekeywords=[7]{(,),[,],.},
  keywordstyle={[7]\color{black}},
  numberstyle=\tiny\color{black},
  rulecolor=\color{black},
  morecomment=**[l][\itshape\color{greencomments}]{;;, ...},
}

\subsection{WebAssembly Overview}
\label{sec:webassemblyover}
WebAssembly (Wasm) was introduced in 2015 as an alternative to JavaScript for web-browser based applications. It superseded \texttt{asm.js}~\cite{asmjs}, a previous attempt by Mozilla which focused on a subset of Javascript code that can be optimized AoT. 

When an application is compiled to Wasm, the resulting binary is called a \emph{module}. Wasm modules contain function definitions, declarations of global variables, tables, and a linear memory address space. All of the application code in Wasm is organized in functions. The conceptual machine in Wasm is stack-based and does not contain registers, therefore all instructions pop their operands from the stack of the machine. However, since application control flow is an explicit part of the module and Wasm operations are typed, it is possible to statically predict the layout of the stack at any point in the program which allows compilers to translate the stack semantics to a register-based instruction set. Similar to other higher-level programming languages, Wasm allows the definition of global variables that are not scoped to a specific function or block. Tables in Wasm modules are used for storing references to functions~\cite{webassembly}. The Wasm ISA currently supports only four data types for variables: (i) \texttt{i32}, 32-bit integers, (ii) \texttt{i64}, 64-bit integers, (iii) \texttt{f32}, 32-bit IEEE 754 floating point numbers, and (iv) \texttt{f64} 64-bit IEEE 754 floating point numbers. For constructing, complex types a combination of these basic types is commonly used.

Wasm provides the capability for data and code to be shared between the module and its embedder using the import/export system.  All of the function definitions that can occur in a Wasm module can be imported from the embedder instead of being defined within it. Similarly, function definitions that are present in the module can be exported
so that the embedder can utilize them (\S\ref{sec:wasi}). 







\begin{lstlisting}[caption={
    Example representation of a compiled C++ application's Wasm module using the \texttt{WASI-SDK} in WebAssembly text format (WAT)~\cite{webassembly_text_format}. Ellipses
    signify sections that are omitted for brevity.
}, float, floatplacement=t, captionpos=b, basicstyle=\ttfamily\tiny,  belowskip=-2 \baselineskip, frame=single, language=WebAssembly, label={lst:watexample}]
(type (;1;) (func (param i32) (result i32)))
...
(type (;5;) (func (param i32 i32) (result i32)))
...
(type (;14;) (func (param i32 i32 i32 i32 i32 i32) 
                        (result i32)))
(type (;15;) (func (param i32 i32 i32 i32) (result i32)))
...
(import "wasi_snapshot_preview1" "path_open" 
    (func $__wasi_path_open (type 22)))
(import "wasi_snapshot_preview1" "fd_close"
    (func $__wasi_fd_close (type 1)))
(import "wasi_snapshot_preview1" "fd_seek"
    (func $__wasi_fd_seek (type 23)))
(import "wasi_snapshot_preview1" "fd_read"
    (func $__wasi_fd_read (type 15)))
(import "wasi_snapshot_preview1" "proc_exit"
    (func $__wasi_proc_exit (type 0)))
...
(export "_start" (func $_start))
(export "memory" (memory 0))
\end{lstlisting}

\subsection{WebAssembly Security and Sandboxing Model}
\label{sec:sec_model}
Wasm utilizes software fault isolation techniques (SFI)~\cite{sfi} to sandbox the executing Wasm module. By default, a Wasm module cannot interact with the host system or perform I/O operations of any kind. Any system interaction that is to be initiated by the Wasm module's code must be done through the functions imported from the embedder (\S\ref{sec:webassemblyover}). As a result, the embedder can act both as a translation layer and as an arbiter to enforce isolation requirements. As a translator, it is possible for the embedder to provide a common interface to the Wasm module even though the underlying system may have different native interfaces, while as an arbiter it is possible for the embedder to restrict access of the Wasm module to system resources based on an application-level security policy. For instance, it is possible for the embedder to allow file I/O only to files that reside in a specific directory to isolate the Wasm module from the rest of the filesystem. While in principle similar to kernel-level system call filtering techniques such as \texttt{Seccomp-BPF}~\cite{seccomp_bpf} on Linux, performing such filtering on the application level allows to define semantically more meaningful policies.

In Wasm, all memory access is confined to a module's \emph{linear memory} which is separate from the code space. Currently, the Wasm specification~\cite{webassembly} supports 32-bit addresses to index the memory that a module has access to. While this limits a single module's memory to 4GiB, it also enables hardware accelerated bound checks of memory accesses at runtime~\cite{runtimebound}. If an embedder is a process with a 64-bit memory address space, it can safely execute an untrusted Wasm module in its memory space without requiring additional isolation by reserving a continuous range of virtual memory for the module to use. Not all pages in this range need to be mapped to physical memory, it is sufficient to only map the required number of pages to fit the amount of memory used by the module at a given point in time. This ensures that a Wasm module can only operate in its own execution environment and cannot corrupt the memory of the embedder, since any out-of-bounds memory access will result in a page fault which can then be handled by it. Moreover, since the memory instructions in Wasm's specification~\cite{webassembly} work with offsets, it is not possible to read and write to arbitrary memory locations in Wasm.

In the assembly produced by C programs, where a function call is expressed as a jump instruction to the address of the function's first instruction, a typical exploit is to change this address to take control of the program's control flow. However, such exploits are not possible with Wasm since it features \emph{control flow integrity} by enforcing structured program control flow. This is because of two reasons. First, in Wasm, a function is represented as an index in a table (\S\ref{sec:webassemblyover}) which adds an additional level of indirection to express the function address. Second, the Wasm specification prevents constructing arbitrary memory addresses~\cite{runtimebound} and the separation of the embedder and the module's memory prevents overwriting function instructions.

\begin{lstlisting}[caption={
      Excerpt of the custom \emph{MPIWasm} \texttt{mpi.h} header file.
}, float, floatplacement=t, captionpos=b, basicstyle=\ttfamily\tiny,  belowskip=-2 \baselineskip, frame=single, language=C, framexrightmargin=0cm, xrightmargin=0cm, 
   label={lst:wasiheader}]
typedef int MPI_Comm;
typedef int MPI_Datatype;
...
int MPI_Init(int* argc, char*** argv);
int MPI_Finalize(void);
int MPI_Send(
    const void* buf, int count, MPI_Datatype datatype,
    int dest, int tag, MPI_Comm comm
);
int MPI_Recv(
    void* buf, int count, MPI_Datatype datatype,
    int source, int tag, MPI_Comm comm, MPI_Status* status
);    
\end{lstlisting}

\subsection{WebAssembly System Interface}
\label{sec:wasi}
Since Wasm was originally designed for web browsers, a system interface that targets POSIX environments and enables execution of Wasm modules on them was not part of the original specification~\cite{webassembly}. To overcome this, the WebAssembly System Interface (WASI) specification~\cite{wasi_sdk} was designed. WASI specifies the interface an embedder needs to implement to execute most POSIX applications. Embedders that implement the WASI specification will be able to run any generic application compiled with the \texttt{WASI-SDK}~\cite{wasi_sdk_new}. The \texttt{WASI-SDK} includes the \texttt{clang} compiler and its own C library based on \emph{musl libc} that call WASI systemcalls imported from the embedder instead of relying on Linux systemcalls~\cite{wasi_libc}. Note that, due to the ubiquity of \emph{glibc}~\cite{glibc} on Linux systems, some applications have come to depend on glibc-specific functions or behavior. Such applications will require modifications before they can be compiled to a WASI-compliant Wasm module.

Listing~\ref{lst:watexample} shows a compiled Wasm module of a C++ application using the \texttt{WASI-SDK} in the WebAssembly text format (WAT). WAT is a human readable format that enables developers to examine the source code of a Wasm module. It can be observed that the module contains several functions with integers as parameter and return types (Lines 1-7) (\S\ref{sec:webassemblyover}), imports WASI functions (Lines 9-18), and exports its \texttt{\_start} (main function) and memory (Lines 20-21). Exporting these two definitions allows the embedder that executes this module to call
its entrypoint function and to read from and write to the module's linear memory. While the import statements on Lines 9-16 enable the Wasm module to open and read from a file, the function \texttt{proc\_exit} is used by the embedder to handle the termination of the application, e.g., by deallocating the memory reserved for the module. For the module to execute, the imported functions need to be implemented by the embedder. 







\section{MPIWasm}
\label{sec:mpiwasm}
In this section, we describe \emph{MPIWasm}, our embedder for executing Wasm modules that utilize functions from the MPI standard in detail.


\begin{lstlisting}[caption={
      WAT representation of module imports that correspond to the functions shown in Listing~\ref{lst:wasiheader}.
}, float, floatplacement=t, captionpos=b, basicstyle=\ttfamily\tiny,  belowskip=-2 \baselineskip, frame=single, language=WebAssembly, framexrightmargin=0cm, xrightmargin=0cm, 
   label={lst:watmpiexample}]
(import "env" "MPI_Init" (
    func $MPI_Init (param i32 i32) (result i32)
))
(import "env" "MPI_Finalize" (func $MPI_Finalize (result i32)))
(import "env" "MPI_Send" (
    func $MPI_Send (param i32 i32 i32 i32 i32 i32) (result i32)
))
(import "env" "MPI_Recv" (
    func $MPI_Recv (param i32 i32 i32 i32 i32 i32 i32) 
                   (result i32)
))
\end{lstlisting}

\subsection{Overview}
\label{sec:mpiwasmoverview}
The purpose of \emph{MPIWasm} is to support the execution of MPI applications compiled to Wasm on HPC systems. To facilitate its adoption and suitability in HPC environments, it (i) supports high-performance execution of MPI-based HPC applications compiled to Wasm (\S\ref{sec:highperf}), (ii) has low-overhead  for MPI calls through zero-copy memory operations (\S\ref{sec:translation}), and (iii) supports high-performance interconnects such as Infiniband~\cite{ib} and Intel OmniPath~\cite{omnipath}. These network interconnects are utilized by MPI libraries on HPC systems for  high-performance inter-rank communication. To enable the immediate support for network interconnects present on modern HPC systems, \emph{MPIWasm} links against the MPI library on the target HPC system at runtime and provides a translation layer between the Wasm module and the host\footnote{We use the term target and host interchangeably for the system on which the \emph{Wasm} module is executing.} MPI library. As a result, the developer doesn't need to be aware about the particular networking libraries or network interconnects present on the target HPC system. Depending on the particular host MPI library such as OpenMPI~\cite{openmpi} or MPICH~\cite{mpich}, \emph{MPIWasm} needs to be built separately. Both of these libraries are currently supported by \emph{MPIWasm}.


Our embedder currently supports the execution of MPI applications written in C/C++ and conforming to the MPI-2.2 standard~\cite{mpi_standard_2}. Integrating the support for MPI-3.1~\cite{mpi_standard_31} is of our interest for the future but is out of scope for this work.  We chose to focus on C/C++ applications due to the stability and maturity of the Wasm backend in the LLVM/Clang~\cite{llvm} project since \texttt{llvm-8}. As the base for \emph{MPIWasm}, we use the open-source Wasm embedder called \texttt{Wasmer}~\cite{wasmer}. Wasmer supports the execution of Wasm modules on three major platforms, i.e., Linux, Windows, and macOS, and supports both \texttt{x86\_64} and \texttt{aarch64} instruction set architectures. Moreover, it implements the WASI specification (\S\ref{sec:wasi}) and provides ergonomic mechanisms to define additional functions that are provided to the module. This dynamic extension of the embedder's functionality enables the addition of MPI functions to the functionality it provides to the Wasm module. For implementing \emph{MPIWasm}, we use the Rust programming language. This is because of two reasons. First, it provides high performance comparable to C/C++ with memory-safety~\cite{sudwoj2020rust}. Second, it has extensive support and documentation for embedding \texttt{Wasmer} and using it as a library. 








\subsection{Compiling C/C++ MPI applications to Wasm}
\label{sec:extending_wasm_system_interface}
Most MPI applications expect POSIX functionality to be available in their execution environment, for instance the ability to read from and write to file descriptors.
WASI (\S\ref{sec:wasi}) defines the WebAssembly exports that enable Wasm modules that target it to call most of the functions defined in the C standard libraries shipped on POSIX systems. Towards this, the \texttt{WASI-SDK}~\cite{wasi_sdk_new} combines the \texttt{clang} compiler and the \texttt{wasi\--libc} C library to enable the compilation of C/C++ applications that only make use of POSIX functions and no additional libraries to Wasm. The compilation of C/C++ MPI applications is not supported by the stock \texttt{WASI-SDK}. To this end, we implement a custom \texttt{mpi.h} MPI header file  and add it to the \texttt{WASI-SDK}. The header file includes the definitions for the different MPI types such as \texttt{MPI\_Op}, \texttt{MPI\_Comm}, and \texttt{MPI\_Datatype} and the definition for the \texttt{MPI\_Status} structure. Moreover, it defines the signatures for the MPI functions according to the MPI-2.2~\cite{mpi_standard_2} standard. An excerpt from the header file is shown in Listing~\ref{lst:wasiheader}. It is a reduced version of a traditional header file found with MPI libraries with most types defined as integers (\S\ref{sec:translation}). By combining our header file with the \texttt{WASI-SDK}, a C/C++ MPI application conforming to the MPI-2.2 standard can be compiled to Wasm. Moreover, to facilitate the ease-of-use and enable adoption, we implement a custom python-based tool that simplifies the entire compilation process for MPI applications. Listing~\ref{lst:watmpiexample} shows the different MPI-specific imports present in a Wasm module corresponding to the functions shown in Listing~\ref{lst:wasiheader}. \emph{MPIWasm} provides definitions for these imports to enable the execution of MPI-based HPC applications. In addition, it supports the WASI specification which enables the POSIX functionality for MPI applications.

\subsection{Executing Wasm Code with High Performance}
\label{sec:highperf}


There exist several strategies for executing Wasm modules. These include using an interpreter~\cite{wasm3}, Ahead-of-Time (AoT) compilation~\cite{wasmtime}, or Just-in-Time (JIT) compilation~\cite{wasmtime}. However, for HPC systems the most useful approach is translating the Wasm instructions (Wasm ISA) to the native instruction set of the host machine before the application is executed, i.e., AoT. Towards this, \emph{MPIwasm} builds on the code generation infrastructure provided by \texttt{Wasmer}~\cite{wasmer}. \texttt{Wasmer} currently supports three compiler backends, i.e., \texttt{Singlepass}~\cite{singlepass}, \texttt{Cranelift}~\cite{cranelift}, and \texttt{LLVM}~\cite{llvm}. The \texttt{SinglePass} compiler is designed to emit machine code in linear time and does not perform many code optimizations. The \texttt{Cranelift} compiler is completely based on Rust and is similar to \texttt{LLVM}. With \texttt{Cranelift}, the WASM instructions are first translated to the intermediate representation (IR) of \texttt{Cranelift}, i.e., (\texttt{Cranelift-IR}) which are then translated to the native instruction set of the host machine by taking microarchitecture-specific optimizations into
account. On the other hand, with \texttt{LLVM} the Wasm ISA is first translated to \texttt{LLVM-IR} followed by the generation of native machine code. \texttt{Cranelift-IR} is similar to \texttt{LLVM-IR} but at a lower level of abstraction which hinders mid-level code optimizations\footnote{A more detailed discussion between \texttt{Cranelift-IR} and \texttt{LLVM-IR} can be found here~\cite{craneliftir}.}. At the end of the compilation process, all three compilers produce a shared object, which can be loaded with a fast \texttt{dlopen} call using the \texttt{libloading} library~\cite{libloading}.

Table~\ref{tab:wasmer_compiler_comparison} shows a comparison of the compile-time and run-time performance of the three different compilers supported by \texttt{Wasmer} for the HPCG benchmark. While \texttt{LLVM} is the slowest to compile the Wasm module, it also
results in the fastest runtime performance for the HPCG application. As a result, we chose \texttt{LLVM} as the compiler backend in \emph{MPIWasm}. To offset the longer compilation times required by \texttt{LLVM} as compared to the other two compilers, we implement a caching mechanism for the generated machine code. Our caching mechanism builds on the \texttt{FileSystemCache}~\cite{filesystemcache} provided by \texttt{Wasmer}. In our implementation, we generate a hash for each Wasm module using the \texttt{Blake-3} hash function~\cite{blake3}. Moreover, we store the generated shared object from \texttt{LLVM} as the generated hash in the local filesystem. As a result, any changes to the Wasm module lead to the generation of a new hash which triggers the recompilation of the module. To this end, repeated execution of the same application on a system with \emph{MPIWasm} will not lead to recompilation overhead for execution. 




    

 \begin{table}[t]
    \caption{Comparing compile duration and performance for the different compiler backends supported by \texttt{Wasmer}~\cite{wasmer} for the HPCG~\cite{hpcg} Wasm module. The Wasm module was generated using our \texttt{WASI-SDK}~(\S\ref{sec:highperf}). The Wasm module is executed using \emph{MPIWasm} on an \texttt{x86\_64} system.}
    \label{tab:wasmer_compiler_comparison}
    \centering
    \begin{adjustbox}{width=8.5cm,center}
    \begin{tabular}{|>{\centering\arraybackslash}c|>{\centering\arraybackslash}c|>{\centering\arraybackslash}c|}
    \hline
        \textbf{Compiler} & \textbf{Compile Duration (ms)} & \textbf{Single-Core Performance (GFLOP/s)} \\\hline
  
        Singlepass~\cite{singlepass} & 52  & 0.3769  \\\hline
        Cranelift~\cite{cranelift} & 150  & 1.3240  \\\hline
        LLVM~\cite{llvm} & 2811  & 1.5426  \\\hline

    \end{tabular}
   
    \end{adjustbox}
\end{table}

\subsection{Filesystem Isolation with MPIWasm}
\label{sec:filesystemisolation}
Since in Wasm all system interactions by the application have to be performed by calling functions implemented by the embedder (\S\ref{sec:sec_model},\S\ref{sec:wasi}), it enables the embedder to place additional restrictions on their use and to employ checks on the arguments supplied to them. In \texttt{Wasmer}, all exported functions that handle file I/O perform their own permission handling that is separate from the one employed by the OS. This  in-process indirection of filesystem accesses allows \texttt{Wasmer} to present a virtual directory tree to the Wasm module that only contains directories that the module is allowed to access. In addition, access rights to individual directories can be more restrictive than the permissions granted to the user that is executing the embedder. For instance, a user can have read and write access to their home directory and all of its subdirectories, but grant read-only access to one specific subdirectory to a Wasm module executed by the embedder. \emph{MPIWasm} exposes this isolation functionality with its \texttt{-d} flag that grants read-write access to the given directory to the Wasm module. Note that the full absolute path to the exposed directories is not presented to the Wasm module. In the virtual directory tree presented to it, all of the subdirectories it has been given access to are direct children of the root
directory. This approach to mapping directory paths avoids exposure of information contained in the full path to the directories, such as a username in the case of a home directory.



\begin{figure}[t]
\includegraphics[width=0.70\columnwidth]{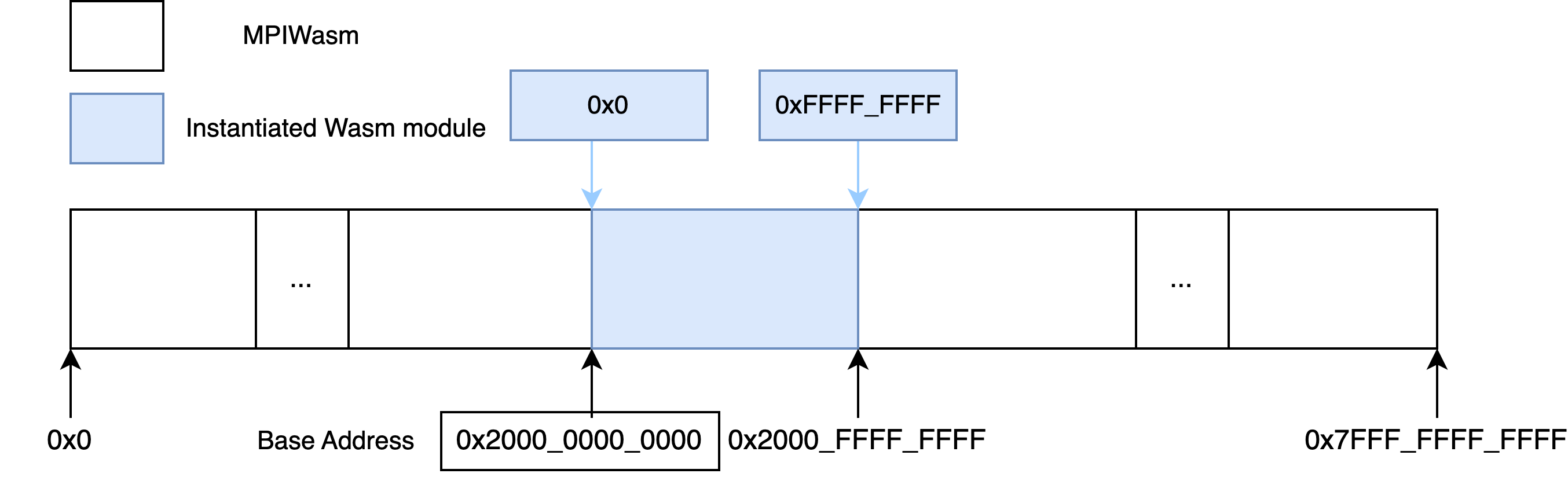}
\caption{Memory address space of \emph{MPIWasm} with an instantiated Wasm module. All memory access instructions to the Wasm module's linear address space are given offsets relative to the base address.}
\label{fig:memadressspacewasm}
\end{figure}

\subsection{Translating from Wasm to Host Memory Address}
\label{sec:memtranslation}
A major part of the Wasm security model is the separation of the host and the module's linear memory address space (\S\ref{sec:sec_model}). Since it is the responsibility of the Wasm embedder to uphold capability restrictions, protecting it's data structures from unintended or malicious access by the modules' code is significantly important. However, this separation presents a challenge for supporting MPI applications, because the MPI API is
based on the library being able to read and write directly to the memory of 
the application. The executing Wasm-based MPI application can only provide memory addresses in its own linear memory address space, while the target MPI library requires addresses in the host memory address space. 

For executing Wasm modules, \emph{MPIWasm} reserves a part of its own address space for use by the Wasm module. As a result, every byte contained in this range can be
addressed either with a memory address in the module's memory space or with a memory address in the embedder's (host's) memory space. Moreover, while instantiating the module's linear memory, \emph{MPIWasm} records its base address. Following this, it is possible to convert an address from the linear address space of the Wasm module to the embedder's address space and vice-versa by treating the address in the linear address space as an offset relative to the module's base address. This is shown in Figure~\ref{fig:memadressspacewasm}. In particular, \emph{MPIWasm} directly converts 32-bit Wasm pointers that refer to the module's linear address space to 64-bit pointers that refer to the embedder's address space and vice-versa. To this end, \emph{MPIWasm} directly utilizes the MPI library present on the host system without copying any data from the module's address space to a different location, i.e., it supports zero-copy memory operations.

Our mechanism for memory address translation does not violate memory-safety because: (i) a malicious Wasm module cannot violate \textit{control flow integrity} (\S\ref{sec:sec_model}) and (ii) since the size of the linear memory is always known, \emph{MPIWasm} can perform runtime bound checks for all memory accesses. As a result, a module cannot access the memory of the embedder or the memory of the underlying operating system unless explicitly given access to it.

\subsection{Translating MPI Datatypes}
\label{sec:translation}

MPI is implemented as a library with the most common being OpenMPI~\cite{openmpi}, MPICH~\cite{mpich}, and MVAPICH~\cite{mvapich}. Hence, it does not guarantee an Application Binary Interface (ABI) and interoperability between libraries. This means that changing the MPI implementation requires recompilation of the entire application code. One of the reasons for ABI incompatibility is that the MPI standard does not specify explicit types for its datatypes such as \texttt{MPI\_Op} and their implementation is completely up to the MPI library. However, since Wasm modules are designed to be portable not just between the different MPI libraries but also between different CPU architectures, it becomes necessary to add an abstraction between the datatypes used by the host's MPI library and the datatypes exposed to the Wasm module by \emph{MPIWasm}. An abstraction is possible since most MPI datatypes such as \texttt{MPI\_Comm}, \texttt{MPI\_Datatype} and \texttt{MPI\_Op} are opaque to the application and only used as arguments to MPI functions. \emph{MPIWasm} defines most MPI datatypes as 32-bit integers from the perspective of the Wasm module (Listing~\ref{lst:wasiheader}) and transparently translates these datatypes to the host equivalents (\S\ref{sec:funcimplementation}). We use integers as datatypes since \emph{MPIWasm} internally uses IDs to identify data structures that it creates on behalf of the module in order to communicate with the host MPI library.

\subsection{Implementing MPI Functions in \emph{MPIWasm}}
\label{sec:funcimplementation}
Wasm imports are referred to by namespace and name of the definition to import. By default, any symbols that are not defined while compiling C/C++ applications to Wasm will be resolved by making them imports of the module in the \texttt{env} namespace. This is also demonstrated in Listing~\ref{lst:watmpiexample} with the function imports related to the MPI standard. \emph{MPIWasm} provides definitions for all these functions with the same name as the original MPI function and exports them in the \texttt{env} namespace. For implementing these functions, we combine the memory address and MPI datatype translations as described in \S\ref{sec:memtranslation} and \S\ref{sec:translation} respectively. Towards this, we maintain a structure called \texttt{Env} that stores the global state required by these translations. This structure includes information about the memory allocated to the Wasm module, it's base pointer (\S\ref{sec:memtranslation}) and information about the different used datatypes such as \texttt{MPI\_Comm} by the module. For directly utilizing the host MPI library, we use the project \texttt{rsmpi}~\cite{rsmpi} in \emph{MPIWasm}. \texttt{rsmpi} provides MPI bindings for Rust and supports OpenMPI~\cite{openmpi} and MPICH~\cite{mpich}. It utilizes the \texttt{rust-bindgen} project to generate foreign function interfaces tailored to specific MPI libraries. Each MPI function in \emph{MPIWasm} directly calls the equivalent function in \texttt{rsmpi} with the appropriate arguments.




While for most functions in the MPI-2.2 standard \emph{MPIWasm} directly defers the execution to the host MPI library, the implementation of the MPI functions \texttt{MPI\_Alloc\_mem} and \texttt{MPI\_Free\_mem} is done differently. With these functions, it is possible to allocate memory for use with other MPI functions. When a Wasm module calls \texttt{MPI\_Alloc\_mem}, it expects a 32-bit memory address in the module's address space, while calling the \texttt{MPI\_Alloc\_mem} function of the host MPI library returns a 64-bit memory address in the embedder's memory address space which is not inside the chunk of memory reserved for the Wasm module. To overcome this, \emph{MPIWasm} only supports \texttt{MPI\_Alloc\_mem} and \texttt{MPI\_Free\_mem} if the Wasm module defines and exports the functions \texttt{malloc} and \texttt{free}. When \texttt{MPI\_Alloc\_mem} is called, \emph{MPIWasm} simply invokes the exported \texttt{malloc} and receives a suitable 32-bit module memory address. This address can then be used
as the return value for \texttt{MPI\_Alloc\_mem}. We implement \texttt{MPI\_Free\_mem} in a similar way.

\subsection{Limitations}
\label{sec:limitations}
The Wasm specification currently assumes little-endian byte order for multi-byte values~\cite{webassembly} in execution environments. By giving direct access to the Wasm module's memory to the host MPI library, we assume that the byte order of values in the module address space and embedder address space is the same. As a result, \emph{MPIWasm} does not support big-endian CPU architectures. This is not a disadvantage since most processor architectures in HPC systems are little-endian. Moreover, due to the current linear 32-bit memory space for a Wasm module, HPC applications compiled to Wasm cannot have more than 4GiB of memory. The support for 64-bit memory addresses is an important milestone for the Wasm specification and is highlighted in the Wasm Memory64 proposal~\cite{wasm_memory64}, but is out of scope for this work.

\section{Experimental Results}
In this section, we present performance results for our embedder \emph{MPIWasm} across different processor architectures. For all our experiments, we follow best practices while reporting results~\cite{hoefler2015scientific}.

\label{sec:results}

\begin{figure*}
 \begin{subfigure}{0.33\textwidth}
    \centering
        \includegraphics[width=0.49\columnwidth]{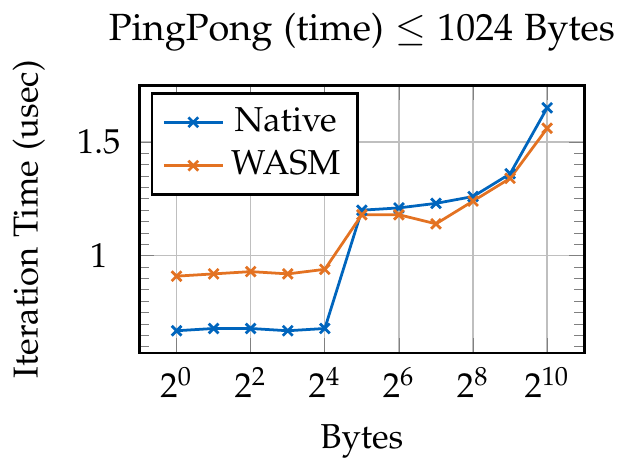}
        \includegraphics[width=0.49\columnwidth]{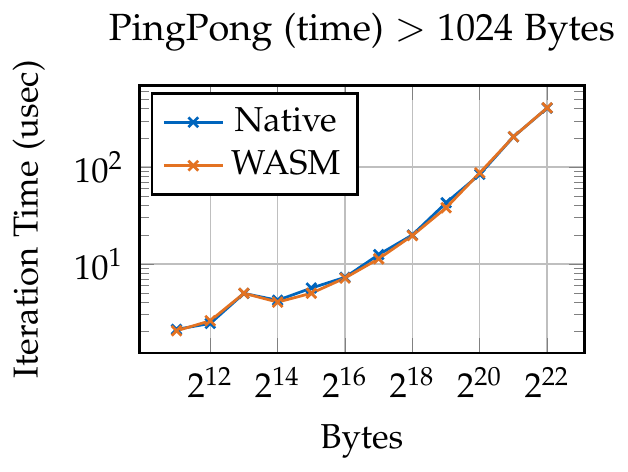}
        \caption{PingPong.}
        \label{fig:pingpong}
    \end{subfigure}
\begin{subfigure}{0.33\textwidth}
    \centering
        \includegraphics[width=0.49\columnwidth]{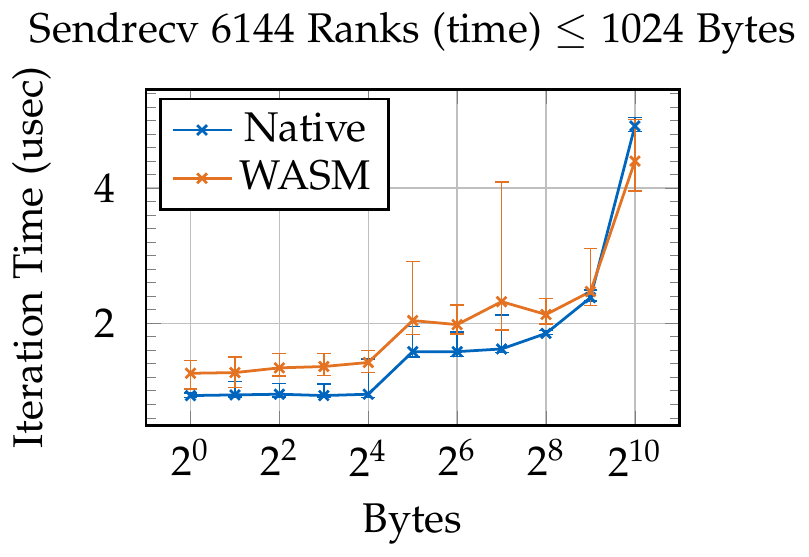}
        \includegraphics[width=0.49\columnwidth]{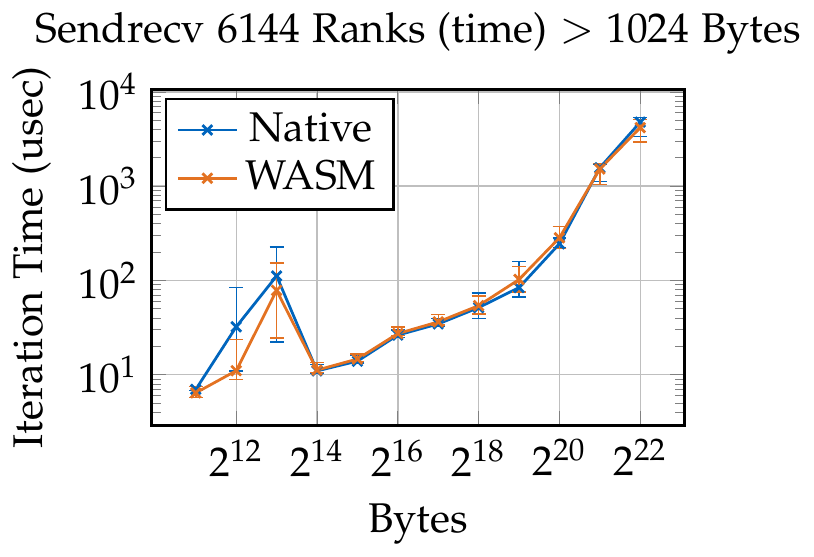}
        \caption{SendRecv.}
        \label{fig:sendrecv}
\end{subfigure}
\begin{subfigure}{0.33\textwidth}
    \centering
        \includegraphics[width=0.49\columnwidth]{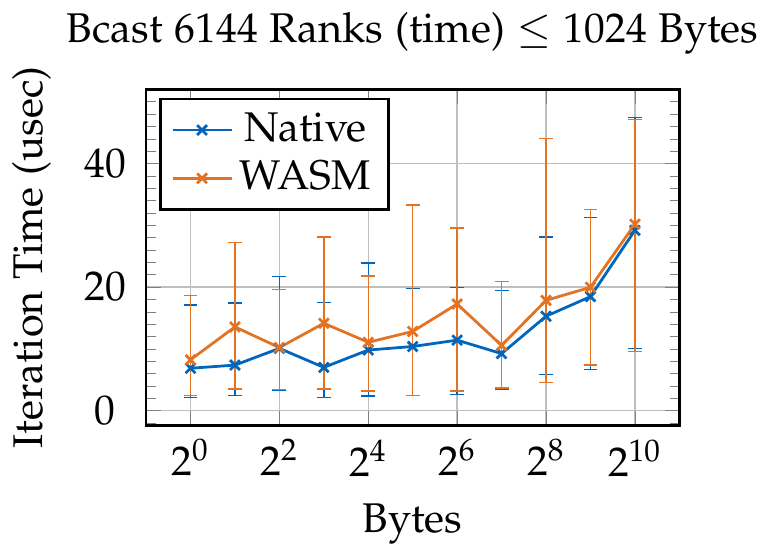}
        \includegraphics[width=0.49\columnwidth]{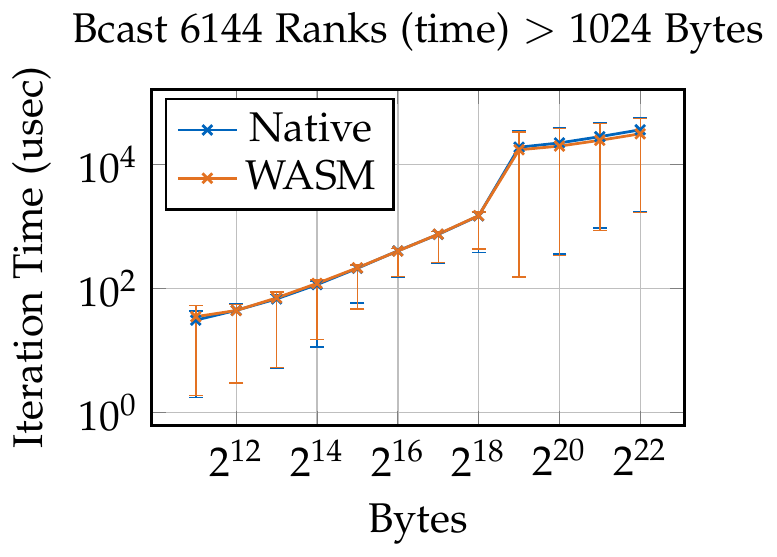}
        \caption{Broadcast.}
        \label{fig:bcast}
\end{subfigure}
\begin{subfigure}{0.33\textwidth}
    \centering
        \includegraphics[width=0.49\columnwidth]{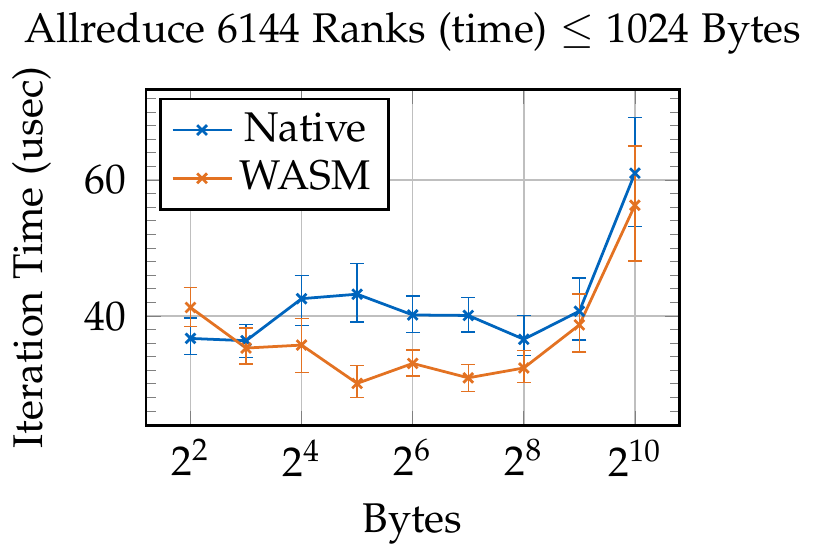}
        \includegraphics[width=0.49\columnwidth]{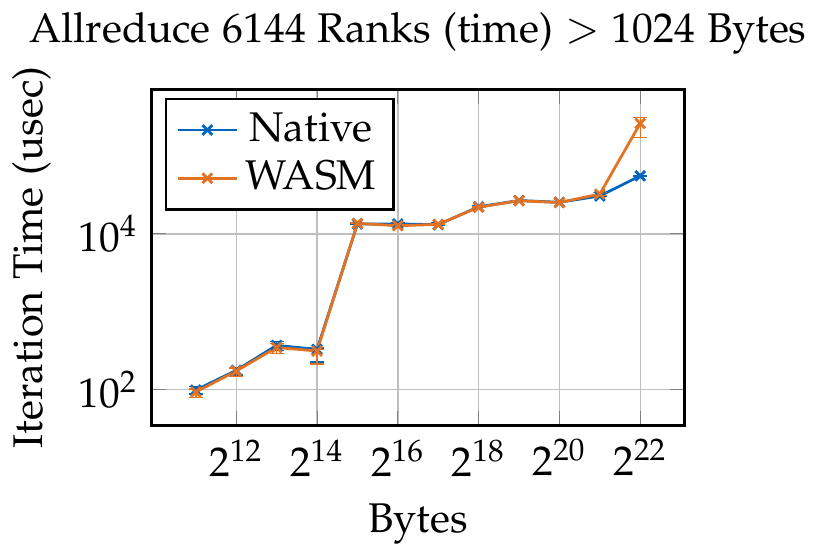}
        \caption{AllReduce.}
        \label{fig:allreduce}
\end{subfigure}
\begin{subfigure}{0.33\textwidth}
    \centering
        \includegraphics[width=0.49\columnwidth]{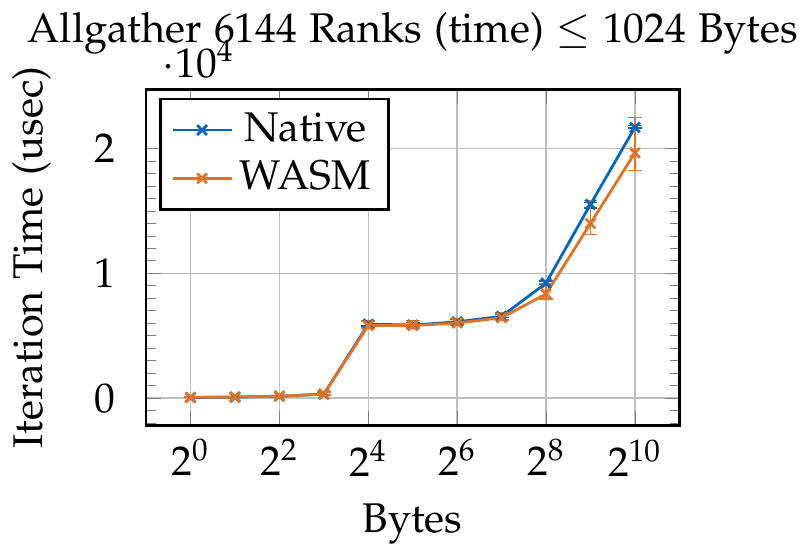}
        \includegraphics[width=0.49\columnwidth]{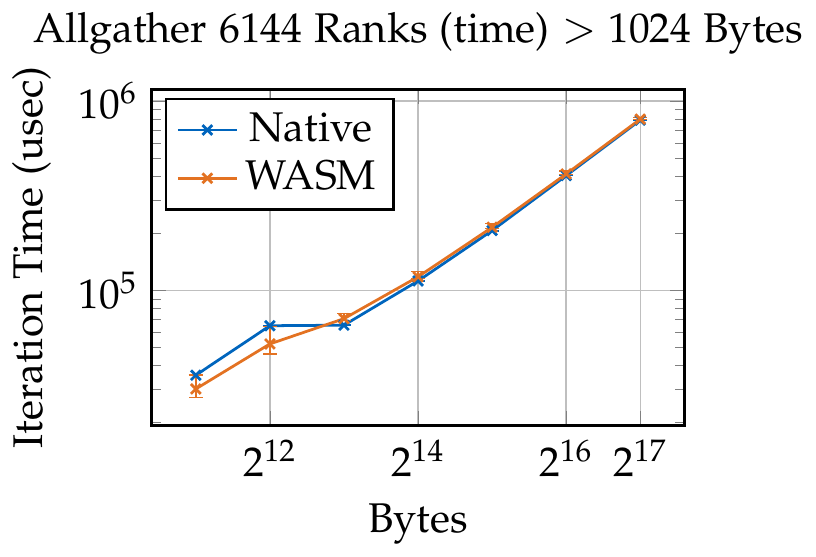}
        \caption{AllGather.}
        \label{fig:allgather}
\end{subfigure}
\begin{subfigure}{0.33\textwidth}
    \centering
        \includegraphics[width=0.49\columnwidth]{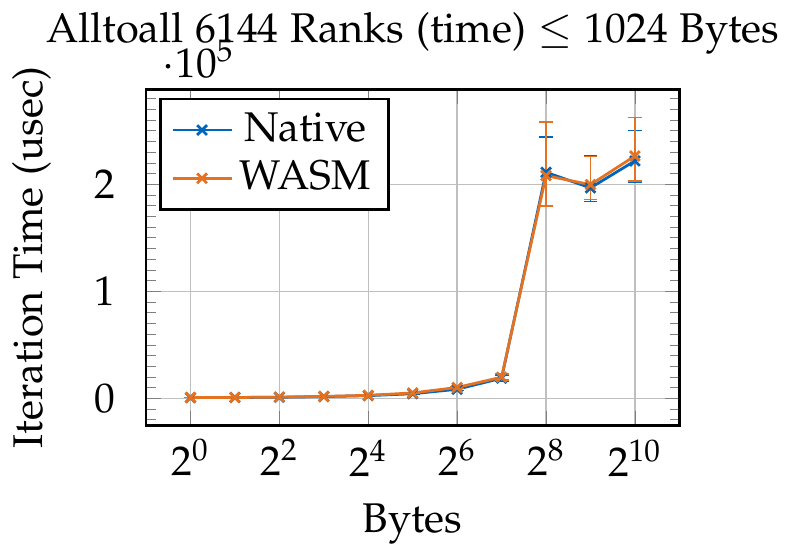}
        \includegraphics[width=0.49\columnwidth]{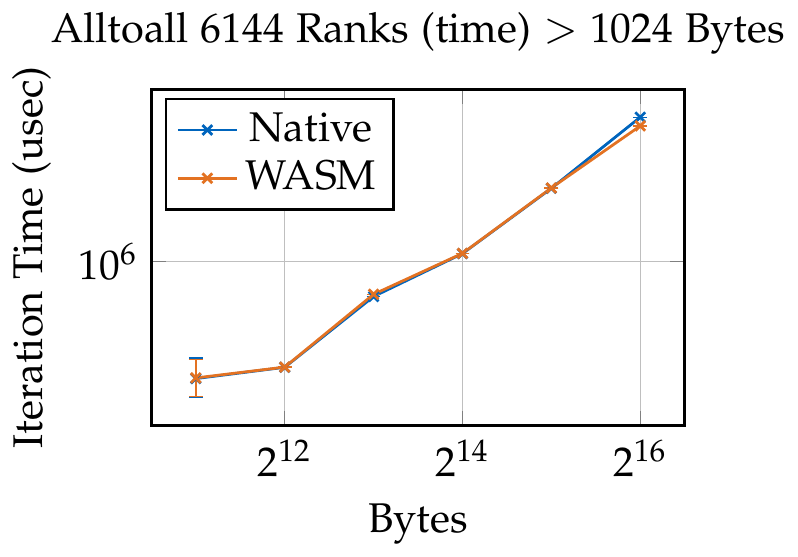}
        \caption{Alltoall.}
        \label{fig:alltoall}
\end{subfigure}
\begin{subfigure}{0.33\textwidth}
    \centering
        \includegraphics[width=0.49\columnwidth]{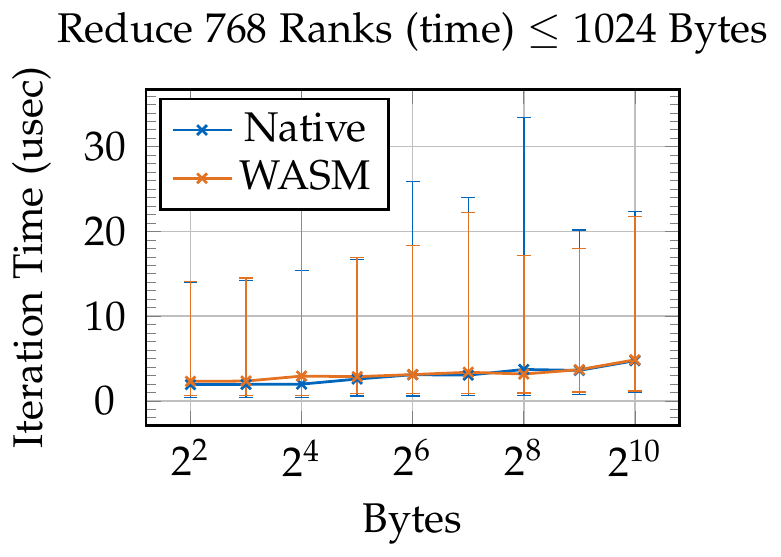}
        \includegraphics[width=0.49\columnwidth]{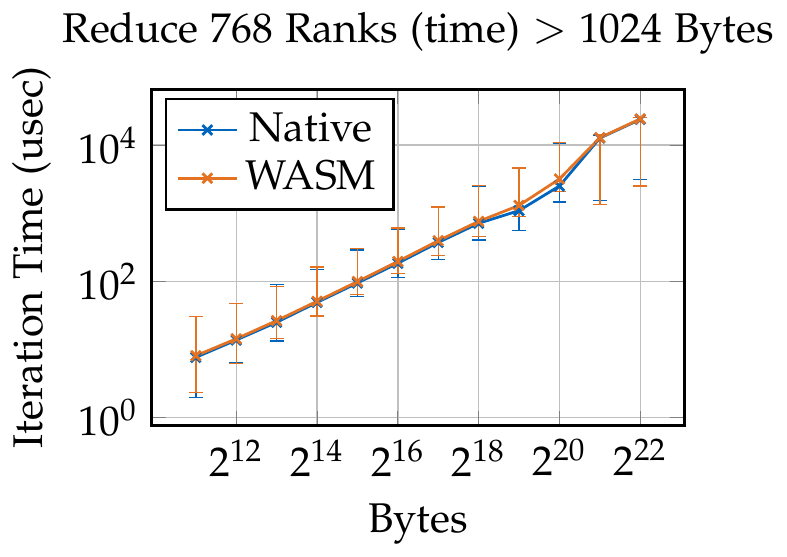}
        \includegraphics[width=0.49\columnwidth]{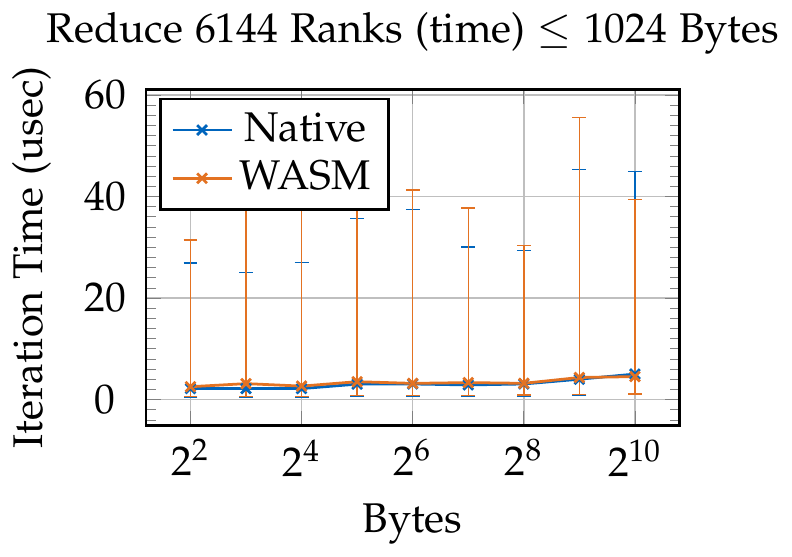}
        \includegraphics[width=0.49\columnwidth]{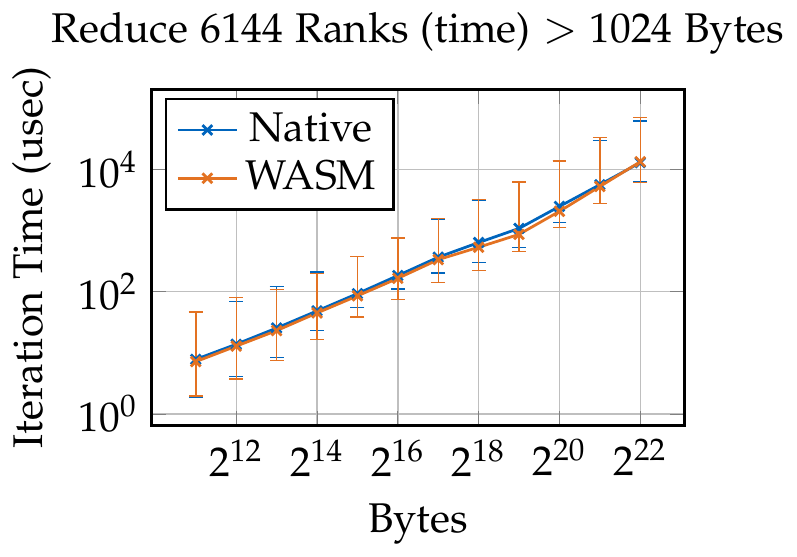}
        \caption{Reduce.}
        \label{fig:reduce}
\end{subfigure}
\begin{subfigure}{0.33\textwidth}
    \centering
        \includegraphics[width=0.49\columnwidth]{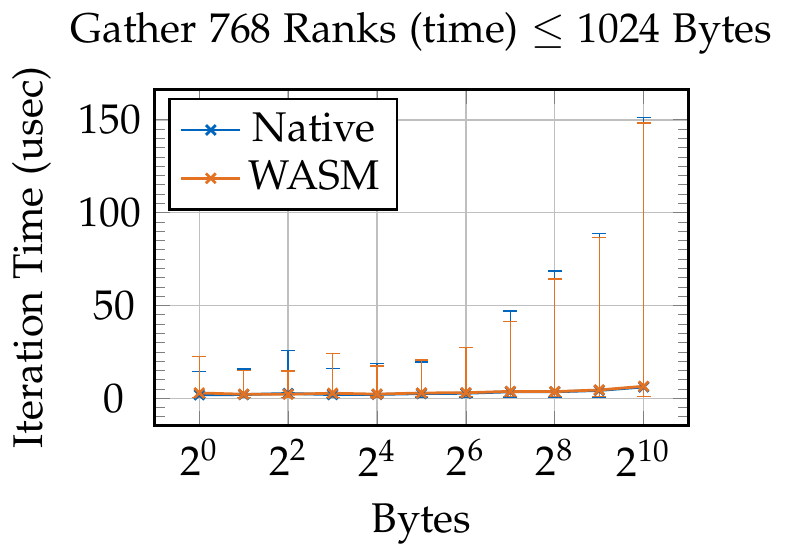}
        \includegraphics[width=0.49\columnwidth]{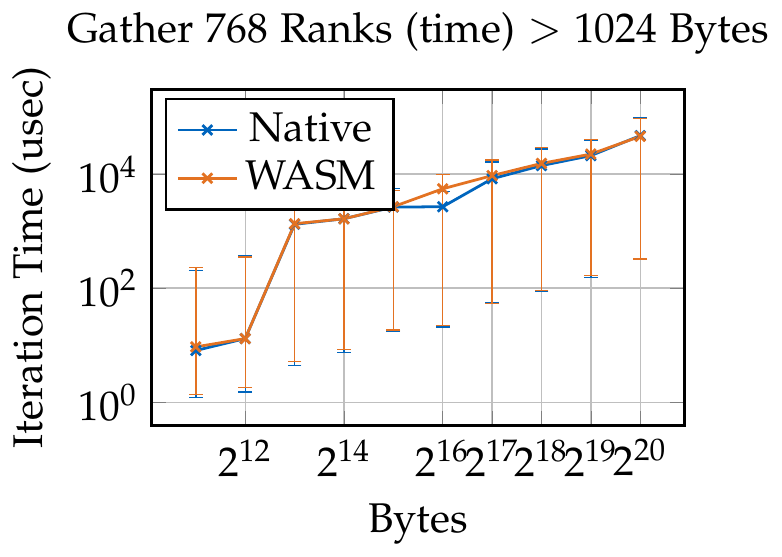}
        \includegraphics[width=0.49\columnwidth]{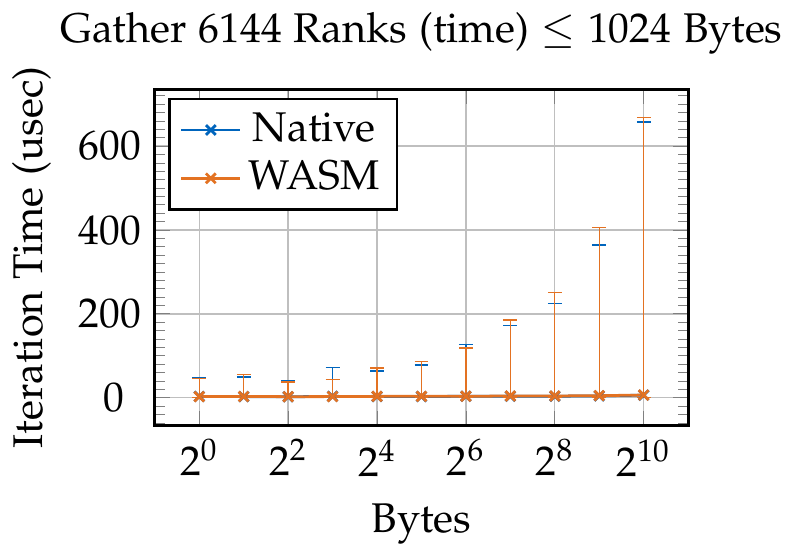}
        \includegraphics[width=0.49\columnwidth]{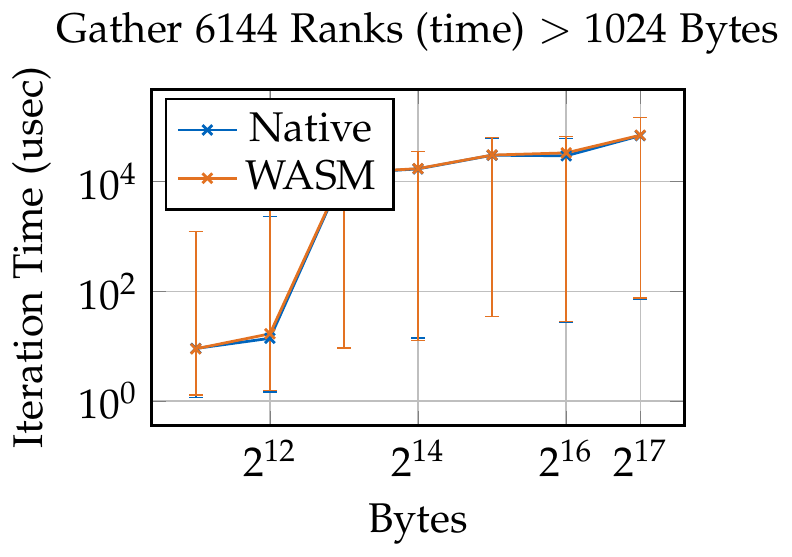}
        \caption{Gather.}
        \label{fig:gather}
\end{subfigure}
\begin{subfigure}{0.33\textwidth}
    \centering
        \includegraphics[width=0.49\columnwidth]{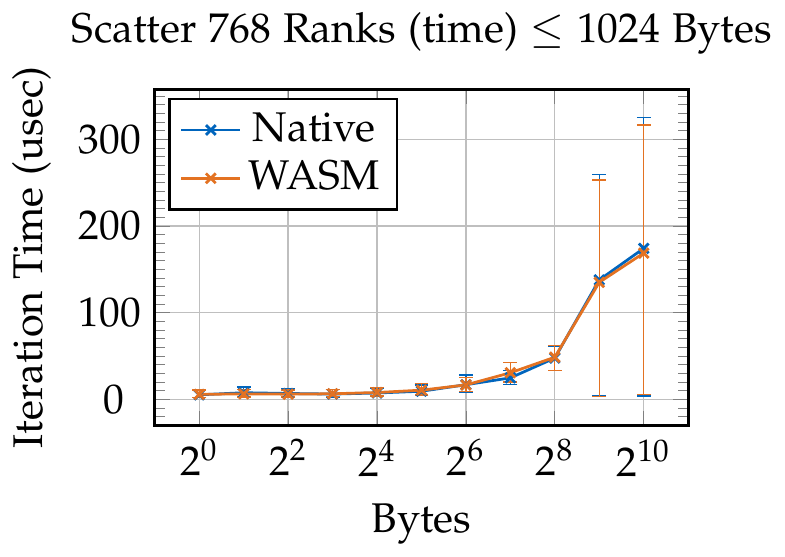}
        \includegraphics[width=0.49\columnwidth]{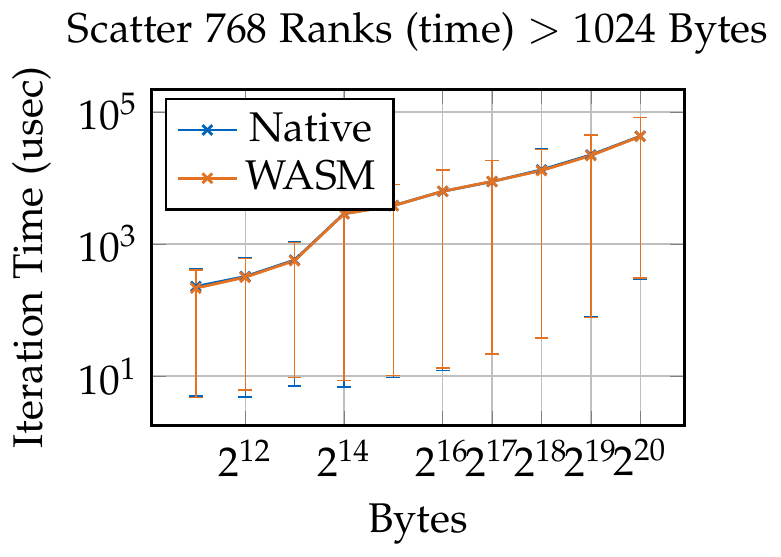}
        \includegraphics[width=0.49\columnwidth]{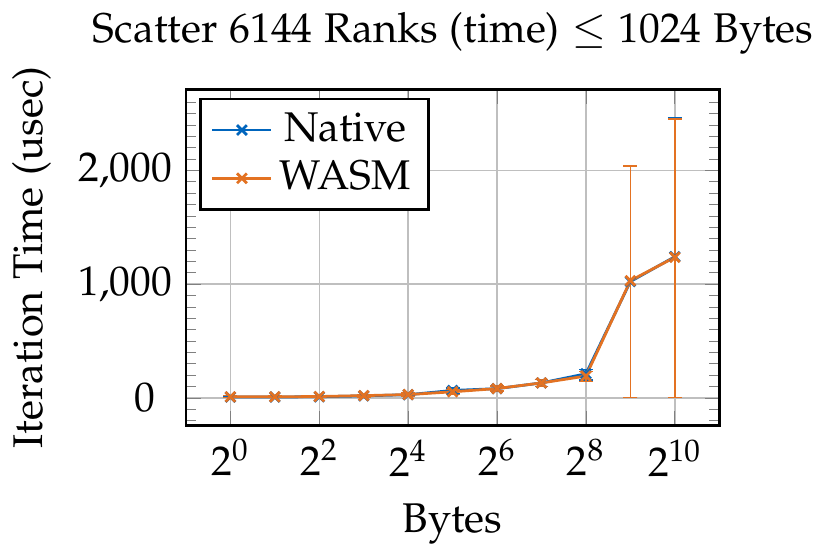}
        \includegraphics[width=0.49\columnwidth]{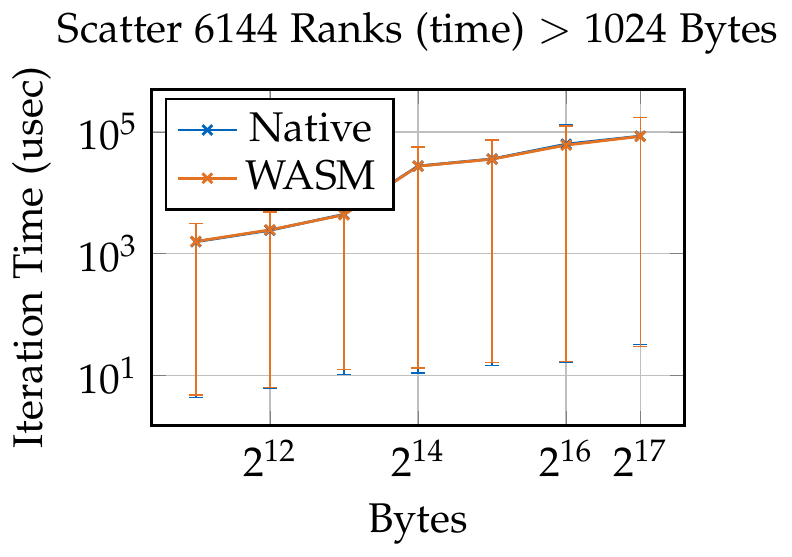}
        \caption{Scatter.}
        \label{fig:scatter}
\end{subfigure}
\caption{Performance comparison of the Intel MPI benchmarks for \emph{MPIWasm} and their native execution on our HPC system.}
\label{fig:impiresults}
\end{figure*}

\begin{figure*}
 \begin{subfigure}{0.33\textwidth}
    \centering
        \includegraphics[width=0.49\columnwidth]{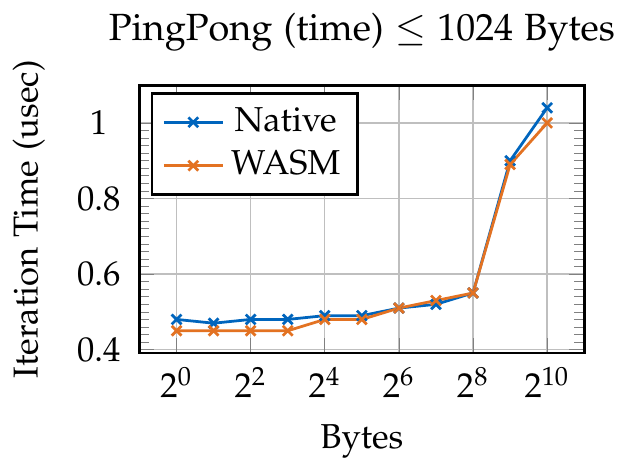}
        \includegraphics[width=0.49\columnwidth]{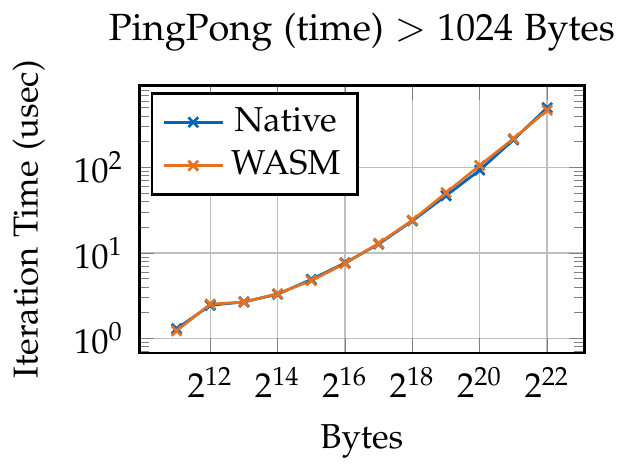}
        \caption{PingPong.}
        \label{fig:pinpongarm}
\end{subfigure}
\begin{subfigure}{0.33\textwidth}
    \centering
        \includegraphics[width=0.49\columnwidth]{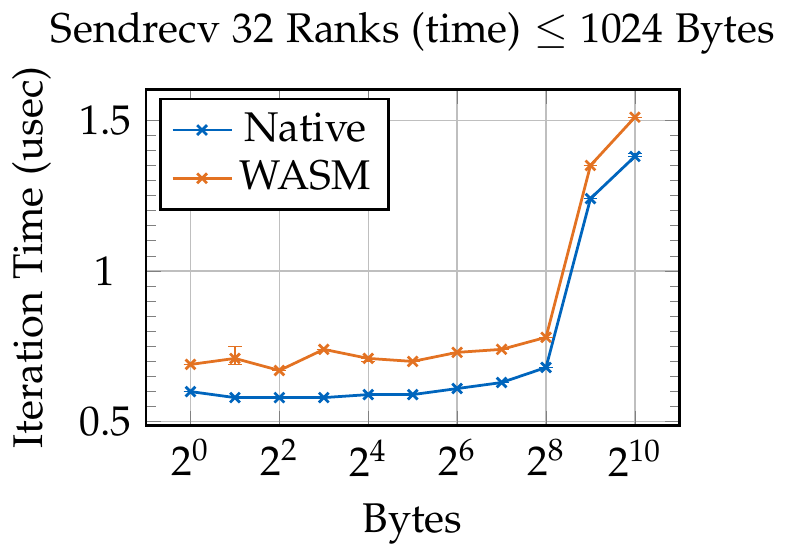}
        \includegraphics[width=0.49\columnwidth]{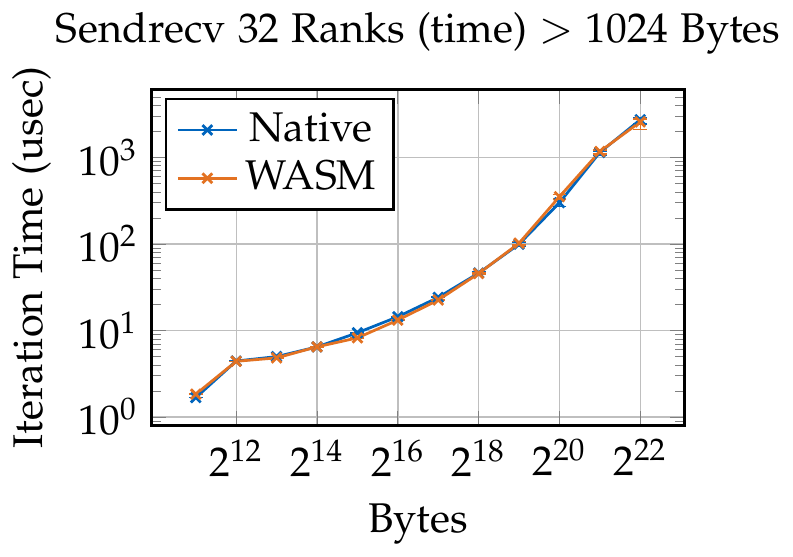}
        \caption{SendRecv.}
        \label{fig:sendrecvarm}
\end{subfigure}
 \begin{subfigure}{0.33\textwidth}
    \centering
        \includegraphics[width=0.49\columnwidth]{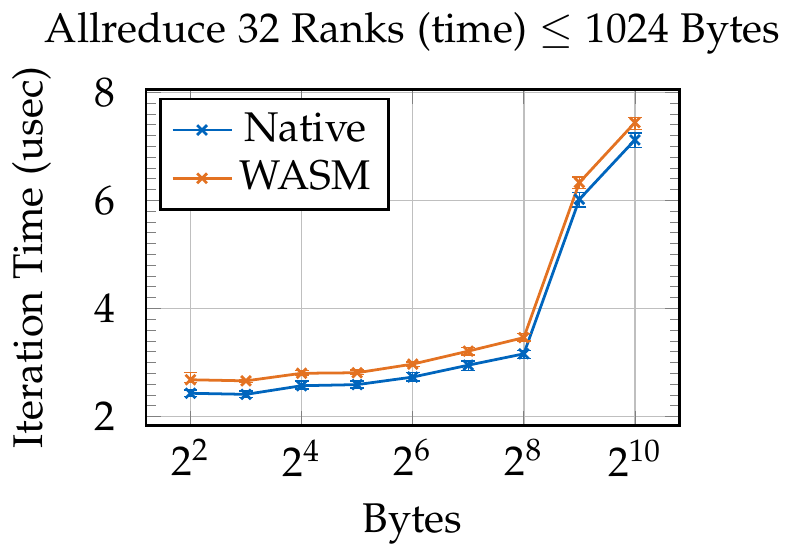}
        \includegraphics[width=0.49\columnwidth]{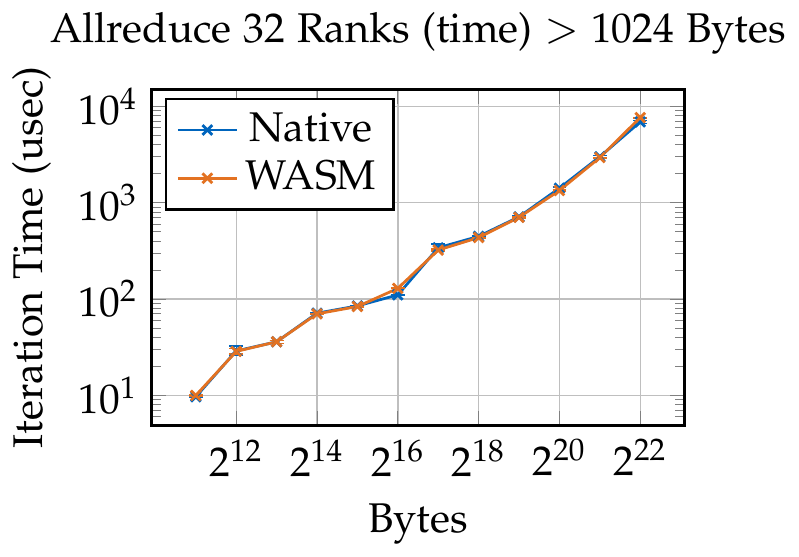}
        \caption{AllReduce.}
        \label{fig:allreducearm}
\end{subfigure}
 \begin{subfigure}{0.33\textwidth}
    \centering
        \includegraphics[width=0.49\columnwidth]{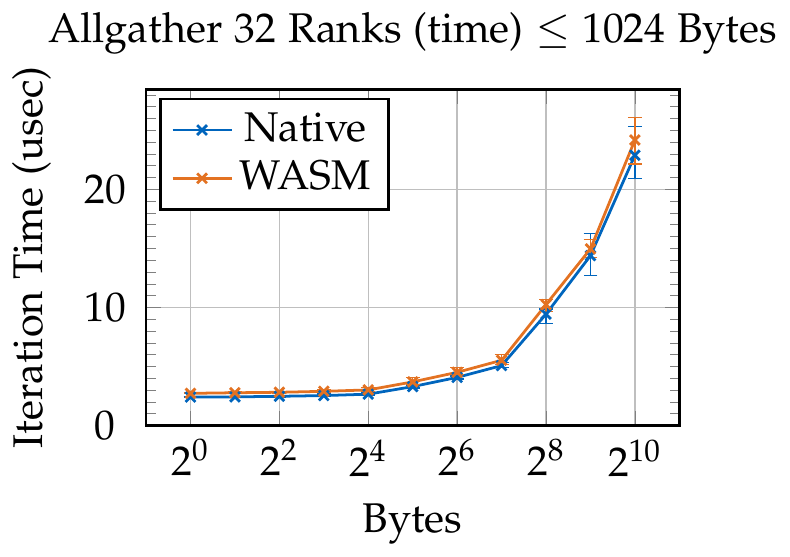}
        \includegraphics[width=0.49\columnwidth]{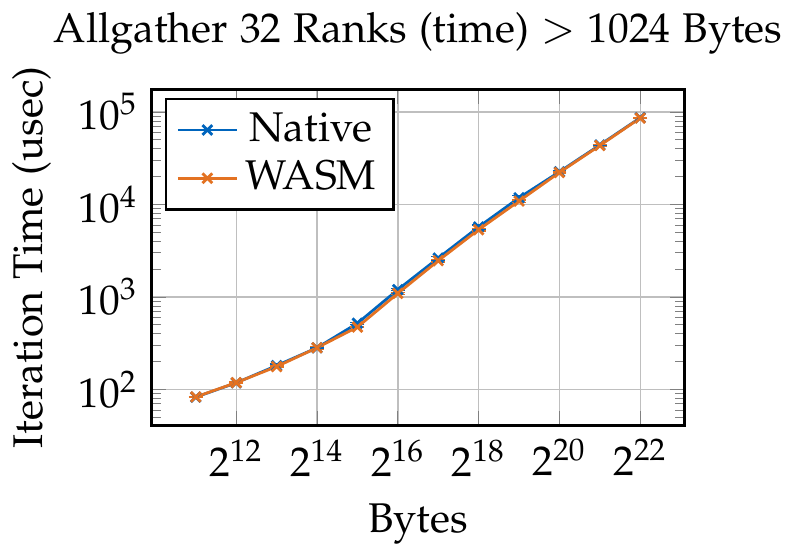}
        \caption{AllGather.}
        \label{fig:allgatherarm}
\end{subfigure}
\begin{subfigure}{0.33\textwidth}
    \centering
        \includegraphics[width=0.49\columnwidth]{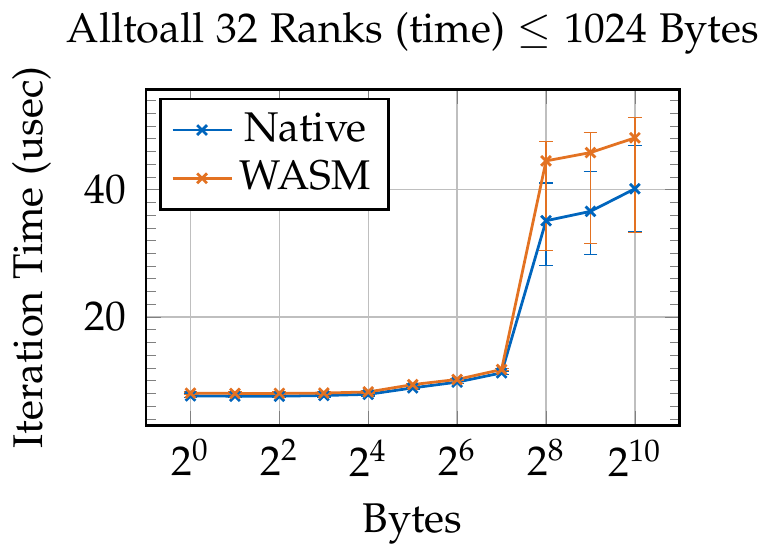}
        \includegraphics[width=0.49\columnwidth]{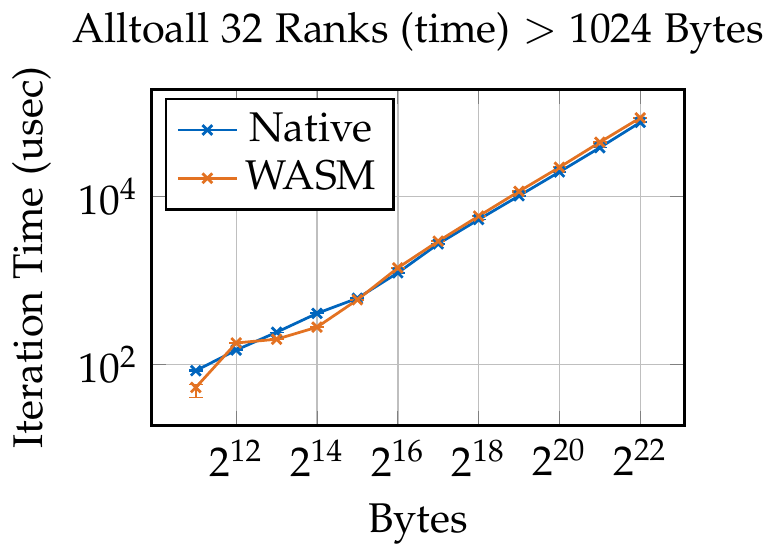}
        \caption{Alltoall.}
        \label{fig:alltoallarm}
\end{subfigure}
\begin{subfigure}{0.33\textwidth}
    \centering
        \includegraphics[width=0.49\columnwidth]{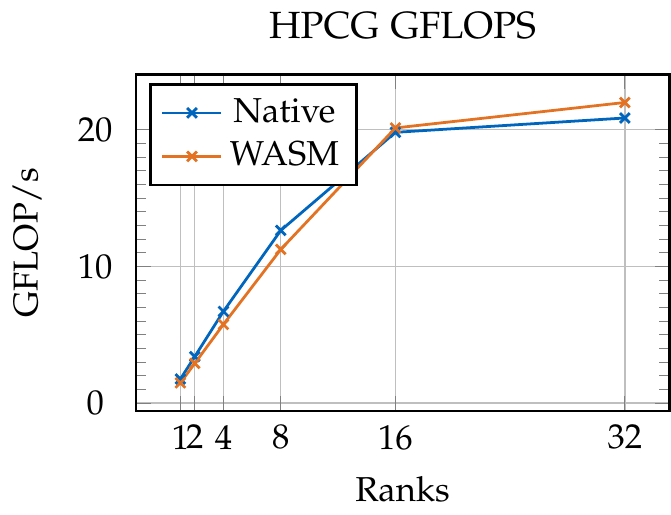}
        \includegraphics[width=0.49\columnwidth]{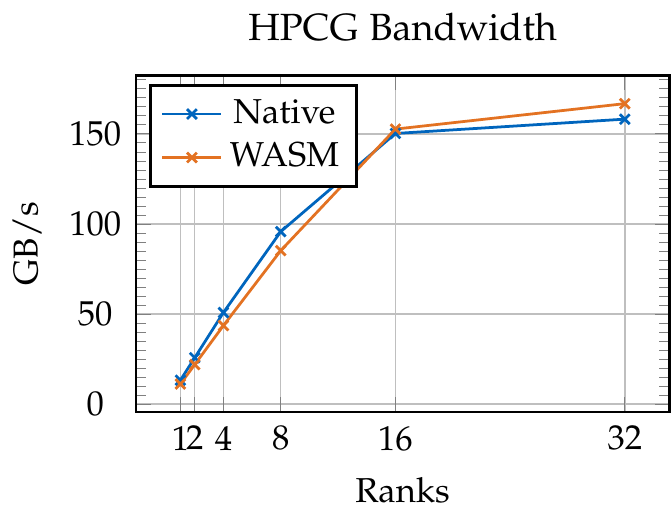}
        \caption{HPCG}
        \label{fig:hpcgarm}
\end{subfigure}
\caption{Performance comparison of selected Intel MPI benchmarks and HPCG for \emph{MPIWasm} against their native execution on the AWS Graviton2 Processor.}
\label{fig:armresults} 
\end{figure*}

\begin{lstlisting}[caption={
      Executing MPI applications compiled to Wasm with \emph{MPIWasm}.
}, float, floatplacement=t, captionpos=b, basicstyle=\ttfamily\tiny,  belowskip=-2 \baselineskip, frame=single, language=bash, framexrightmargin=0cm, xrightmargin=0cm, 
   label={lst:executingwithwasm}]
 mpirun -np <number-of-processes> ./mpiWasm mpi-app.wasm <args>
\end{lstlisting}

\subsection{System Description}
\label{sec:sysdescription}

For analyzing the performance of our implemented Wasm embedder, we use two systems. First, a production HPC cluster located at our institute, i.e., \texttt{SuperMUC-NG}. Second, an AWS EC2 virtual machine (VM) instance with the Graviton2 processor~\cite{graviton}. Our HPC cluster contains eight islands comprising a total of $6480$ compute nodes based on the Intel Skylake-SP architecture. Each compute node has two sockets, comprising two Intel Xeon Platinum 8174 processors, with 24 cores each and a total of 96GiB of main memory. The nominal operating core frequency for each core is 3.10 GHz. Hyper-Threading and Turbo Boost are disabled on the system. The internal interconnect on our system is a fast Intel OmniPath~\cite{omnipath} network with a bandwidth of 100 Gbit/s. Moreover, our cluster provides a general parallel filesystem based on the Lenovo DSS-G for IBM Spectrum Scale~\cite{lenovodsg} with an aggregate bandwidth of $200$ GiB/s.  For our experiments, we use up to 128 nodes of the HPC system, i.e., 6144 cores. On the other hand, the AWS Graviton2 processor based on the 64-bit \texttt{ARMv8-A} Neoverse-N1~\cite{neoverse} architecture consists of 32 cores each with a nominal frequency of $2.50$ GHz and a total main memory of 64GiB. We limit our experiments to one node for the Graviton2 processor.

    
    


    

  \begin{table}[t]
\caption{Comparing the size of native dynamically-linked, statically-linked, and Wasm binaries for the different MPI applications. The native applications are compiled for the \texttt{x86\_64} architecture.}
\centering
 \begin{adjustbox}{width=8cm,center}
    \begin{tabular}{|>{\centering\arraybackslash}c|>{\centering\arraybackslash}c|>{\centering\arraybackslash}c|>{\centering\arraybackslash}c|}
      \hline
        \textbf{Application} & \textbf{Native Size Dynamic (KiB)} & \textbf{Native Size Static (MiB)} & \textbf{Wasm Size (KiB)}   \\ \hline 
       Intel MPI Benchmarks~\cite{intel_mpi_benchmarks}. & 1087 & 27  & 893   \\ \hline
        HPCG~\cite{hpcg}.  & 164 & 26 & 722   \\ \hline
        IOR~\cite{ior_benchmark}. & 364 & 16  & 315.32  \\ \hline
        IS~\cite{Bailey:1991:NPB:125826.125925}. & 36 & 15  & 57.88   \\ \hline
        DT~\cite{Bailey:1991:NPB:125826.125925}. & 40 & 15 & 49.51   \\ \hline
        
   
  \end{tabular}
    
  \end{adjustbox}
  \vspace{-0.5cm}
\label{table:binary_sizes}
\end{table}

\subsection{HPC Benchmarks}
\label{sec:benchmarksapplications}

For our experiments with \emph{MPIWasm}, we use the Intel MPI Benchmarks~\cite{intel_mpi_benchmarks}, two benchmarks from the the NASA Advanced Supercomputing (NAS) Parallel Benchmark (NPB) suite~\cite{Bailey:1991:NPB:125826.125925}, the IOR benchmark~\cite{ior_benchmark}, and the High Performance Compute Gradient (HPCG) benchmark~\cite{hpcg}.

The Intel MPI benchmarks perform a set of MPI performance measurements for point-to-point and global communication operations for a range of message sizes. We use them since they characterize the performance of a cluster and are an indication of the efficiency of the used MPI implementation. The NPB suite includes a set of benchmarks that aim to evaluate the overall performance of HPC clusters. Due to the support for compiling Fortran to Wasm being in the early stages (\S\ref{sec:futurehpcwasm}), only the Integer Sort (\texttt{IS}) and Data Transfer (\texttt{DT}) benchmarks from this suite were used since they are written in pure C. The \texttt{IS} benchmark performs bucketed parallel sorting of integers across all participating processes, while the \texttt{DT} benchmark tests the communication and the performance of 64-bit floating point operations of a HPC cluster by sending data through a topology of nodes. We use the topologies \texttt{Black-Hole} (bh), \texttt{White-Hole} (wh), and \texttt{Shuffle} (sh) for the \texttt{DT} benchmark. For our experiments, we use the classes C and B for the \texttt{IS} and \texttt{DT} benchmarks respectively. The \texttt{IOR} Benchmark measures the filesystem I/O performance available to MPI processes. It supports multiple backends that utilize different APIs to perform system I/O. For our experiments with \emph{MPIWasm}, we use the POSIX API backend since the POSIX filesystem APIs are included in the WASI specification (\S\ref{sec:wasi}, \S\ref{sec:extending_wasm_system_interface}). The HPCG benchmark aims to evaluate the real-world performance of HPC systems by solving a system of linear equations with the conjugate gradient method. For our experiments, we use the default available problem size for HPCG.  Note that, in our experiments we use the versions \texttt{2019 Update 6} and \texttt{3.3.1} for the Intel MPI and NAS parallel benchmarks respectively.

\subsection{Experiment Setup}
\label{sec:expsetup}

For all our experiments, we execute the benchmarks in a pure-MPI configuration without shared memory parallelization with OpenMP, as it is currently not supported by \emph{MPIWasm}. We use \texttt{OpenMPI-4.0} as the MPI library since it is available on our HPC system and can be easily installed on the AWS Graviton2 nodes. For compiling the native applications on our HPC system (\S\ref{sec:sysdescription}), we use the \texttt{clang-11} compiler, while for the AWS Graviton2 node, we use the \texttt{gcc7.1} compiler. In both cases, the applications were compiled with the \texttt{-O3} optimization flag. For compiling the different benchmarks to Wasm, we use the \texttt{clang-11} compiler along with our customized \texttt{WASI-SDK} with  \texttt{-O3 -msimd128} flags for both test systems~(\S\ref{sec:extending_wasm_system_interface}). The \texttt{-msimd128} flag enables the generation of SIMD instructions in Wasm. We compile the applications to Wasm only once on our local systems and execute them directly with \emph{MPIWasm} on the different test systems. We build \emph{MPIWasm} on our local system for the different platforms, i.e., \texttt{x86\_64} and \texttt{aarch64} with  \texttt{OpenMPI} to generate bindings for \texttt{rsmpi} (\S\ref{sec:funcimplementation}). Following this, we directly execute the applications compiled to Wasm on the test systems as shown in Listing~\ref{lst:executingwithwasm}. Each MPI rank corresponds to one instance of the embedder with it's own Wasm module. The native applications were executed directly using \texttt{mpirun}.







\begin{figure*}

 \begin{subfigure}{0.33\textwidth}
    \centering
        \includegraphics[width=0.49\columnwidth]{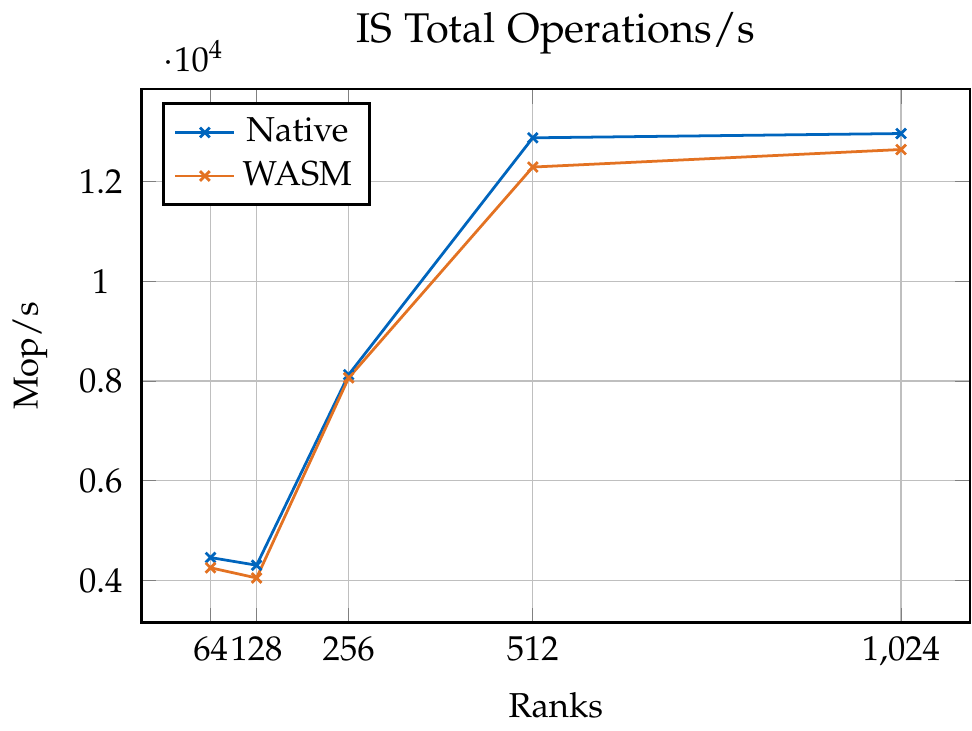}
        \includegraphics[width=0.49\columnwidth]{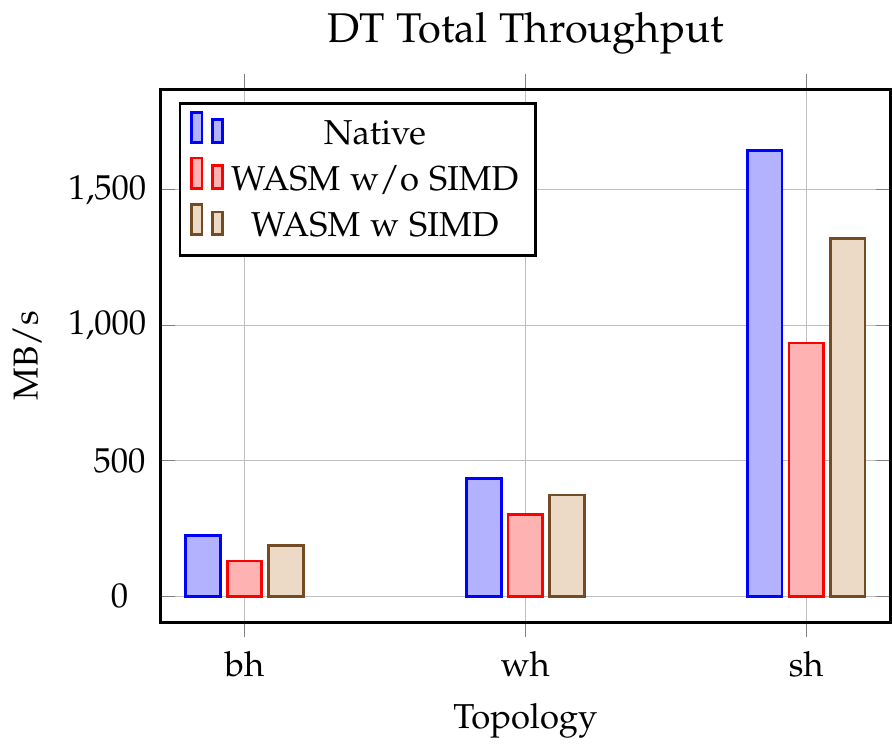}
        \caption{NPB.}
        \label{fig:npb}
\end{subfigure}
\begin{subfigure}{0.33\textwidth}
    \centering
        \includegraphics[width=0.49\columnwidth]{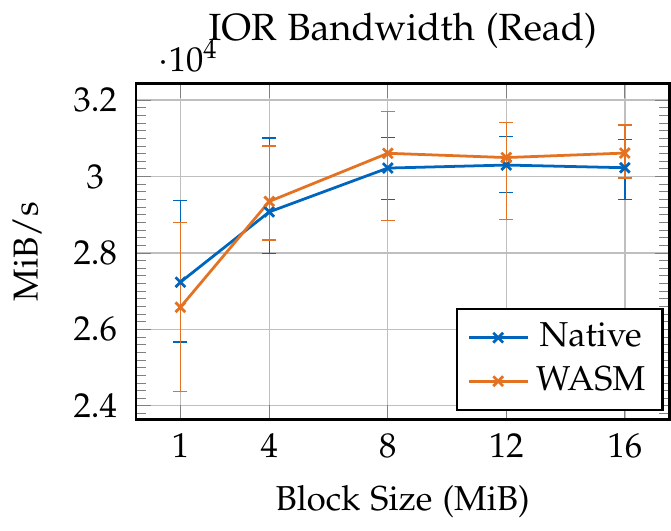}
        \includegraphics[width=0.49\columnwidth]{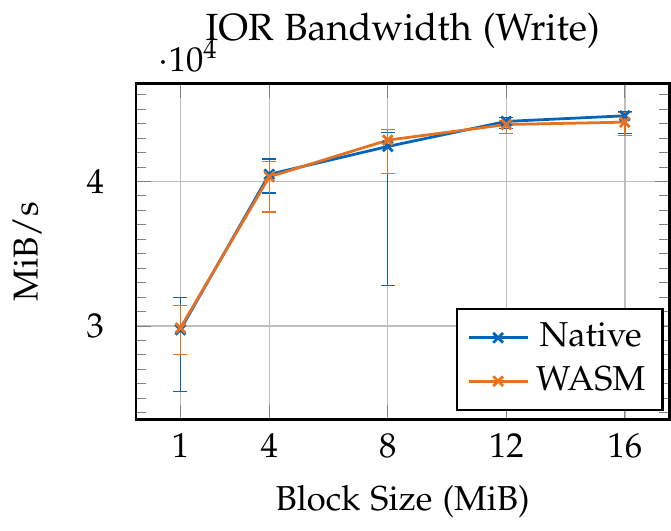}
        \caption{IOR.}
        \label{fig:ior}
\end{subfigure}
 \begin{subfigure}{0.33\textwidth}
    \centering
        \includegraphics[width=0.49\columnwidth]{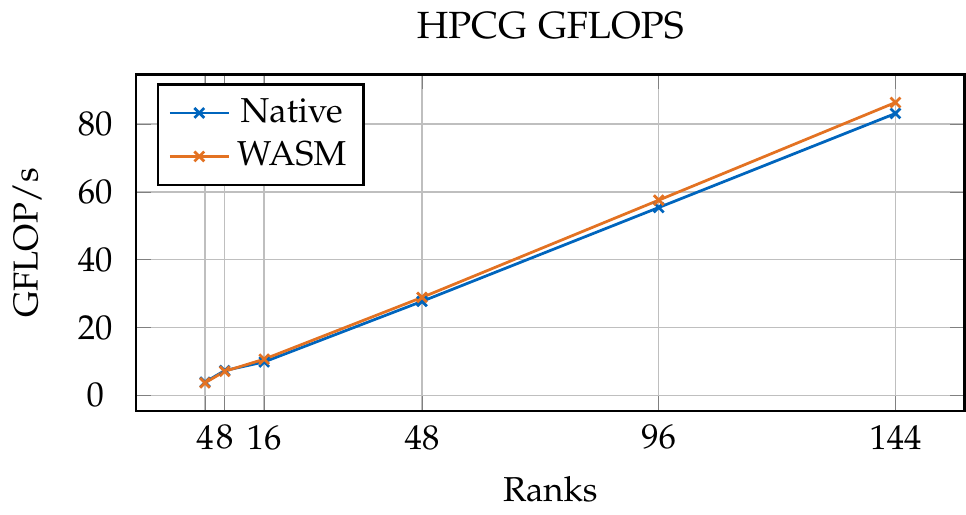}
        \includegraphics[width=0.49\columnwidth]{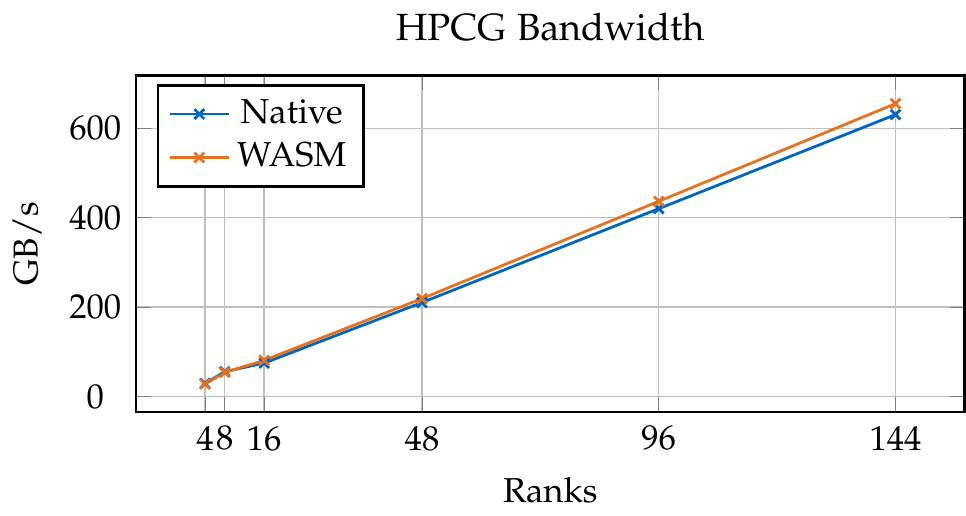}
        \includegraphics[width=0.49\columnwidth]{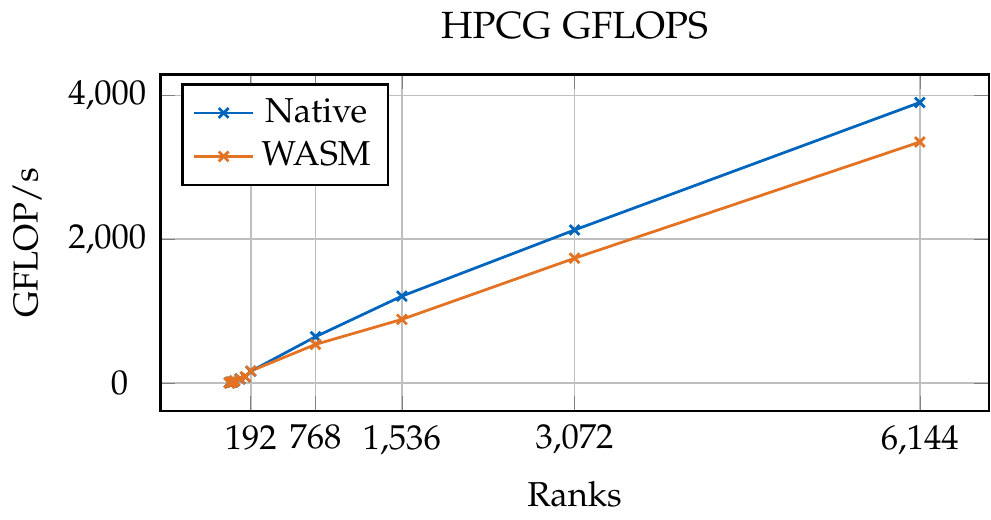}
        \includegraphics[width=0.49\columnwidth]{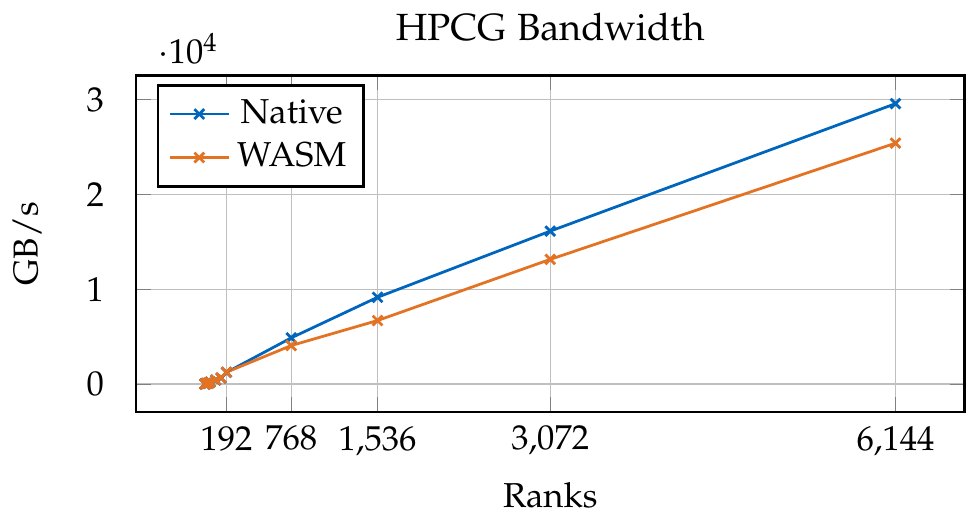}
        \caption{HPCG.}
        \label{fig:hpcg}
\end{subfigure}
\caption{Performance comparison of standardized HPC benchmarks for \emph{MPIWasm} against their native execution on our HPC system.}
\label{fig:applicationresults} 
\end{figure*}

\subsection{Comparing Wasm Binary Size}
\label{sec:binarysizecomparison}
Table~\ref{table:binary_sizes} shows the comparison between the absolute binary sizes for the different applications. The static versions of the binaries are generated by supplying the \texttt{-static} flag to the \texttt{clang-11} compiler (\S\ref{sec:expsetup}) and linking the different applications with the static versions of the required libraries such as \texttt{libmpi.a}, \texttt{libopen-rte.a}, and \texttt{libz.a}. To this end, we made necessary changes to the \texttt{Make}~\cite{make} and \texttt{CMake}~\cite{cmake} files used by the different applications (\S\ref{sec:benchmarksapplications}). While the stack-based instruction set and compact binary format give Wasm the potential to produce smaller binaries for the same applications as compared to the native dynamically-linked binaries \cite{webassembly}, three out of five applications that we used had a bigger binary size when compiled to WebAssembly in comparison to the equivalent dynamically-linked native binary. While Wasm can benefit from a smaller representation on a function-by-function basis, in practice dynamically-linked native binaries can offset that advantage by being able to rely on commonly used libraries to be present on the system. For instance, a native binary can dynamically link against \texttt{glibc}, while a Wasm binary must statically include functions from \texttt{wasi-libc} (\S\ref{sec:wasi}). However, in contrast to containers, Wasm binaries are significantly smaller making them more feasible for application distribution in HPC environments. In addition, Wasm binaries are $139.5$x smaller on average than the statically-linked binaries of the different applications. This is because the \texttt{linker}, i.e., \texttt{lld} copies all library routines from the different libraries used by an application into the binary during static linking.

\subsection{Benchmarking MPIWasm}
\label{sec:perfcomparison}
Figure~\ref{fig:impiresults} and Figure~\ref{fig:armresults}  show the iteration times for the different Intel MPI benchmarks for their native execution as compared to their execution with \emph{MPIWasm} on our HPC system and the AWS Graviton2 processor respectively. For execution with \emph{MPIWasm}, the iteration times don't include the time required for compiling the Wasm modules to native machine code (\S\ref{sec:highperf}). To avoid repetition, we omit some results for the Graviton2 processor. Error bars in the graphs represent minimum and maximum values for iteration timings as reported by the Intel MPI Benchmarks, while points in the graphs represent the average timings as reported by the benchmarks.

For the \texttt{PingPong} benchmark using \emph{MPIWasm} leads to a geometric mean (GM) average slowdown of $0.05$x for the \texttt{x86\_64} system and a GM average speedup of $1.01$x for the \texttt{aarch64} system across all message sizes (Figures~\ref{fig:pingpong},~\ref{fig:pinpongarm}). We calculate this value by dividing the metric \texttt{t\_avg\_us} reported by the benchmarks for their native execution by the value reported for execution with \emph{MPIWasm}, followed by a GM of the obtained values. For computing slowdown, we subtract one from the obtained GM value. We observe a maximum bandwidth of $12.80$ GiB/s and $10.98$ GiB/s for the native execution of the \texttt{PingPong} benchmark on the HPC and Graviton2 processor respectively. On the other hand, with \emph{MPIWasm}, we observe a maximum bandwidth of $13.44$ GiB/s and $10.61$ GiB/s on the two systems. For the \texttt{SendRecv} benchmark, we observe a GM average slowdown of $0.06$x and $0.07$x with \emph{MPIWasm} across all message sizes on the \texttt{x86\_64} and \texttt{aarch64} systems respectively (Figures~\ref{fig:sendrecv},~\ref{fig:sendrecvarm}). For the native version of the benchmark, we observe a maximum bandwidth of $7.24$ GiB/s and $11.01$ GiB/s on the two systems, while with \emph{MPIWasm}, we observe a maximum bandwidth of $7.50$ GiB/s and $10.83$ GiB/s. For the collective communication \texttt{Broadcast} routine, we observe an average GM slowdown of $0.13$x with \emph{MPIWasm} across all message sizes for $128$ nodes as shown in Figure~\ref{fig:bcast}. We observe an average GM slowdown of $0.06$x and $0.10$x with  \emph{MPIWasm} across all message sizes for the collective communication \texttt{AllReduce} routine as shown in Figures~\ref{fig:allreduce} and~\ref{fig:allreducearm}. For \texttt{AllGather} with \emph{MPIWasm}, we observe an average GM slowdown of $0.06$x and $0.09$x across all message sizes for the HPC system and Graviton2 processor respectively (Figure~\ref{fig:allgather},~\ref{fig:allgatherarm}).  Similarly, for the \texttt{Alltoall} collective communication routine, we observe an average GM slowdown of $0.10$x for the two systems across all message sizes with \emph{MPIWasm} as shown in Figures~\ref{fig:alltoall} and~\ref{fig:alltoallarm}. For $16$ nodes of our HPC system, we observe an average GM slowdown of $0.12$x, $0.14$x, and $0.05$x across message sizes for the routines \texttt{Reduce}, \texttt{Gather}, and \texttt{Scatter} as shown in Figures~\ref{fig:reduce},~\ref{fig:gather}, and~\ref{fig:scatter}. On the other hand, for $128$ nodes, we observe an average GM slowdown of $0.05$x, $0.10$x, and $0.08$x for the three routines. The results for testing \emph{MPIWasm} with the Intel MPI Benchmarks compiled to Wasm demonstrate that neither the mechanism for calling host functions in \texttt{Wasmer}~\cite{wasmer} nor the  translation layer implemented in \emph{MPIWasm} induce significant overhead for MPI communication (\S\ref{sec:memtranslation},\S\ref{sec:translation},\S\ref{sec:funcimplementation}). We expand on the translation overhead in \emph{MPIWasm} in \S\ref{sec:transoverhead}. Overall, our results indicate that \emph{MPIWasm} delivers close to native performance for the different MPI routines on both \texttt{x86\_64} and \texttt{aarch64} architectures.

The performance of \emph{MPIWasm} on our HPC system for the \texttt{IS} and \texttt{DT} benchmarks is shown in Figure~\ref{fig:npb}. For the \texttt{IS} benchmark with \emph{MPIWasm}, we observe $8260$ average mega operations per second across all processes as compared to $8546$ average mega operations per second for the native execution. For the \texttt{DT} benchmark with different topologies (\S\ref{sec:benchmarksapplications}), execution with \emph{MPIWasm} leads to decreased throughput as compared to the native execution. The \texttt{DT} benchmark performs a significant number of pairwise comparison operations which benefit greatly from vectorization with SIMD instructions. To demonstrate the effect of SIMD for the \texttt{DT} benchmark, we compile it to Wasm by disabling and enabling the generation of SIMD instructions. The Wasm version of the \texttt{DT} benchmark with SIMD leads to $1.36$x better throughput as compared to the Wasm version without SIMD (Figure~\ref{fig:npb}). The difference in performance as compared to the native version of the \texttt{DT} benchmark can be attributed to the support for only 128-bit SIMD instructions in the Wasm specification~\cite{webassembly} as compared to 512-bit SIMD instructions present in modern Intel processors~\cite{schone2019energy,architecture, demystifying} (\S\ref{sec:highperf}). Support for higher-width SIMD in Wasm is an important milestone in its road-map but out of scope for this work (\S\ref{sec:futurehpcwasm}).

Figure~\ref{fig:ior} shows total aggregated read and write bandwidth available to all MPI processes for the \texttt{IOR} benchmark. Points in the graph represent the average bandwidth reported by the benchmark, while error bars in the graph represent the maximum and minimum bandwidth observed over all iterations of the benchmark with the same block size. With four nodes of our HPC system, the upper bound for achievable bandwidth in our setup (\S\ref{sec:sysdescription}) with \texttt{IOR} is 400 GBit/s ($\approx$ 47684 MiB/s). With \emph{MPIWasm}, we observe similar read ($29411$ MiB/s) and write ($40206$ MiB/s) bandwidth averaged across all block sizes as compared to the native execution of the benchmark. Testing the filesystem I/O performance of the \emph{MPIWasm} demonstrates that the userspace permission handling and virtual directory tree implemented by Wasmer to provide filesystem isolation (\S\ref{sec:filesystemisolation}) has no significant impact on the achievable bandwidth when performing I/O with the POSIX filesystem API. For the \texttt{HPCG} benchmark, we observe similar performance when executed with \emph{MPIWasm} as compared to it's native execution on the HPC system and the Graviton2 processor up to 192 MPI processes (Figures~\ref{fig:hpcg} and~\ref{fig:hpcgarm}). On increasing the number of processes, the native execution of the \texttt{HPCG} benchmark outperforms the execution with \textit{MPIWasm} as shown in Figure~\ref{fig:hpcg}. For $6144$ MPI processes, we observe a $14$\% reduction in \texttt{GFLOP/s} on execution with \textit{MPIWasm}. This behavior can be attributed to the significantly frequent amount of communication required by the \texttt{HPCG }benchmark~\cite{perfhpcg}. \texttt{HPCG} repeatedly uses the \texttt{Allreduce} routine to reduce a single variable of size double over all MPI processes to finalize vector-vector dot operations. With increasing number of processes, the number of times the \texttt{Allreduce} routine is called also increases. For instance, executing \texttt{HPCG} with $768$ processes results in four times more calls to \texttt{Allreduce} as compared to the execution with $192$ processes. As a result, the repeated datatype translations in \emph{MPIWasm} increase the cost for invoking the collective communication routine leading to performance degradation (\S\ref{sec:transoverhead}).

\subsection{Analyzing Datatype Translation Overhead}
\label{sec:transoverhead}
\begin{figure}[t]
        
\includegraphics[width=0.8\columnwidth]{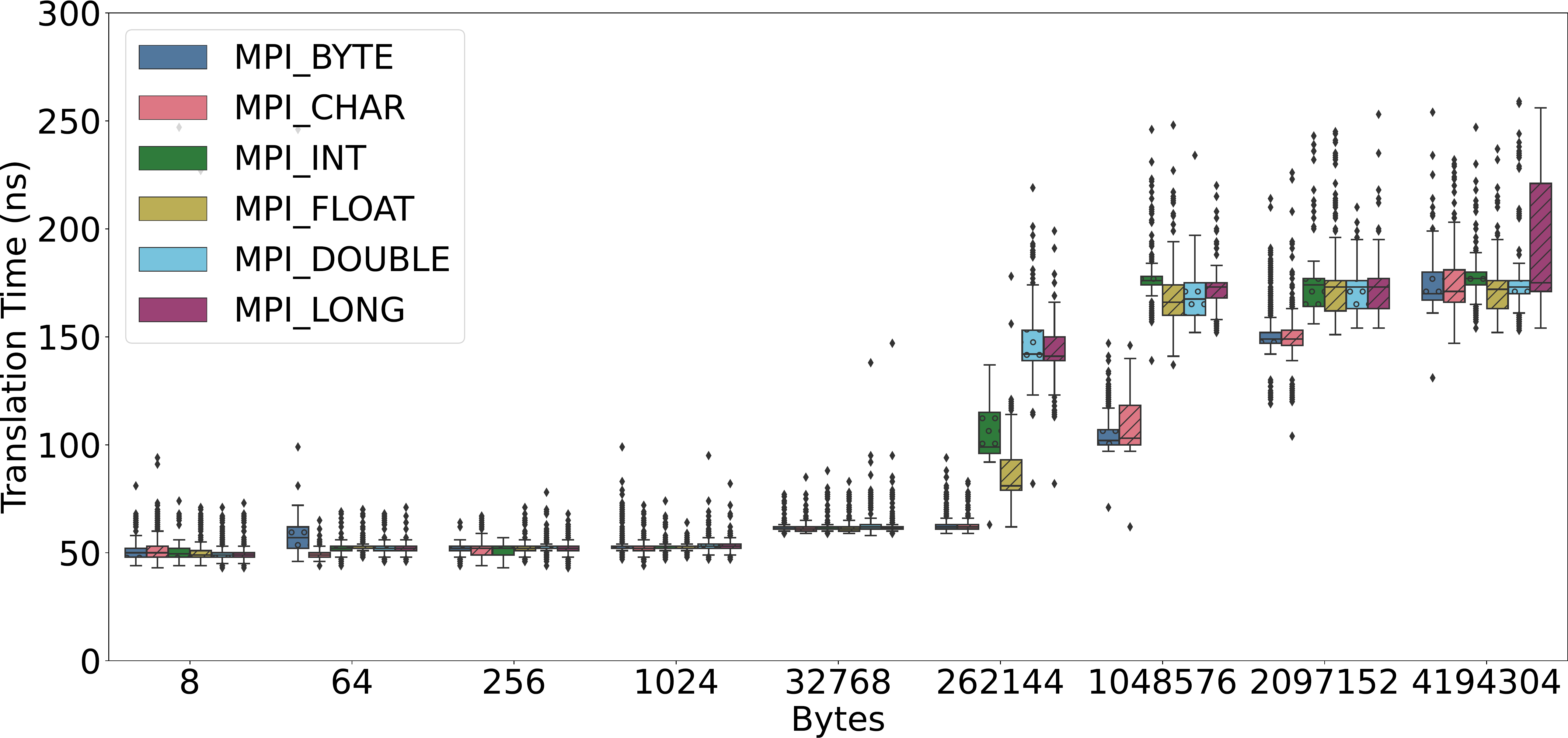}
\caption{Comparing the datatype translation overhead in \emph{MPIWasm}.}
        
\label{fig:tranlationoverhead}
\end{figure}

To measure the datatype translation overhead in \emph{MPIWasm}, we implement a custom \texttt{PingPong} application that sends/receives messages of varying sizes between two processes and iterates over the different MPI datatypes, i.e., \texttt{BYTE}, \texttt{CHAR}, \texttt{INT}, \texttt{FLOAT}, \texttt{DOUBLE}, and \texttt{LONG}. Following this, we instrument the \texttt{Send} routine in \emph{MPIWasm} to determine the latency for translating the different datatypes. Finally, we execute the application on our HPC system. Figure~\ref{fig:tranlationoverhead} shows the translation overhead for different datatypes and message sizes in \emph{MPIWasm}. We observe an average overhead of  $85.44$ns, $84.72$ns, $99.78$ns, $96.32$ns, $103.35$ns, and $104.79$ns across all message sizes for the MPI datatypes \texttt{BYTE}, \texttt{CHAR}, \texttt{INT}, \texttt{FLOAT}, \texttt{DOUBLE}, and \texttt{LONG} respectively. We observe an increase in the translation overhead for message sizes greater than $256$KB. This can be attributed to an increased latency for acquiring read locks from the \texttt{Env} structure that maintains the global state for translations in \emph{MPIWasm} (\S\ref{sec:funcimplementation}).

\section{Into the Future: Wasm and HPC}
\label{sec:futurehpcwasm}

In this section, we highlight and discuss the different extensions proposed by the Wasm community to the current Wasm specification~\cite{webassembly} that can be implemented in an embedder for HPC applications to enhance performance and portability.

\textbf{Controlled Threading for Wasm modules}. The Wasm Threads proposal~\cite{wasm_threads_proposal} lays the foundation for utilizing Wasm for  multithreaded algorithms. It enables Wasm modules to define shared memories, informing the embedder that the module expects the memory to be accessed
by multiple threads. To enable safe multithreaded access to shared memory, the proposal also defines atomic Wasm instructions that can be used to implement locks and atomic data structures in the functions of the module. To enable HPC applications to make use of the functionality added by the threads proposal, an API that allows Wasm modules to create additional threads on its own needs to be added to the embedder. Implementing the POSIX threads~\cite{posixthreads} and OpenMP~\cite{dagum1998} APIs in the embedder would enable compatibility with the threading code in existing HPC applications.

\textbf{Wasm Extended SIMD}. The current Wasm SIMD proposal~\cite{webassembly} only specifies 128-bit SIMD instructions, while modern CPUs support higher-width-SIMD, for instance the \texttt{AVX-512} instruction set extensions for x86\_64, which specifies 512-bit SIMD instructions. Towards this, the Wasm Flexible Vector proposal~\cite{wasm_flexible_vectors} aims to provide support for SIMD instructions that are wider than 128-bit. Moreover, the Wasm relaxed SIMD instructions~\cite{wasm_relaxed_simd} aim to make it possible to utilize hardware SIMD instructions that are not well defined, i.e., they differ in rounding behavior from the Wasm specification. Implementing these proposals in the embedder would allow compiled Wasm modules for HPC applications that contain vectorizable code to make better use of SIMD instructions available in modern CPU architectures.

\textbf{Wasm PGAS}. Partitioned Global Address Space (PGAS) is a programming model for parallel distributed memory applications that introduces a memory address space that spans the local memory of multiple processes. With a memory address from this global address space a process can read from and write to the memory of other processes. Since Wasm already specifies the concept of defining and importing memories, the embedder could be extended to provide non-local memory to the Wasm module. To support this use-case, the Wasm Multi-Memory proposal~\cite{wasm_multi_memory} needs to be implemented, which allows a Wasm module to define or import more than one memory.

\textbf{Dynamic Linking of Wasm Modules}. While there is existing work on establishing an ABI for dynamic linking between Wasm modules~\cite{wasm_dynamic_linking}, it has not been standardized yet. Supporting dynamic linking would significantly decrease the size of Wasm binaries for more complex applications as they would no longer need to statically link parts of \texttt{wasi-libc}. For HPC, it would enable commonly used libraries such as \texttt{BLAS} to be provided by \emph{MPIWasm}. Combining dynamic linking with efforts to provide repositories for Wasm modules such as \texttt{WAPM}~\cite{wapm} could enable automatic dependency management for Wasm applications.

\textbf{Compiling Fortran applications to Wasm} Currently, the support for compiling Fortran-based applications to Wasm is very nascent with only one known attempt based on \texttt{Dragon\-egg}~\cite{dragonegg}. However, the implementation of the Memory64 proposal~\cite{wasm_memory64} should enable the usage of existing Fortran LLVM compilers such as \texttt{F18} for easily compiling Fortran-based applications to Wasm~\cite{fortranwasm}.


\textbf{Wasm and Accelerators} The module execution hints proposal~\cite{wasmgpus} highlights the changes required in the Wasm specification to enable the support for executing Wasm modules on hardware accelerators such as GPUs. Implementing the proposal in the embedder would enable compatibility with existing GPU-based HPC applications.

\begin{figure}[t]
\begin{subfigure}{0.85\columnwidth}
        \includegraphics[width=0.45\columnwidth]{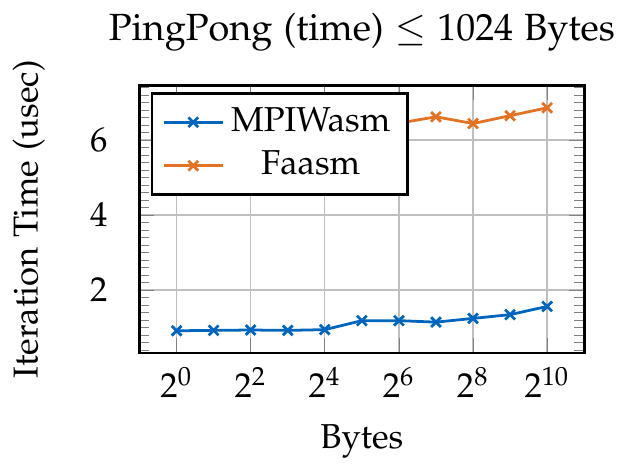}
        \includegraphics[width=0.45\columnwidth]{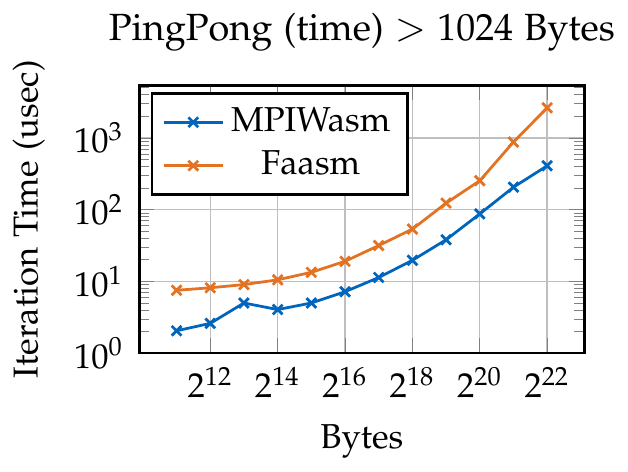}
        
\end{subfigure}
\caption{Comparing the performance of \emph{MPIWasm} and Faasm~\cite{faasm}.}
        
\label{fig:pingpongfaasm}
\vspace{-0.74cm}
\end{figure}

\section{Related Work}
\label{sec:related_work}
\textbf{Solutions for Packaging and distributing HPC applications}. 
Recently several HPC-focused tools such as Charliecloud~\cite{priedhorsky2017charliecloud} and  Singularity~\cite{kurtzer2017singularity} have been introduced for distributing HPC applications through containerization. In contrast, we utilize the universal binary instruction format Wasm to package and distribute HPC applications. Moreover, while containerization requires building HPC application containers for different platforms, HPC applications can be compiled once to Wasm and executed on any platform using a supporting Wasm embedder.

\textbf{MPI and WebAssembly}. To the best of our knowledge, \emph{Faasm}~\cite{faasm} is the only compute platform that enables the execution of MPI applications compiled to Wasm. It is based on a gRPC-based distributed messaging library called \texttt{Faabric} and contains a Wasm runtime, a workload scheduler, and a distributed state store. In order to run an application on Faasm, it needs to be compiled to Wasm and uploaded to the shared function storage. Following this, the application can be invoked using events such as HTTP requests. For supporting MPI applications, Faasm implements only a subset of the MPI-1 specification on top of its messaging library and its own workload scheduler. Moreover, it also does not support user-defined communicators required by the Intel MPI benchmarks~(\S\ref{sec:sysdescription}). In contrast, we take an inverted approach, where \emph{MPIWasm} builds on top of existing native MPI libraries and provides a way for Wasm modules to call functions from them efficiently. Figure~\ref{fig:pingpongfaasm} shows the performance comparison between \emph{MPIWasm} and Faasm for the \texttt{PingPong} benchmark (\S\ref{sec:benchmarksapplications}). With \emph{MPIWasm}, we achieve a GM average speedup of $4.28$x across all message sizes as compared to Faasm.


\section{Conclusion and Future Work}
\label{sec:conclusion}
In this paper, we took the first step towards bringing WebAssembly to the HPC ecosystem and presented \emph{MPIWasm}, a Wasm embedder that enables high performance execution of MPI applications compiled to Wasm across different processor architectures. In the future, we plan to extend \emph{MPIWasm} to support acceralators such as GPUs found on HPC systems~\cite{wasmgpus}.

\section{Acknowledgement}
We thank our shepherd Milind Chabbi for his help in preparing the final version of this paper. Furthermore, we thank the anonymous reviewers for their insightful comments and valuable feedback that significantly improved the quality of this paper. This work was supported by the funding of the German Federal Ministry of Education and Research (BMBF) in the scope of the Software Campus program.

\bibliographystyle{ACM-Reference-Format}
\bibliography{serverless}


\begin{thebibliography}{92}


\ifx \showCODEN    \undefined \def \showCODEN     #1{\unskip}     \fi
\ifx \showDOI      \undefined \def \showDOI       #1{#1}\fi
\ifx \showISBNx    \undefined \def \showISBNx     #1{\unskip}     \fi
\ifx \showISBNxiii \undefined \def \showISBNxiii  #1{\unskip}     \fi
\ifx \showISSN     \undefined \def \showISSN      #1{\unskip}     \fi
\ifx \showLCCN     \undefined \def \showLCCN      #1{\unskip}     \fi
\ifx \shownote     \undefined \def \shownote      #1{#1}          \fi
\ifx \showarticletitle \undefined \def \showarticletitle #1{#1}   \fi
\ifx \showURL      \undefined \def \showURL       {\relax}        \fi
\providecommand\bibfield[2]{#2}
\providecommand\bibinfo[2]{#2}
\providecommand\natexlab[1]{#1}
\providecommand\showeprint[2][]{arXiv:#2}

\bibitem[\protect\citeauthoryear{??}{gra}{[n.d.]}]%
        {graviton}
 \bibinfo{year}{[n.d.]}\natexlab{}.
\newblock \bibinfo{booktitle}{\emph{AWS Graviton 2 Processors}}.
\newblock
\urldef\tempurl%
\url{https://aws.amazon.com/ec2/graviton/}
\showURL{%
\tempurl}


\bibitem[\protect\citeauthoryear{??}{bla}{[n.d.]}]%
        {blake3}
 \bibinfo{year}{[n.d.]}\natexlab{}.
\newblock \bibinfo{booktitle}{\emph{Blake-3 Hash function}}.
\newblock
\urldef\tempurl%
\url{https://github.com/BLAKE3-team/BLAKE3}
\showURL{%
\tempurl}


\bibitem[\protect\citeauthoryear{??}{cra}{[n.d.]}]%
        {cranelift}
 \bibinfo{year}{[n.d.]}\natexlab{}.
\newblock \bibinfo{booktitle}{\emph{Cranelift Compiler}}.
\newblock
\urldef\tempurl%
\url{https://github.com/bytecodealliance/wasmtime/tree/main/cranelift}
\showURL{%
\tempurl}


\bibitem[\protect\citeauthoryear{??}{omn}{[n.d.]}]%
        {omnipath}
 \bibinfo{year}{[n.d.]}\natexlab{}.
\newblock \bibinfo{booktitle}{\emph{Intel Omni-Path Fabric}}.
\newblock
\urldef\tempurl%
\url{https://www.intel.com/content/www/us/en/high-performance-computing-fabrics/omni-path-fabric-software-components.html}
\showURL{%
\tempurl}


\bibitem[\protect\citeauthoryear{??}{psm}{[n.d.]}]%
        {psm2}
 \bibinfo{year}{[n.d.]}\natexlab{}.
\newblock \bibinfo{booktitle}{\emph{Intel Performance Scaled Messaging 2}}.
\newblock
\urldef\tempurl%
\url{https://github.com/cornelisnetworks/opa-psm2/blob/master/README}
\showURL{%
\tempurl}


\bibitem[\protect\citeauthoryear{??}{mpi}{[n.d.]a}]%
        {mpi_standard}
 \bibinfo{year}{[n.d.]}\natexlab{a}.
\newblock \bibinfo{booktitle}{\emph{{MPI}: A {Message-Passing} Interface
  Standard}}.
\newblock
\urldef\tempurl%
\url{https://www.mpi-forum.org/docs/mpi-3.1/mpi31-report.pdf}
\showURL{%
\tempurl}


\bibitem[\protect\citeauthoryear{??}{rsm}{[n.d.]}]%
        {rsmpi}
 \bibinfo{year}{[n.d.]}\natexlab{}.
\newblock \bibinfo{booktitle}{\emph{MPI bindings for Rust}}.
\newblock
\urldef\tempurl%
\url{https://github.com/rsmpi/rsmpi}
\showURL{%
\tempurl}


\bibitem[\protect\citeauthoryear{??}{mpi}{[n.d.]b}]%
        {mpich}
 \bibinfo{year}{[n.d.]}\natexlab{b}.
\newblock \bibinfo{booktitle}{\emph{{MPICH}: {High-Performance} portable
  {MPI}}}.
\newblock
\urldef\tempurl%
\url{https://www.mpich.org/}
\showURL{%
\tempurl}


\bibitem[\protect\citeauthoryear{??}{mva}{[n.d.]}]%
        {mvapich}
 \bibinfo{year}{[n.d.]}\natexlab{}.
\newblock \bibinfo{booktitle}{\emph{{MVAPICH}:{MPI} over InfiniBand, Omni-Path,
  Ethernet/iWARP, and RoCE}}.
\newblock
\urldef\tempurl%
\url{http://mvapich.cse.ohio-state.edu/}
\showURL{%
\tempurl}


\bibitem[\protect\citeauthoryear{??}{ope}{[n.d.]}]%
        {openmpi}
 \bibinfo{year}{[n.d.]}\natexlab{}.
\newblock \bibinfo{booktitle}{\emph{{OpenMPI}: {Open-Source} {High-Performance}
  computing}}.
\newblock
\urldef\tempurl%
\url{https://www.open-mpi.org/}
\showURL{%
\tempurl}


\bibitem[\protect\citeauthoryear{??}{top}{[n.d.]}]%
        {top500_june}
 \bibinfo{year}{[n.d.]}\natexlab{}.
\newblock \bibinfo{booktitle}{\emph{The {Top500} list}}.
\newblock
\urldef\tempurl%
\url{https://www.top500.org/lists/top500/2022/06/}
\showURL{%
\tempurl}


\bibitem[\protect\citeauthoryear{??}{was}{[n.d.]}]%
        {wasm_dynamic_linking}
 \bibinfo{year}{[n.d.]}\natexlab{}.
\newblock \bibinfo{booktitle}{\emph{WebAssembly Dynamic Linking}}.
\newblock
\urldef\tempurl%
\url{https://github.com/WebAssembly/tool-conventions/blob/master/DynamicLinking.md}
\showURL{%
\tempurl}


\bibitem[\protect\citeauthoryear{??}{was}{2020a}]%
        {wasm_flexible_vectors}
 \bibinfo{year}{2020}\natexlab{a}.
\newblock \bibinfo{booktitle}{\emph{Flexible Vectors Proposal for
  WebAssembly}}.
\newblock
\urldef\tempurl%
\url{https://github.com/WebAssembly/flexible-vectors/blob/master/proposals/flexible-vectors/Overview.md}
\showURL{%
\tempurl}


\bibitem[\protect\citeauthoryear{??}{was}{2020b}]%
        {wasm_multi_memory}
 \bibinfo{year}{2020}\natexlab{b}.
\newblock \bibinfo{booktitle}{\emph{Multi Memory Proposal for WebAssembly}}.
\newblock
\urldef\tempurl%
\url{https://github.com/WebAssembly/multi-memory/blob/master/proposals/multi-memory/Overview.md}
\showURL{%
\tempurl}


\bibitem[\protect\citeauthoryear{??}{pos}{2020}]%
        {posixthreads}
 \bibinfo{year}{2020}\natexlab{}.
\newblock \bibinfo{booktitle}{\emph{POSIX Threads}}.
\newblock
\urldef\tempurl%
\url{https://man7.org/linux/man-pages/man7/pthreads.7.html}
\showURL{%
\tempurl}


\bibitem[\protect\citeauthoryear{??}{was}{2020c}]%
        {wasm_threads_proposal}
 \bibinfo{year}{2020}\natexlab{c}.
\newblock \bibinfo{booktitle}{\emph{Threading proposal for WebAssembly}}.
\newblock
\urldef\tempurl%
\url{https://github.com/WebAssembly/threads/blob/master/proposals/threads/Overview.md}
\showURL{%
\tempurl}


\bibitem[\protect\citeauthoryear{??}{dra}{2021}]%
        {dragonegg}
 \bibinfo{year}{2021}\natexlab{}.
\newblock \bibinfo{booktitle}{\emph{Compiling Fortran applications to Wasm}}.
\newblock
\urldef\tempurl%
\url{https://github.com/StarGate01/Full-Stack-Fortran}
\showURL{%
\tempurl}


\bibitem[\protect\citeauthoryear{??}{for}{2021}]%
        {fortranwasm}
 \bibinfo{year}{2021}\natexlab{}.
\newblock \bibinfo{booktitle}{\emph{Fortran and WebAssembly}}.
\newblock
\urldef\tempurl%
\url{https://chrz.de/2020/04/21/fortran-in-the-browser/}
\showURL{%
\tempurl}


\bibitem[\protect\citeauthoryear{??}{len}{2021}]%
        {lenovodsg}
 \bibinfo{year}{2021}\natexlab{}.
\newblock \bibinfo{booktitle}{\emph{Lenovo DSS-G}}.
\newblock
\urldef\tempurl%
\url{https://tinyurl.com/ytmjjauy}
\showURL{%
\tempurl}


\bibitem[\protect\citeauthoryear{??}{was}{2021a}]%
        {wasm_memory64}
 \bibinfo{year}{2021}\natexlab{a}.
\newblock \bibinfo{booktitle}{\emph{Memory64 Proposal for WebAssembly}}.
\newblock
\urldef\tempurl%
\url{https://github.com/WebAssembly/memory64/blob/master/proposals/memory64/Overview.md}
\showURL{%
\tempurl}


\bibitem[\protect\citeauthoryear{??}{was}{2021b}]%
        {wasm_relaxed_simd}
 \bibinfo{year}{2021}\natexlab{b}.
\newblock \bibinfo{booktitle}{\emph{Relaxed SIMD Proposal for WebAssembly}}.
\newblock
\urldef\tempurl%
\url{https://github.com/WebAssembly/relaxed-simd/blob/main/proposals/relaxed-simd/Overview.md}
\showURL{%
\tempurl}


\bibitem[\protect\citeauthoryear{??}{was}{2021c}]%
        {wasi_libc}
 \bibinfo{year}{2021}\natexlab{c}.
\newblock \bibinfo{booktitle}{\emph{WASI Libc}}.
\newblock
\urldef\tempurl%
\url{https://github.com/WebAssembly/wasi-libc}
\showURL{%
\tempurl}


\bibitem[\protect\citeauthoryear{??}{was}{2021d}]%
        {wasmgpus}
 \bibinfo{year}{2021}\natexlab{d}.
\newblock \bibinfo{booktitle}{\emph{Wasm and GPUs}}.
\newblock
\urldef\tempurl%
\url{https://github.com/WebAssembly/design/issues/1448}
\showURL{%
\tempurl}


\bibitem[\protect\citeauthoryear{??}{neo}{2022}]%
        {neoverse}
 \bibinfo{year}{2022}\natexlab{}.
\newblock \bibinfo{booktitle}{\emph{ARM Neoverse-N1 Architecture}}.
\newblock
\urldef\tempurl%
\url{https://www.arm.com/products/silicon-ip-cpu/neoverse/neoverse-n1}
\showURL{%
\tempurl}


\bibitem[\protect\citeauthoryear{??}{cra}{2022}]%
        {craneliftir}
 \bibinfo{year}{2022}\natexlab{}.
\newblock \bibinfo{booktitle}{\emph{Comparing Cranelift-IR and LLVM-IR}}.
\newblock
\urldef\tempurl%
\url{https://github.com/bytecodealliance/wasmtime/blob/main/cranelift/docs/compare-llvm.md}
\showURL{%
\tempurl}


\bibitem[\protect\citeauthoryear{??}{gli}{2022}]%
        {glibc}
 \bibinfo{year}{2022}\natexlab{}.
\newblock \bibinfo{booktitle}{\emph{The GNU C Library (glibc)}}.
\newblock
\urldef\tempurl%
\url{https://www.gnu.org/software/libc/}
\showURL{%
\tempurl}


\bibitem[\protect\citeauthoryear{??}{lib}{2022}]%
        {libloading}
 \bibinfo{year}{2022}\natexlab{}.
\newblock \bibinfo{booktitle}{\emph{The libloading library}}.
\newblock
\urldef\tempurl%
\url{https://docs.rs/libloading/latest/libloading/}
\showURL{%
\tempurl}


\bibitem[\protect\citeauthoryear{??}{was}{2022}]%
        {wasi_sdk_new}
 \bibinfo{year}{2022}\natexlab{}.
\newblock \bibinfo{booktitle}{\emph{WASI SDK}}.
\newblock
\urldef\tempurl%
\url{https://github.com/WebAssembly/wasi-sdk}
\showURL{%
\tempurl}


\bibitem[\protect\citeauthoryear{{14 Tech Companies Embracing Container
  Technology.}}{{14 Tech Companies Embracing Container Technology.}}{[n.d.]}]%
        {cloud_container}
\bibfield{author}{\bibinfo{person}{{14 Tech Companies Embracing Container
  Technology.}}} \bibinfo{year}{[n.d.]}\natexlab{}.
\newblock
  \bibinfo{howpublished}{\url{https://learn.g2.com/container-technology}}.
\newblock
\newblock
\shownote{Accessed 09/29/2021.}


\bibitem[\protect\citeauthoryear{Abraham, Paul, Khan, and Butt}{Abraham
  et~al\mbox{.}}{2020}]%
        {abraham2020use}
\bibfield{author}{\bibinfo{person}{Subil Abraham}, \bibinfo{person}{Arnab~K
  Paul}, \bibinfo{person}{Redwan Ibne~Seraj Khan}, {and} \bibinfo{person}{Ali~R
  Butt}.} \bibinfo{year}{2020}\natexlab{}.
\newblock \showarticletitle{On the use of containers in high performance
  computing environments}. In \bibinfo{booktitle}{\emph{2020 IEEE 13th
  International Conference on Cloud Computing (CLOUD)}}. IEEE,
  \bibinfo{pages}{284--293}.
\newblock
\urldef\tempurl%
\url{https://doi.org/10.1109/CLOUD49709.2020.00048}
\showDOI{\tempurl}


\bibitem[\protect\citeauthoryear{Bailey, Barszcz, Barton, Browning, Carter,
  Dagum, Fatoohi, Frederickson, Lasinski, Schreiber, Simon, Venkatakrishnan,
  and Weeratunga}{Bailey et~al\mbox{.}}{1991}]%
        {Bailey:1991:NPB:125826.125925}
\bibfield{author}{\bibinfo{person}{D.~H. Bailey}, \bibinfo{person}{E. Barszcz},
  \bibinfo{person}{J.~T. Barton}, \bibinfo{person}{D.~S. Browning},
  \bibinfo{person}{R.~L. Carter}, \bibinfo{person}{L. Dagum},
  \bibinfo{person}{R.~A. Fatoohi}, \bibinfo{person}{P.~O. Frederickson},
  \bibinfo{person}{T.~A. Lasinski}, \bibinfo{person}{R.~S. Schreiber},
  \bibinfo{person}{H.~D. Simon}, \bibinfo{person}{V. Venkatakrishnan}, {and}
  \bibinfo{person}{S.~K. Weeratunga}.} \bibinfo{year}{1991}\natexlab{}.
\newblock \showarticletitle{The NAS Parallel Benchmarks\&Mdash;Summary and
  Preliminary Results}. In \bibinfo{booktitle}{\emph{Proceedings of the 1991
  ACM/IEEE Conference on Supercomputing}} (Albuquerque, New Mexico, USA)
  \emph{(\bibinfo{series}{Supercomputing '91})}. \bibinfo{publisher}{ACM},
  \bibinfo{address}{New York, NY, USA}, \bibinfo{pages}{158--165}.
\newblock
\showISBNx{0-89791-459-7}
\urldef\tempurl%
\url{https://doi.org/10.1145/125826.125925}
\showDOI{\tempurl}


\bibitem[\protect\citeauthoryear{Bastien, Lively, and Ahn}{Bastien
  et~al\mbox{.}}{2019}]%
        {clang_wasm}
\bibfield{author}{\bibinfo{person}{JF Bastien}, \bibinfo{person}{Thomas
  Lively}, {and} \bibinfo{person}{Heejin Ahn}.}
  \bibinfo{year}{2019}\natexlab{}.
\newblock \bibinfo{booktitle}{\emph{LLVM WebAssembly Backend}}.
\newblock
\urldef\tempurl%
\url{https://github.com/llvm/llvm-project/blob/main/llvm/lib/Target/WebAssembly/README.txt}
\showURL{%
\tempurl}


\bibitem[\protect\citeauthoryear{Benedicic, Cruz, Madonna, and
  Mariotti}{Benedicic et~al\mbox{.}}{2019}]%
        {benedicic2019sarus}
\bibfield{author}{\bibinfo{person}{Lucas Benedicic}, \bibinfo{person}{Felipe~A
  Cruz}, \bibinfo{person}{Alberto Madonna}, {and} \bibinfo{person}{Kean
  Mariotti}.} \bibinfo{year}{2019}\natexlab{}.
\newblock \showarticletitle{Sarus: Highly scalable docker containers for hpc
  systems}. In \bibinfo{booktitle}{\emph{International Conference on High
  Performance Computing}}. Springer, \bibinfo{pages}{46--60}.
\newblock
\urldef\tempurl%
\url{https://doi.org/10.1007/978-3-030-34356-9_5}
\showDOI{\tempurl}


\bibitem[\protect\citeauthoryear{Bernholdt, Boehm, Bosilca, Gorentla~Venkata,
  Grant, Naughton, Pritchard, Schulz, and Vallee}{Bernholdt
  et~al\mbox{.}}{2017}]%
        {bernholdt2017survey}
\bibfield{author}{\bibinfo{person}{David~E Bernholdt}, \bibinfo{person}{Swen
  Boehm}, \bibinfo{person}{George Bosilca}, \bibinfo{person}{Manjunath
  Gorentla~Venkata}, \bibinfo{person}{Ryan~E Grant}, \bibinfo{person}{Thomas
  Naughton}, \bibinfo{person}{Howard~P Pritchard}, \bibinfo{person}{Martin
  Schulz}, {and} \bibinfo{person}{Geoffroy~R Vallee}.}
  \bibinfo{year}{2017}\natexlab{}.
\newblock \showarticletitle{A survey of MPI usage in the US exascale computing
  project}.
\newblock \bibinfo{journal}{\emph{Concurrency and Computation: Practice and
  Experience}} (\bibinfo{year}{2017}).
\newblock
\urldef\tempurl%
\url{https://doi.org/10.1002/cpe.4851}
\showURL{%
\tempurl}


\bibitem[\protect\citeauthoryear{{Blazor}}{{Blazor}}{[n.d.]}]%
        {Blazor}
\bibfield{author}{\bibinfo{person}{{Blazor}}.}
  \bibinfo{year}{[n.d.]}\natexlab{}.
\newblock
  \bibinfo{howpublished}{\url{https://dotnet.microsoft.com/apps/aspnet/web-apps/blazor}}.
\newblock
\newblock
\shownote{Accessed 09/27/2021.}


\bibitem[\protect\citeauthoryear{{Brian Austin et al.}}{{Brian Austin et
  al.}}{[n.d.]}]%
        {nersc}
\bibfield{author}{\bibinfo{person}{{Brian Austin et al.}}}
  \bibinfo{year}{[n.d.]}\natexlab{}.
\newblock \bibinfo{title}{NERSC-10 workload analysis}.
\newblock
  \bibinfo{howpublished}{\url{https://portal.nersc.gov/project/m888/nersc10/workload/N10_Workload_Analysis.latest.pdf}}.
\newblock
\newblock
\shownote{Accessed 09/27/2021.}


\bibitem[\protect\citeauthoryear{{Bytecode Alliance}}{{Bytecode
  Alliance}}{[n.d.]}]%
        {wasmtime}
\bibfield{author}{\bibinfo{person}{{Bytecode Alliance}}.}
  \bibinfo{year}{[n.d.]}\natexlab{}.
\newblock \bibinfo{booktitle}{\emph{Wasmtime - A standalone runtime for
  WebAssembly}}.
\newblock
\urldef\tempurl%
\url{https://github.com/bytecodealliance/wasmtime}
\showURL{%
\tempurl}


\bibitem[\protect\citeauthoryear{Chadha, Jindal, and Gerndt}{Chadha
  et~al\mbox{.}}{2021}]%
        {architecture}
\bibfield{author}{\bibinfo{person}{Mohak Chadha}, \bibinfo{person}{Anshul
  Jindal}, {and} \bibinfo{person}{Michael Gerndt}.}
  \bibinfo{year}{2021}\natexlab{}.
\newblock \showarticletitle{Architecture-Specific Performance Optimization of
  Compute-Intensive FaaS Functions}. In \bibinfo{booktitle}{\emph{2021 IEEE
  14th International Conference on Cloud Computing (CLOUD)}}.
  \bibinfo{pages}{478--483}.
\newblock
\urldef\tempurl%
\url{https://doi.org/10.1109/CLOUD53861.2021.00062}
\showDOI{\tempurl}


\bibitem[\protect\citeauthoryear{{CMake}}{{CMake}}{[n.d.]}]%
        {cmake}
\bibfield{author}{\bibinfo{person}{{CMake}}.}
  \bibinfo{year}{[n.d.]}\natexlab{}.
\newblock \bibinfo{howpublished}{\url{https://cmake.org/}}.
\newblock
\newblock
\shownote{Accessed on 08/12/2022.}


\bibitem[\protect\citeauthoryear{{Compiling Rust to WebAssembly}}{{Compiling
  Rust to WebAssembly}}{[n.d.]}]%
        {rusttowasm}
\bibfield{author}{\bibinfo{person}{{Compiling Rust to WebAssembly}}.}
  \bibinfo{year}{[n.d.]}\natexlab{}.
\newblock
  \bibinfo{howpublished}{\url{https://developer.mozilla.org/en-US/docs/WebAssembly/Rust_to_wasm}}.
\newblock
\newblock
\shownote{Accessed 09/27/2021.}


\bibitem[\protect\citeauthoryear{Dagum and Enon}{Dagum and Enon}{1998}]%
        {dagum1998}
\bibfield{author}{\bibinfo{person}{Leonardo Dagum} {and}
  \bibinfo{person}{Rameshm Enon}.} \bibinfo{year}{1998}\natexlab{}.
\newblock \showarticletitle{OpenMP: an industry standard API for shared-memory
  programming}.
\newblock \bibinfo{journal}{\emph{Computational Science \& Engineering, IEEE}}
  \bibinfo{volume}{5}, \bibinfo{number}{1} (\bibinfo{year}{1998}),
  \bibinfo{pages}{46--55}.
\newblock


\bibitem[\protect\citeauthoryear{Denis}{Denis}{[n.d.]}]%
        {runtimebound}
\bibfield{author}{\bibinfo{person}{Frank Denis}.}
  \bibinfo{year}{[n.d.]}\natexlab{}.
\newblock \bibinfo{booktitle}{\emph{Memory management in WebAssembly: guide for
  C and Rust programmers}}.
\newblock
\urldef\tempurl%
\url{https://www.fastly.com/blog/webassembly-memory-management-guide-for-c-rust-programmers}
\showURL{%
\tempurl}


\bibitem[\protect\citeauthoryear{{Docker}}{{Docker}}{[n.d.]}]%
        {docker}
\bibfield{author}{\bibinfo{person}{{Docker}}.}
  \bibinfo{year}{[n.d.]}\natexlab{}.
\newblock \bibinfo{howpublished}{\url{https://www.docker.com/}}.
\newblock
\newblock
\shownote{Accessed 07/07/2022.}


\bibitem[\protect\citeauthoryear{{Docker Buildx}}{{Docker Buildx}}{[n.d.]}]%
        {dockerbuildx}
\bibfield{author}{\bibinfo{person}{{Docker Buildx}}.}
  \bibinfo{year}{[n.d.]}\natexlab{}.
\newblock
  \bibinfo{howpublished}{\url{https://docs.docker.com/buildx/working-with-buildx/}}.
\newblock
\newblock
\shownote{Accessed 07/07/2022.}


\bibitem[\protect\citeauthoryear{{Emscripten}}{{Emscripten}}{[n.d.]}]%
        {emscripten}
\bibfield{author}{\bibinfo{person}{{Emscripten}}.}
  \bibinfo{year}{[n.d.]}\natexlab{}.
\newblock \bibinfo{howpublished}{\url{https://emscripten.org/}}.
\newblock
\newblock
\shownote{Accessed 09/27/2021.}


\bibitem[\protect\citeauthoryear{{FileSystemCache in Wasmer}}{{FileSystemCache
  in Wasmer}}{[n.d.]}]%
        {filesystemcache}
\bibfield{author}{\bibinfo{person}{{FileSystemCache in Wasmer}}.}
  \bibinfo{year}{[n.d.]}\natexlab{}.
\newblock
  \bibinfo{howpublished}{\url{https://docs.rs/wasmer-cache/3.0.2/wasmer_cache/struct.FileSystemCache.html}}.
\newblock
\newblock
\shownote{Accessed 12/11/2022.}


\bibitem[\protect\citeauthoryear{Gadepalli, McBride, Peach, Cherkasova, and
  Parmer}{Gadepalli et~al\mbox{.}}{2020}]%
        {sledge}
\bibfield{author}{\bibinfo{person}{Phani~Kishore Gadepalli},
  \bibinfo{person}{Sean McBride}, \bibinfo{person}{Gregor Peach},
  \bibinfo{person}{Ludmila Cherkasova}, {and} \bibinfo{person}{Gabriel
  Parmer}.} \bibinfo{year}{2020}\natexlab{}.
\newblock \showarticletitle{Sledge: A Serverless-First, Light-Weight Wasm
  Runtime for the Edge}. In \bibinfo{booktitle}{\emph{Proceedings of the 21st
  International Middleware Conference}} (Delft, Netherlands)
  \emph{(\bibinfo{series}{Middleware '20})}. \bibinfo{publisher}{Association
  for Computing Machinery}, \bibinfo{address}{New York, NY, USA},
  \bibinfo{pages}{265–279}.
\newblock
\showISBNx{9781450381536}
\urldef\tempurl%
\url{https://doi.org/10.1145/3423211.3425680}
\showDOI{\tempurl}


\bibitem[\protect\citeauthoryear{Gantikow, Walter, and Reich}{Gantikow
  et~al\mbox{.}}{2020}]%
        {gantikow2020rootless}
\bibfield{author}{\bibinfo{person}{Holger Gantikow}, \bibinfo{person}{Steffen
  Walter}, {and} \bibinfo{person}{Christoph Reich}.}
  \bibinfo{year}{2020}\natexlab{}.
\newblock \showarticletitle{Rootless Containers with Podman for HPC}. In
  \bibinfo{booktitle}{\emph{International Conference on High Performance
  Computing}}. Springer, \bibinfo{pages}{343--354}.
\newblock
\urldef\tempurl%
\url{https://doi.org/10.1007/978-3-030-59851-8_23}
\showDOI{\tempurl}


\bibitem[\protect\citeauthoryear{Gerhardt, Bhimji, Canon, Fasel, Jacobsen,
  Mustafa, Porter, and Tsulaia}{Gerhardt et~al\mbox{.}}{2017}]%
        {gerhardt2017shifter}
\bibfield{author}{\bibinfo{person}{Lisa Gerhardt}, \bibinfo{person}{Wahid
  Bhimji}, \bibinfo{person}{Shane Canon}, \bibinfo{person}{Markus Fasel},
  \bibinfo{person}{Doug Jacobsen}, \bibinfo{person}{Mustafa Mustafa},
  \bibinfo{person}{Jeff Porter}, {and} \bibinfo{person}{Vakho Tsulaia}.}
  \bibinfo{year}{2017}\natexlab{}.
\newblock \showarticletitle{Shifter: Containers for hpc}. In
  \bibinfo{booktitle}{\emph{Journal of physics: Conference series}},
  Vol.~\bibinfo{volume}{898}. IOP Publishing, \bibinfo{pages}{082021}.
\newblock
\urldef\tempurl%
\url{https://doi.org/10.1088/1742-6596/898/8/082021}
\showDOI{\tempurl}


\bibitem[\protect\citeauthoryear{{GNU Make}}{{GNU Make}}{[n.d.]}]%
        {make}
\bibfield{author}{\bibinfo{person}{{GNU Make}}.}
  \bibinfo{year}{[n.d.]}\natexlab{}.
\newblock \bibinfo{howpublished}{\url{https://www.gnu.org/software/make/}}.
\newblock
\newblock
\shownote{Accessed on 08/12/2022.}


\bibitem[\protect\citeauthoryear{Gurdeep~Singh and Scholliers}{Gurdeep~Singh
  and Scholliers}{2019}]%
        {warduino}
\bibfield{author}{\bibinfo{person}{Robbert Gurdeep~Singh} {and}
  \bibinfo{person}{Christophe Scholliers}.} \bibinfo{year}{2019}\natexlab{}.
\newblock \showarticletitle{WARDuino: A Dynamic WebAssembly Virtual Machine for
  Programming Microcontrollers}. In \bibinfo{booktitle}{\emph{Proceedings of
  the 16th ACM SIGPLAN International Conference on Managed Programming
  Languages and Runtimes}} (Athens, Greece) \emph{(\bibinfo{series}{MPLR
  2019})}. \bibinfo{publisher}{Association for Computing Machinery},
  \bibinfo{address}{New York, NY, USA}, \bibinfo{pages}{27–36}.
\newblock
\showISBNx{9781450369770}
\urldef\tempurl%
\url{https://doi.org/10.1145/3357390.3361029}
\showDOI{\tempurl}


\bibitem[\protect\citeauthoryear{Haas, Rossberg, Schuff, Titzer, Holman,
  Gohman, Wagner, Zakai, and Bastien}{Haas et~al\mbox{.}}{2017}]%
        {webassembly}
\bibfield{author}{\bibinfo{person}{Andreas Haas}, \bibinfo{person}{Andreas
  Rossberg}, \bibinfo{person}{Derek~L. Schuff}, \bibinfo{person}{Ben~L.
  Titzer}, \bibinfo{person}{Michael Holman}, \bibinfo{person}{Dan Gohman},
  \bibinfo{person}{Luke Wagner}, \bibinfo{person}{Alon Zakai}, {and}
  \bibinfo{person}{JF Bastien}.} \bibinfo{year}{2017}\natexlab{}.
\newblock \showarticletitle{Bringing the Web up to Speed with WebAssembly}.
\newblock \bibinfo{journal}{\emph{SIGPLAN Not.}} \bibinfo{volume}{52},
  \bibinfo{number}{6} (\bibinfo{date}{June} \bibinfo{year}{2017}),
  \bibinfo{pages}{185–200}.
\newblock
\showISSN{0362-1340}
\urldef\tempurl%
\url{https://doi.org/10.1145/3140587.3062363}
\showDOI{\tempurl}


\bibitem[\protect\citeauthoryear{Hall and Ramachandran}{Hall and
  Ramachandran}{2019}]%
        {serverlessedge}
\bibfield{author}{\bibinfo{person}{Adam Hall} {and} \bibinfo{person}{Umakishore
  Ramachandran}.} \bibinfo{year}{2019}\natexlab{}.
\newblock \showarticletitle{An Execution Model for Serverless Functions at the
  Edge}. In \bibinfo{booktitle}{\emph{Proceedings of the International
  Conference on Internet of Things Design and Implementation}} (Montreal,
  Quebec, Canada) \emph{(\bibinfo{series}{IoTDI '19})}.
  \bibinfo{publisher}{Association for Computing Machinery},
  \bibinfo{address}{New York, NY, USA}, \bibinfo{pages}{225–236}.
\newblock
\showISBNx{9781450362832}
\urldef\tempurl%
\url{https://doi.org/10.1145/3302505.3310084}
\showDOI{\tempurl}


\bibitem[\protect\citeauthoryear{Hoefler and Belli}{Hoefler and Belli}{2015}]%
        {hoefler2015scientific}
\bibfield{author}{\bibinfo{person}{Torsten Hoefler} {and}
  \bibinfo{person}{Roberto Belli}.} \bibinfo{year}{2015}\natexlab{}.
\newblock \showarticletitle{Scientific benchmarking of parallel computing
  systems: twelve ways to tell the masses when reporting performance results}.
  In \bibinfo{booktitle}{\emph{Proceedings of the international conference for
  high performance computing, networking, storage and analysis}}.
  \bibinfo{pages}{1--12}.
\newblock
\urldef\tempurl%
\url{https://doi.org/10.1145/2807591.2807644}
\showDOI{\tempurl}


\bibitem[\protect\citeauthoryear{{Intel Corporation}}{{Intel
  Corporation}}{2018}]%
        {intel_mpi_benchmarks}
\bibfield{author}{\bibinfo{person}{{Intel Corporation}}.}
  \bibinfo{year}{2018}\natexlab{}.
\newblock \bibinfo{booktitle}{\emph{Introducing Intel® MPI Benchmarks}}.
\newblock
\urldef\tempurl%
\url{https://software.intel.com/content/www/us/en/develop/articles/intel-mpi-benchmarks.html}
\showURL{%
\tempurl}


\bibitem[\protect\citeauthoryear{Jangda, Powers, Berger, and Guha}{Jangda
  et~al\mbox{.}}{2019}]%
        {notsofast}
\bibfield{author}{\bibinfo{person}{Abhinav Jangda}, \bibinfo{person}{Bobby
  Powers}, \bibinfo{person}{Emery~D. Berger}, {and} \bibinfo{person}{Arjun
  Guha}.} \bibinfo{year}{2019}\natexlab{}.
\newblock \showarticletitle{Not So Fast: Analyzing the Performance of
  {WebAssembly} vs. Native Code}. In \bibinfo{booktitle}{\emph{2019 USENIX
  Annual Technical Conference (USENIX ATC 19)}}. \bibinfo{publisher}{USENIX
  Association}, \bibinfo{address}{Renton, WA}, \bibinfo{pages}{107--120}.
\newblock
\showISBNx{978-1-939133-03-8}
\urldef\tempurl%
\url{https://www.usenix.org/conference/atc19/presentation/jangda}
\showURL{%
\tempurl}


\bibitem[\protect\citeauthoryear{Kiener, Chadha, and Gerndt}{Kiener
  et~al\mbox{.}}{2021}]%
        {demystifying}
\bibfield{author}{\bibinfo{person}{Michael Kiener}, \bibinfo{person}{Mohak
  Chadha}, {and} \bibinfo{person}{Michael Gerndt}.}
  \bibinfo{year}{2021}\natexlab{}.
\newblock \showarticletitle{Towards Demystifying Intra-Function Parallelism in
  Serverless Computing}. In \bibinfo{booktitle}{\emph{Proceedings of the
  Seventh International Workshop on Serverless Computing (WoSC7) 2021}}
  (Virtual Event, Canada) \emph{(\bibinfo{series}{WoSC '21})}.
  \bibinfo{publisher}{Association for Computing Machinery},
  \bibinfo{address}{New York, NY, USA}, \bibinfo{pages}{42–49}.
\newblock
\showISBNx{9781450391726}
\urldef\tempurl%
\url{https://doi.org/10.1145/3493651.3493672}
\showDOI{\tempurl}


\bibitem[\protect\citeauthoryear{Kurtzer, Sochat, and Bauer}{Kurtzer
  et~al\mbox{.}}{2017}]%
        {kurtzer2017singularity}
\bibfield{author}{\bibinfo{person}{Gregory~M Kurtzer}, \bibinfo{person}{Vanessa
  Sochat}, {and} \bibinfo{person}{Michael~W Bauer}.}
  \bibinfo{year}{2017}\natexlab{}.
\newblock \showarticletitle{Singularity: Scientific containers for mobility of
  compute}.
\newblock \bibinfo{journal}{\emph{PloS one}} \bibinfo{volume}{12},
  \bibinfo{number}{5} (\bibinfo{year}{2017}), \bibinfo{pages}{e0177459}.
\newblock
\urldef\tempurl%
\url{https://doi.org/10.1371/journal.pone.0177459}
\showDOI{\tempurl}


\bibitem[\protect\citeauthoryear{{Lattner} and {Adve}}{{Lattner} and
  {Adve}}{2004}]%
        {llvm}
\bibfield{author}{\bibinfo{person}{C. {Lattner}} {and} \bibinfo{person}{V.
  {Adve}}.} \bibinfo{year}{2004}\natexlab{}.
\newblock \showarticletitle{LLVM: a compilation framework for lifelong program
  analysis transformation}. In \bibinfo{booktitle}{\emph{International
  Symposium on Code Generation and Optimization, 2004. CGO 2004.}}
  \bibinfo{pages}{75--86}.
\newblock
\urldef\tempurl%
\url{https://doi.org/10.1109/CGO.2004.1281665}
\showDOI{\tempurl}


\bibitem[\protect\citeauthoryear{{Lawrence Livermore National
  Laboratory}}{{Lawrence Livermore National Laboratory}}{2021}]%
        {ior_benchmark}
\bibfield{author}{\bibinfo{person}{{Lawrence Livermore National Laboratory}}.}
  \bibinfo{year}{2021}\natexlab{}.
\newblock \bibinfo{booktitle}{\emph{HPC IO Benchmark Repository}}.
\newblock
\urldef\tempurl%
\url{https://github.com/hpc/ior}
\showURL{%
\tempurl}


\bibitem[\protect\citeauthoryear{{Linux.}}{{Linux.}}{[n.d.]}]%
        {fakeroot}
\bibfield{author}{\bibinfo{person}{{Linux.}}}
  \bibinfo{year}{[n.d.]}\natexlab{}.
\newblock \bibinfo{title}{Fakeroot.}
\newblock \bibinfo{howpublished}{\url{https://wiki.debian.org/FakeRoot}}.
\newblock
\newblock
\shownote{Accessed 09/27/2021.}


\bibitem[\protect\citeauthoryear{{Linux Kernel Support for Miscellaneous Binary
  Formats}}{{Linux Kernel Support for Miscellaneous Binary Formats}}{[n.d.]}]%
        {binft_misc}
\bibfield{author}{\bibinfo{person}{{Linux Kernel Support for Miscellaneous
  Binary Formats}}.} \bibinfo{year}{[n.d.]}\natexlab{}.
\newblock
  \bibinfo{howpublished}{\url{https://www.kernel.org/doc/html/latest/admin-guide/binfmt-misc.html}}.
\newblock
\newblock
\shownote{Accessed 07/07/2022.}


\bibitem[\protect\citeauthoryear{Marjanovi{\'{c}}, Gracia, and
  Glass}{Marjanovi{\'{c}} et~al\mbox{.}}{2015}]%
        {perfhpcg}
\bibfield{author}{\bibinfo{person}{Vladimir Marjanovi{\'{c}}},
  \bibinfo{person}{Jos{\'e} Gracia}, {and} \bibinfo{person}{Colin~W. Glass}.}
  \bibinfo{year}{2015}\natexlab{}.
\newblock \showarticletitle{Performance Modeling of the HPCG Benchmark}. In
  \bibinfo{booktitle}{\emph{High Performance Computing Systems. Performance
  Modeling, Benchmarking, and Simulation}},
  \bibfield{editor}{\bibinfo{person}{Stephen~A. Jarvis},
  \bibinfo{person}{Steven~A. Wright}, {and} \bibinfo{person}{Simon~D. Hammond}}
  (Eds.). \bibinfo{publisher}{Springer International Publishing},
  \bibinfo{address}{Cham}, \bibinfo{pages}{172--192}.
\newblock
\showISBNx{978-3-319-17248-4}


\bibitem[\protect\citeauthoryear{{Mellanox}}{{Mellanox}}{[n.d.]}]%
        {ib}
\bibfield{author}{\bibinfo{person}{{Mellanox}}.}
  \bibinfo{year}{[n.d.]}\natexlab{}.
\newblock \bibinfo{booktitle}{\emph{Infiniband}}.
\newblock
\urldef\tempurl%
\url{https://wiki.archlinux.org/title/InfiniBand}
\showURL{%
\tempurl}


\bibitem[\protect\citeauthoryear{{Message Passing Interface Forum}}{{Message
  Passing Interface Forum}}{2009}]%
        {mpi_standard_2}
\bibfield{author}{\bibinfo{person}{{Message Passing Interface Forum}}.}
  \bibinfo{year}{2009}\natexlab{}.
\newblock \bibinfo{booktitle}{\emph{MPI: A Message-Passing Interface Standard
  Version 2.2}}.
\newblock
\urldef\tempurl%
\url{https://www.mpi-forum.org/docs/mpi-2.2/mpi22-report.pdf}
\showURL{%
\tempurl}


\bibitem[\protect\citeauthoryear{{Message Passing Interface Forum}}{{Message
  Passing Interface Forum}}{2015}]%
        {mpi_standard_31}
\bibfield{author}{\bibinfo{person}{{Message Passing Interface Forum}}.}
  \bibinfo{year}{2015}\natexlab{}.
\newblock \bibinfo{booktitle}{\emph{MPI: A Message-Passing Interface Standard
  Version 3.1}}.
\newblock
\urldef\tempurl%
\url{https://www.mpi-forum.org/docs/mpi-3.1/mpi31-report.pdf}
\showURL{%
\tempurl}


\bibitem[\protect\citeauthoryear{{Mozilla}}{{Mozilla}}{[n.d.]}]%
        {asmjs}
\bibfield{author}{\bibinfo{person}{{Mozilla}}.}
  \bibinfo{year}{[n.d.]}\natexlab{}.
\newblock \bibinfo{booktitle}{\emph{asm.js}}.
\newblock
\urldef\tempurl%
\url{http://asmjs.org/}
\showURL{%
\tempurl}


\bibitem[\protect\citeauthoryear{{Mozilla}}{{Mozilla}}{2021}]%
        {webassembly_text_format}
\bibfield{author}{\bibinfo{person}{{Mozilla}}.}
  \bibinfo{year}{2021}\natexlab{}.
\newblock \bibinfo{booktitle}{\emph{Understanding WebAssembly text format}}.
\newblock
\urldef\tempurl%
\url{https://developer.mozilla.org/en-US/docs/WebAssembly/Understanding_the_text_format}
\showURL{%
\tempurl}


\bibitem[\protect\citeauthoryear{{Open Container Initiative (OCI)}}{{Open
  Container Initiative (OCI)}}{[n.d.]}]%
        {oci}
\bibfield{author}{\bibinfo{person}{{Open Container Initiative (OCI)}}.}
  \bibinfo{year}{[n.d.]}\natexlab{}.
\newblock \bibinfo{howpublished}{\url{https://opencontainers.org/}}.
\newblock
\newblock
\shownote{Accessed 07/07/2022.}


\bibitem[\protect\citeauthoryear{{Podman.}}{{Podman.}}{[n.d.]}]%
        {podmanlimitations}
\bibfield{author}{\bibinfo{person}{{Podman.}}}
  \bibinfo{year}{[n.d.]}\natexlab{}.
\newblock \bibinfo{title}{Filesystem Considerations in Rootless Mode.}
\newblock \bibinfo{howpublished}{\url{https://tinyurl.com/2p8efntt}}.
\newblock
\newblock
\shownote{Accessed 09/27/2021.}


\bibitem[\protect\citeauthoryear{Priedhorsky, Canon, Randles, and
  Younge}{Priedhorsky et~al\mbox{.}}{2021}]%
        {minimizingpriv}
\bibfield{author}{\bibinfo{person}{Reid Priedhorsky}, \bibinfo{person}{R.~Shane
  Canon}, \bibinfo{person}{Timothy Randles}, {and} \bibinfo{person}{Andrew~J.
  Younge}.} \bibinfo{year}{2021}\natexlab{}.
\newblock \showarticletitle{Minimizing Privilege for Building HPC Containers}.
  In \bibinfo{booktitle}{\emph{Proceedings of the International Conference for
  High Performance Computing, Networking, Storage and Analysis}} (St. Louis,
  Missouri) \emph{(\bibinfo{series}{SC '21})}. \bibinfo{publisher}{Association
  for Computing Machinery}, \bibinfo{address}{New York, NY, USA}, Article
  \bibinfo{articleno}{32}, \bibinfo{numpages}{14}~pages.
\newblock
\showISBNx{9781450384421}
\urldef\tempurl%
\url{https://doi.org/10.1145/3458817.3476187}
\showDOI{\tempurl}


\bibitem[\protect\citeauthoryear{Priedhorsky and Randles}{Priedhorsky and
  Randles}{2017}]%
        {priedhorsky2017charliecloud}
\bibfield{author}{\bibinfo{person}{Reid Priedhorsky} {and} \bibinfo{person}{Tim
  Randles}.} \bibinfo{year}{2017}\natexlab{}.
\newblock \showarticletitle{Charliecloud: Unprivileged containers for
  user-defined software stacks in hpc}. In
  \bibinfo{booktitle}{\emph{Proceedings of the international conference for
  high performance computing, networking, storage and analysis}}.
  \bibinfo{pages}{1--10}.
\newblock
\urldef\tempurl%
\url{https://doi.org/10.1145/3126908.3126925}
\showDOI{\tempurl}


\bibitem[\protect\citeauthoryear{Report, Heroux, Dongarra, and Luszczek}{Report
  et~al\mbox{.}}{2013}]%
        {hpcg}
\bibfield{author}{\bibinfo{person}{S. Report}, \bibinfo{person}{M. Heroux},
  \bibinfo{person}{J. Dongarra}, {and} \bibinfo{person}{P. Luszczek}.}
  \bibinfo{year}{2013}\natexlab{}.
\newblock \showarticletitle{HPCG Technical Specification}.
\newblock


\bibitem[\protect\citeauthoryear{Rudyy, Garcia-Gasulla, Mantovani, Santiago,
  Sirvent, and V{\'a}zquez}{Rudyy et~al\mbox{.}}{2019}]%
        {rudyy2019containers}
\bibfield{author}{\bibinfo{person}{Oleksandr Rudyy}, \bibinfo{person}{Marta
  Garcia-Gasulla}, \bibinfo{person}{Filippo Mantovani},
  \bibinfo{person}{Alfonso Santiago}, \bibinfo{person}{Ra{\"u}l Sirvent}, {and}
  \bibinfo{person}{Mariano V{\'a}zquez}.} \bibinfo{year}{2019}\natexlab{}.
\newblock \showarticletitle{Containers in hpc: A scalability and portability
  study in production biological simulations}. In
  \bibinfo{booktitle}{\emph{2019 IEEE International Parallel and Distributed
  Processing Symposium (IPDPS)}}. IEEE, \bibinfo{pages}{567--577}.
\newblock
\urldef\tempurl%
\url{https://doi.org/10.1109/IPDPS.2019.00066}
\showDOI{\tempurl}


\bibitem[\protect\citeauthoryear{Ruhela, Harrell, Evans, Zynda, Fonner, Vaughn,
  Minyard, and Cazes}{Ruhela et~al\mbox{.}}{2021}]%
        {ruhela2021characterizing}
\bibfield{author}{\bibinfo{person}{Amit Ruhela}, \bibinfo{person}{Stephen~Lien
  Harrell}, \bibinfo{person}{Richard~Todd Evans}, \bibinfo{person}{Gregory~J
  Zynda}, \bibinfo{person}{John Fonner}, \bibinfo{person}{Matt Vaughn},
  \bibinfo{person}{Tommy Minyard}, {and} \bibinfo{person}{John Cazes}.}
  \bibinfo{year}{2021}\natexlab{}.
\newblock \showarticletitle{Characterizing Containerized HPC Applications
  Performance at Petascale on CPU and GPU Architectures}. In
  \bibinfo{booktitle}{\emph{International Conference on High Performance
  Computing}}. Springer, \bibinfo{pages}{411--430}.
\newblock
\urldef\tempurl%
\url{https://doi.org/10.1007/978-3-030-78713-4_22}
\showDOI{\tempurl}


\bibitem[\protect\citeauthoryear{Sch{\"o}ne, Ilsche, Bielert, Gocht, and
  Hackenberg}{Sch{\"o}ne et~al\mbox{.}}{2019}]%
        {schone2019energy}
\bibfield{author}{\bibinfo{person}{Robert Sch{\"o}ne}, \bibinfo{person}{Thomas
  Ilsche}, \bibinfo{person}{Mario Bielert}, \bibinfo{person}{Andreas Gocht},
  {and} \bibinfo{person}{Daniel Hackenberg}.} \bibinfo{year}{2019}\natexlab{}.
\newblock \showarticletitle{Energy efficiency features of the Intel Skylake-SP
  processor and their impact on performance}. In \bibinfo{booktitle}{\emph{2019
  International Conference on High Performance Computing \& Simulation
  (HPCS)}}. IEEE, \bibinfo{pages}{399--406}.
\newblock
\urldef\tempurl%
\url{https://doi.org/10.1109/HPCS48598.2019.9188239}
\showDOI{\tempurl}


\bibitem[\protect\citeauthoryear{Schwinge}{Schwinge}{2019}]%
        {gcc_wasm}
\bibfield{author}{\bibinfo{person}{Thomas Schwinge}.}
  \bibinfo{year}{2019}\natexlab{}.
\newblock \bibinfo{booktitle}{\emph{GCC Mailing List: Re: More compatibility
  it's possible?}}
\newblock
\urldef\tempurl%
\url{https://gcc.gnu.org/legacy-ml/gcc/2019-11/msg00095.html}
\showURL{%
\tempurl}


\bibitem[\protect\citeauthoryear{Shillaker and Pietzuch}{Shillaker and
  Pietzuch}{2020}]%
        {faasm}
\bibfield{author}{\bibinfo{person}{Simon Shillaker} {and}
  \bibinfo{person}{Peter Pietzuch}.} \bibinfo{year}{2020}\natexlab{}.
\newblock \showarticletitle{Faasm: Lightweight Isolation for Efficient Stateful
  Serverless Computing}. In \bibinfo{booktitle}{\emph{2020 {USENIX} Annual
  Technical Conference ({USENIX} {ATC} 20)}}. \bibinfo{publisher}{{USENIX}
  Association}, \bibinfo{pages}{419--433}.
\newblock
\showISBNx{978-1-939133-14-4}
\urldef\tempurl%
\url{https://www.usenix.org/conference/atc20/presentation/shillaker}
\showURL{%
\tempurl}


\bibitem[\protect\citeauthoryear{Shymanskyy and Massey}{Shymanskyy and
  Massey}{[n.d.]}]%
        {wasm3}
\bibfield{author}{\bibinfo{person}{Volodymyr Shymanskyy} {and}
  \bibinfo{person}{Steven Massey}.} \bibinfo{year}{[n.d.]}\natexlab{}.
\newblock \bibinfo{booktitle}{\emph{Wasm3}}.
\newblock
\urldef\tempurl%
\url{https://github.com/wasm3/wasm3}
\showURL{%
\tempurl}


\bibitem[\protect\citeauthoryear{{Singularity.}}{{Singularity.}}{[n.d.]}]%
        {singularitylimitatsion}
\bibfield{author}{\bibinfo{person}{{Singularity.}}}
  \bibinfo{year}{[n.d.]}\natexlab{}.
\newblock \bibinfo{title}{Filesystem Considerations in Rootless Mode.}
\newblock
  \bibinfo{howpublished}{\url{https://apptainer.org/admin-docs/master/installation.html}}.
\newblock
\newblock
\shownote{Accessed 09/27/2021.}


\bibitem[\protect\citeauthoryear{{Solomon Hykes}}{{Solomon Hykes}}{[n.d.]}]%
        {alttocontainers}
\bibfield{author}{\bibinfo{person}{{Solomon Hykes}}.}
  \bibinfo{year}{[n.d.]}\natexlab{}.
\newblock \bibinfo{title}{Wasm+WASI: An alternative to Linux Containers}.
\newblock
  \bibinfo{howpublished}{\url{https://twitter.com/solomonstre/status/1111004913222324225?lang=en}}.
\newblock
\newblock
\shownote{Accessed 09/27/2021.}


\bibitem[\protect\citeauthoryear{Sudwoj}{Sudwoj}{2020}]%
        {sudwoj2020rust}
\bibfield{author}{\bibinfo{person}{Michal Sudwoj}.}
  \bibinfo{year}{2020}\natexlab{}.
\newblock \emph{\bibinfo{title}{Rust programming language in the
  high-performance computing environment}}.
\newblock {B.S.} thesis. \bibinfo{school}{ETH Zurich}.
\newblock
\urldef\tempurl%
\url{https://www.research-collection.ethz.ch/handle/20.500.11850/474922}
\showURL{%
\tempurl}


\bibitem[\protect\citeauthoryear{{The kernel development community}}{{The
  kernel development community}}{[n.d.]}]%
        {seccomp_bpf}
\bibfield{author}{\bibinfo{person}{{The kernel development community}}.}
  \bibinfo{year}{[n.d.]}\natexlab{}.
\newblock \bibinfo{booktitle}{\emph{Seccomp BPF (SECure COMPuting with
  filters)}}.
\newblock
\urldef\tempurl%
\url{https://www.kernel.org/doc/html/v4.16/userspace-api/seccomp_filter.html}
\showURL{%
\tempurl}


\bibitem[\protect\citeauthoryear{Torrez, Randles, and Priedhorsky}{Torrez
  et~al\mbox{.}}{2019}]%
        {containerhpcnoperf}
\bibfield{author}{\bibinfo{person}{Alfred Torrez}, \bibinfo{person}{Timothy
  Randles}, {and} \bibinfo{person}{Reid Priedhorsky}.}
  \bibinfo{year}{2019}\natexlab{}.
\newblock \showarticletitle{HPC Container Runtimes have Minimal or No
  Performance Impact}. In \bibinfo{booktitle}{\emph{2019 IEEE/ACM International
  Workshop on Containers and New Orchestration Paradigms for Isolated
  Environments in HPC (CANOPIE-HPC)}}. \bibinfo{pages}{37--42}.
\newblock
\urldef\tempurl%
\url{https://doi.org/10.1109/CANOPIE-HPC49598.2019.00010}
\showDOI{\tempurl}


\bibitem[\protect\citeauthoryear{Wahbe, Lucco, Anderson, and Graham}{Wahbe
  et~al\mbox{.}}{1993}]%
        {sfi}
\bibfield{author}{\bibinfo{person}{Robert Wahbe}, \bibinfo{person}{Steven
  Lucco}, \bibinfo{person}{Thomas~E. Anderson}, {and} \bibinfo{person}{Susan~L.
  Graham}.} \bibinfo{year}{1993}\natexlab{}.
\newblock \showarticletitle{Efficient Software-Based Fault Isolation}.
\newblock \bibinfo{journal}{\emph{SIGOPS Oper. Syst. Rev.}}
  \bibinfo{volume}{27}, \bibinfo{number}{5} (\bibinfo{date}{dec}
  \bibinfo{year}{1993}), \bibinfo{pages}{203–216}.
\newblock
\showISSN{0163-5980}
\urldef\tempurl%
\url{https://doi.org/10.1145/173668.168635}
\showDOI{\tempurl}


\bibitem[\protect\citeauthoryear{{Wasmer, Inc.}}{{Wasmer, Inc.}}{[n.d.]a}]%
        {singlepass}
\bibfield{author}{\bibinfo{person}{{Wasmer, Inc.}}}
  \bibinfo{year}{[n.d.]}\natexlab{a}.
\newblock \bibinfo{booktitle}{\emph{Single Pass Compiler in Wasmer}}.
\newblock
\urldef\tempurl%
\url{https://docs.rs/wasmer-compiler-singlepass/latest/wasmer_compiler_singlepass/}
\showURL{%
\tempurl}


\bibitem[\protect\citeauthoryear{{Wasmer, Inc.}}{{Wasmer, Inc.}}{[n.d.]b}]%
        {wasmer}
\bibfield{author}{\bibinfo{person}{{Wasmer, Inc.}}}
  \bibinfo{year}{[n.d.]}\natexlab{b}.
\newblock \bibinfo{booktitle}{\emph{Universal WebAssembly Runtime}}.
\newblock
\urldef\tempurl%
\url{https://wasmer.io/}
\showURL{%
\tempurl}


\bibitem[\protect\citeauthoryear{{Wasmer, Inc.}}{{Wasmer, Inc.}}{[n.d.]c}]%
        {wapm}
\bibfield{author}{\bibinfo{person}{{Wasmer, Inc.}}}
  \bibinfo{year}{[n.d.]}\natexlab{c}.
\newblock \bibinfo{booktitle}{\emph{wapm is the WebAssembly Package Manager}}.
\newblock
\urldef\tempurl%
\url{https://wapm.io}
\showURL{%
\tempurl}


\bibitem[\protect\citeauthoryear{{WebAssembly Community Group}}{{WebAssembly
  Community Group}}{2021}]%
        {wasi_sdk}
\bibfield{author}{\bibinfo{person}{{WebAssembly Community Group}}.}
  \bibinfo{year}{2021}\natexlab{}.
\newblock \bibinfo{booktitle}{\emph{WebAssembly System Interface}}.
\newblock
\urldef\tempurl%
\url{https://github.com/WebAssembly/WASI}
\showURL{%
\tempurl}


\bibitem[\protect\citeauthoryear{Xavier, Neves, Rossi, Ferreto, Lange, and
  De~Rose}{Xavier et~al\mbox{.}}{2013}]%
        {xavier2013performance}
\bibfield{author}{\bibinfo{person}{Miguel~G Xavier}, \bibinfo{person}{Marcelo~V
  Neves}, \bibinfo{person}{Fabio~D Rossi}, \bibinfo{person}{Tiago~C Ferreto},
  \bibinfo{person}{Timoteo Lange}, {and} \bibinfo{person}{Cesar~AF De~Rose}.}
  \bibinfo{year}{2013}\natexlab{}.
\newblock \showarticletitle{Performance evaluation of container-based
  virtualization for high performance computing environments}. In
  \bibinfo{booktitle}{\emph{2013 21st Euromicro International Conference on
  Parallel, Distributed, and Network-Based Processing}}. IEEE,
  \bibinfo{pages}{233--240}.
\newblock
\urldef\tempurl%
\url{https://doi.org/10.1109/PDP.2013.41}
\showDOI{\tempurl}


\bibitem[\protect\citeauthoryear{Yan, Tu, Zhao, Zhou, and Wang}{Yan
  et~al\mbox{.}}{2021}]%
        {understandingperf}
\bibfield{author}{\bibinfo{person}{Yutian Yan}, \bibinfo{person}{Tengfei Tu},
  \bibinfo{person}{Lijian Zhao}, \bibinfo{person}{Yuchen Zhou}, {and}
  \bibinfo{person}{Weihang Wang}.} \bibinfo{year}{2021}\natexlab{}.
\newblock \showarticletitle{Understanding the Performance of Webassembly
  Applications}. In \bibinfo{booktitle}{\emph{Proceedings of the 21st ACM
  Internet Measurement Conference}} (Virtual Event) \emph{(\bibinfo{series}{IMC
  '21})}. \bibinfo{publisher}{Association for Computing Machinery},
  \bibinfo{address}{New York, NY, USA}, \bibinfo{pages}{533–549}.
\newblock
\showISBNx{9781450391290}
\urldef\tempurl%
\url{https://doi.org/10.1145/3487552.3487827}
\showDOI{\tempurl}


\bibitem[\protect\citeauthoryear{Yoo, Jette, and Grondona}{Yoo
  et~al\mbox{.}}{2003}]%
        {yoo2003slurm}
\bibfield{author}{\bibinfo{person}{Andy~B Yoo}, \bibinfo{person}{Morris~A
  Jette}, {and} \bibinfo{person}{Mark Grondona}.}
  \bibinfo{year}{2003}\natexlab{}.
\newblock \showarticletitle{Slurm: Simple linux utility for resource
  management}. In \bibinfo{booktitle}{\emph{Workshop on Job Scheduling
  Strategies for Parallel Processing}}. Springer, \bibinfo{pages}{44--60}.
\newblock


\end{thebibliography}

\appendix

\section{Artifact Appendix}

\subsection{Description}
\emph{MPIWasm} is an embedder for MPI-based HPC applications based on \texttt{Wasmer}~\cite{wasmer}. It enables the high performance execution of these applications compiled to WebAssembly (Wasm) and serves two purposes:
\begin{enumerate}
    \item Delivering close to native application performance, i.e., when applications are executed directly on a host machine without using Wasm. 
    \item Enabling the distribution of MPI-based HPC applications as Wasm binaries.
\end{enumerate}
Our artifact contains the source code for \emph{MPIWasm}, toolchain for compiling C/C++ based MPI applications to Wasm, the Wasm binaries for the different standardized HPC benchmarks used in this paper, pre-built versions of our embedder for different operating systems, and scripts for parsing experiment data and generating plots. The artifact is available at:
\\
\centerline{\url{https://doi.org/10.5281/zenodo.7468121}}
\\
\centerline{or}
\\
\centerline{\url{https://github.com/kky-fury/MPIWasm}}

\subsection{Getting Started}
For testing our Wasm embedder for executing MPI applications compiled to WebAssembly, we provide a pre-built docker image for the \texttt{linux/amd64} platform with all the required dependencies. 
\begin{lstlisting}[caption={
      Getting started with \emph{MPIWasm}.
}, captionpos=b, basicstyle=\tiny,  frame=single, language=bash, framexrightmargin=-1cm, xrightmargin=-1cm, 
   label={lst:startingwithwasm}]
sudo docker run -it kkyfury/ppoppae:v2 /bin/bash
#Executing the HPCG benchmark compiled to Wasm
mpirun --allow-run-as-root -np 4 ./target/release/embedder \
    examples/xhpcg.wasm
#Executing the IntelMPI benchmarks compiled to Wasm
mpirun --allow-run-as-root -np 4 ./target/release/embedder \ 
    examples/imb.wasm 
\end{lstlisting}
Towards this, a user can follow the steps described in Listing~\ref{lst:startingwithwasm}. Following this, \emph{MPIWasm} should successfully execute the \texttt{HPCG} and \texttt{IntelMPI} benchmarks. We provide sample output for the two benchmarks in the provided artifact. The user can increase/decrease the number of processes (\texttt{-np}) for executing the benchmarks. However, depending on the system where the container is executing, the user might need to provide the \texttt{-oversubscribe} flag to \texttt{mpirun}. 

\subsection{Running Experiments with \emph{MPIWasm}}
This section describes how to run experiments with our embedder to obtain plots similar to the ones in this paper.

\subsubsection{Running small-scale experiments}
To run small-scale experiments inside the docker container, we provide an end-to-end script. This script:
\begin{enumerate}
    \item Executes the \texttt{HPCG}, \texttt{IS}, and \texttt{IntelMPI} benchmarks for their native execution and when they are executed using \emph{MPIWasm}. 
    \item Parses the obtained results and generates the relevant plots.
\end{enumerate}

\begin{lstlisting}[caption={
      Running small-scale experiments with \emph{MPIWasm}.
}, captionpos=b, basicstyle=\tiny,  frame=single, language=bash, framexrightmargin=-1cm, xrightmargin=-1cm, 
   label={lst:smallexperiments}]
sudo docker run -it kkyfury/ppoppae:v2 /bin/bash
 cd run_experiments
    ./runme.sh
\end{lstlisting}
For running the experiments, the user can follow the steps described in Listing~\ref{lst:smallexperiments}. The script can take around 10-15 minutes to finish execution. After completion, you can see the generated data in the \textit{run\_experiments/experiment\_data} folder. The generated plots can be found in the \textit{run\_experiments/Plots} folder. We provide sample plots for the different benchmarks in our artifact. However, on executing benchmarks inside the container, the performance difference between the native execution of the application and \emph{MPIWasm} can be around 8-12\%.

\subsubsection{Running large-scale experiments on an HPC system}
For running large-scale experiments with our embedder, a user needs to do the following:

\begin{enumerate}
    \item Build a version of the embedder for your HPC system depending on the particular architecture, operating system, and the MPI library on the system. MPIWasm currently supports the \texttt{OpenMPI} library with limited support for \texttt{MPICH} and \texttt{MVAPICH}. We provide examples for building \emph{MPIWasm} for different operating systems and architectures in our artifact.
    \item Execute the MPI applications using the built embedder on the HPC system. This can be done via submitting jobs to a RJMS software on an HPC system such as \texttt{SLURM}~\cite{yoo2003slurm}. We provide sample job scripts for our HPC system, i.e., SuperMUC-NG that uses SLURM in our artifact.
    \item After executing the applications, the user can utilize the different parsers provided in our artifact to parse the benchmark data. Following this, the results can be visualized using the plotting helper provided in the artifact.
\end{enumerate}


\subsection{Compiling C/C++ applications to Wasm}
We have setup a docker container with the required dependencies for compiling different MPI applications conformant to the MPI-2.2 standard to Wasm. The artifact also includes \texttt{HPCG}, \text{IntelMPI}, and \texttt{IS} benchmarks as examples.

\begin{lstlisting}[caption={
      Compiling applications to Wasm.
}, captionpos=b, basicstyle=\tiny,  frame=single, language=bash, framexrightmargin=-1cm, xrightmargin=-1cm, 
   label={lst:compilingtowasm}]
sudo docker run -it kkyfury/wasitoolchain:v1 /bin/bash
#Compiling HPCG
cd /work/example/hpcg-benchmark
./wasi-cmake.sh
cd cmake-build-wasi
make
\end{lstlisting}
Listing~\ref{lst:compilingtowasm} describes the steps a user can follow for compiling the \texttt{HPCG} benchmark to Wasm. For steps to compile the other benchmarks to Wasm, please look at the base \texttt{Readme.md} file provided with the artifact. All the different applications compiled to Wasm that we used in this paper are also present in the artifact.

\subsection{Using \emph{MPIWasm}}
For detailed usage instructions, please look at the base \texttt{Readme.md} file provided with the artifact.

\subsection{Modifying \emph{MPIWasm}}
\label{sec:modifyingmpiwasm}
For modifying our embedder, we recommend using our provided \texttt{docker-compose} file in the artifact. This docker-compose file mounts the volume with the embedder's source code inside the container. As a result, any changes to it's source code will be reflected inside it. For our embedder, we currently support the following operating systems:

\begin{enumerate}
    \item \texttt{CentOS-8.2}
    \item \texttt{Opensuse-15-1}
    \item \texttt{Ubuntu-20-04}
    \item \texttt{MacOS-monterey}
\end{enumerate}

\begin{lstlisting}[caption={
      Building \emph{MPIWasm}.
}, captionpos=b, basicstyle=\tiny,  frame=single, language=bash, framexrightmargin=-1cm, xrightmargin=-1cm, 
   label={lst:buildingMPIwasm}]
cd wasi-mpi-rs
docker compose run centos-8-2
cargo build --release
#After the build process, you can see the built embedder in 
#    the /target/release/ folder.
\end{lstlisting}

Listing~\ref{lst:buildingMPIwasm} describes the steps for building \emph{MPIWasm} for \texttt{CentOS-8.2} after  modifications. The instructions for building the embedder for other operating systems are provided in the base \texttt{Readme.md} file inside the artifact. After the build process, the embedder can be copied to the user's local filesystem using the \texttt{docker-cp} command as shown in Listing~\ref{lst:copyingMPIwasm}.

\begin{lstlisting}[caption={
      Copying \emph{MPIWasm}.
}, captionpos=b, basicstyle=\tiny,  frame=single, language=bash, framexrightmargin=-1cm, xrightmargin=-1cm, 
   label={lst:copyingMPIwasm}]
docker cp <container-id>:/s/target/release/embedder \
    <destination-path-user-filesystem>
\end{lstlisting}

We provide provide the base image dockerfiles for the different supported operating systems inside the artifact. These example dockerfiles can be easily extended to support other different linux distributions.

\subsection{Support for \texttt{aarch64}}
Our embedder also supports execution on \texttt{linux/arm64} platforms. We provide pre-built versions of our embedder for \texttt{arm64} for the different supported operating systems in the artifact.

\subsubsection{Building images for \texttt{aarch64}}
If the user is building the docker image on an \texttt{x86\_64} system then \texttt{docker-buildx} is required. Note that, in this case, building the image might take around 12 hours. 

\begin{lstlisting}[caption={
      Building \emph{MPIWasm} for \texttt{aarch64} on \texttt{x86\_64}.
}, captionpos=b, basicstyle=\tiny,  frame=single, language=bash, framexrightmargin=-1cm, xrightmargin=-1cm, 
   label={lst:buildingMPIwasmarm}]
sudo docker buildx create --name mybuilder --use --bootstrap
cd wasi-mpi-rs/.gitlab/ci/images/
sudo docker buildx build --push -f ubuntu-20-04.Dockerfile \
        --platform linux/arm64 \ 
        -t kkyfury/ubuntumodifiedbase:v1 .
cd ../../../
sudo docker buildx build --push -f Dockerfile \
        --platform linux/arm64 \ 
        -t kkyfury/embedderarm:v1 .
\end{lstlisting}
Listing~\ref{lst:buildingMPIwasmarm} describes the steps required for building \emph{MPIWasm} for arm systems with \texttt{docker-buildx}. The user should change the docker image tags according to their docker registry account, i.e., replace kkyfury with your registry username.
On the other hand, if the user is using an \texttt{aardch64} system then please follow the instructions described in~\S\ref{sec:modifyingmpiwasm}.

\end{document}